\documentclass[useAMS,usenatbib]{mnras}
\usepackage{graphicx}
\usepackage{epstopdf}
\usepackage{subcaption}
\usepackage{caption}
\usepackage{amsmath}
\usepackage{amssymb}
\usepackage{gensymb}
\usepackage[normal]{threeparttable}
\usepackage{calc}
\usepackage{pdflscape}
\usepackage{rotating}
\usepackage{afterpage}
\usepackage{color}

\title[H{\sc i} absorption in nearby compact radio galaxies]{H{\sc i} absorption in nearby compact radio galaxies}
\author[]{M. Glowacki$^{1,2,3}$\thanks{E-mail:
marcin@physics.usyd.edu.au}, J. R. Allison$^{2,3}$, E. M. Sadler$^{1,2}$, V. A. Moss$^{1,2}$, S. J. Curran$^{4}$, \newauthor A. Musaeva$^{1}$, C. Deng$^{1}$, R. Parry$^{1}$, M. C. Sligo$^{1}$ \\
$^{1}$Sydney Institute for Astronomy, School of Physics A28, University of Sydney, NSW 2006, Australia\\
$^{2}$ARC Centre of Excellence for All-sky Astrophysics (CAASTRO), Australia\\
$^{3}$CSIRO Astronomy \& Space Science, PO Box 76, Epping NSW 1710, Australia\\
$^{4}$School of Chemical and Physical Sciences, Victoria University of Wellington, Wellington 6140, New Zealand}

\begin{document}


\pagerange{\pageref{firstpage}--\pageref{lastpage}} \pubyear{2002}

\maketitle

\label{firstpage}

\begin{abstract}
H{\sc i} absorption studies yield information on both AGN feeding and feedback processes. This AGN activity interacts with the neutral gas in compact radio sources, which are believed to represent the young or recently re-triggered AGN population. We present the results of a survey for H{\sc i} absorption in a sample of 66 compact radio sources at 0.040~$<$~z~$<$~0.096 with the Australia Telescope Compact Array. In total, we obtained seven detections, five of which are new, with a large range of peak optical depths (3\%~to~87\%). Of the detections, 71\% exhibit asymmetric, broad ($\Delta{v}_{\mathrm{FWHM}}$~$>$~100~km~s$^{-1}$) features, indicative of disturbed gas kinematics. Such broad, shallow and offset features are also found within low-excitation radio galaxies which is attributed to disturbed circumnuclear gas, consistent with early-type galaxies typically devoid of a gas-rich disk. Comparing mid-infrared colours of our galaxies with H{\sc i} detections indicates that narrow and deep absorption features are preferentially found in late-type and high-excitation radio galaxies in our sample. These features are attributed to gas in galactic disks. By combining XMM-Newton archival data with 21-cm data, we find support that absorbed X-ray sources may be good tracers of HI content within the host galaxy. This sample extends previous H{\sc i} surveys in compact radio galaxies to lower radio luminosities and provides a basis for future work exploring the higher redshift universe.
\end{abstract}


\section{Introduction}

Compact radio galaxies are thought to represent the youngest or most recently triggered active galactic nuclei \cite[AGN][]{Fanti1995,Readhead1996,Owsianik1998}. Studies are ongoing on the significant effects AGN feedback have on the host galaxy, such as whether this activity triggers or suppresses star-formation, and what its role is in galaxy evolution \cite[see review by][]{Heckman2014}. This can be examined via the kinematics of circumnuclear neutral gas, sometimes exhibiting fast ($>$~1000~km~s$^{-1}$) and massive (up to 50~M$_{\bigodot}$~yr$^{-1}$) outflows created by interactions from radio-loud AGN \cite[e.g.][]{Morganti2005}.

The 21-cm line of atomic hydrogen (HI) is a useful tool for tracing the neutral gas within galaxies in emission to study the diffuse neutral gas in galaxies \citep[see][and references therein]{Giovanelli2016}. However, due to flux limitations, it is currently difficult to detect H{\sc i} in emission beyond redshifts of z $>$ 0.2 \cite[e.g.][]{Verheijen2007, Catinella2008, Freudling2011}. 

There is no such redshift limit for detecting the transition in absorption, as the line strength only depends on the brightness of the background source, and the spin temperature and column density of H{\sc i} along the line of sight. Therefore, H{\sc i} can be detected towards sufficiently strong continuum radio sources regardless of gas redshift \cite[e.g.][]{Morganti2005}. Absorption from H{\sc i} gas located in front of an AGN is a good tracer for the kinematics of gas, and hence H{\sc i} associated absorption surveys allow us to investigate how AGN feedback directly interacts with the clumpy medium in the host galaxy in compact radio sources \cite[e.g.][]{Wagner2012}.

It was suggested that intrinsically compact sources - primarily, gigahertz peaked spectrum (GPS) ($\leq$~1~kpc in size) and compact steep-spectrum (CSS) radio sources ($\sim$1 - 10 kpc) - have higher detection rates of H{\sc i} than extended sources \cite[e.g. with detection rates of $\sim$20 - 40\% in][]{Vermeulen2003, Gupta2006, Chandola2011}. Furthermore, \cite{Curran2016b} shows that the H{\sc i} detection rate is proportional to the turnover frequency, which is inversely proportional to source size. \cite{Morganti2001} searched for H{\sc i} in 23 nearby (z $<$ 0.22) radio galaxies and found a detection rate of H{\sc i} of 23\% in the compact sources, higher than that of a subset of extended Fanaroff-Riley type I and II (FR-I, FR-II) sources (10\%). \cite*{Pihlstrom2003} also suggest an anti-correlation between the H{\sc i} column density and source size. \cite{Curran2013} attributed this to a source size anti-correlation with optical depth, proposing this is due to an effect of geometry with a larger covering factor for smaller source sizes. 

These studies reveal evidence for disturbed neutral gas kinematics near young AGN. \cite{Gereb2015} found that asymmetric, broad absorption features appear more often in compact sources, suggesting a greater interplay between the AGN feedback of the host galaxy and the neutral gas environment. Such interplay, where the ultra-violet luminosity of the active nucleus has a direct impact on the neutral gas in the host, has been suggested by \citet{cww+08}, who first noted a decrease in the detection of associated H{\sc i} absorption with redshift. They attributed this to the flux limitations of optical surveys, biasing towards increasingly UV luminous sources \cite[further demonstrated by][]{cw10}. The detection rate reaches zero at ionising photon rates of $Q_\text{H{\sc i}}\sim3\times10^{56}$~sec$^{-1}$ ($L_{\rm UV}\sim10^{23}$ W Hz$^{-1}$ at $\lambda = 912$~\AA), at which point all of the neutral gas in the host galaxy is believed to be ionised \citep{Curran2012}. This critical UV luminosity limit has been observed several times since,
through the non-detection of H{\sc i} above this value \citep{cwm+10,cwsb12,cwt+12,caw+16,gd11,Allison2012,gmmo14,Aditya2016}.
 
Here we present the results of the search of H{\sc i} absorption complementrary to \cite{Allison2012}, who searched for H{\sc i} in 29 targets as part of a sample of 66 nearby (0.04~$<$~z~$<$~0.08) compact sources selected from the Australia Telescope 20 GHz survey \cite[AT20G;][]{Murphy2010}, with the Australian Telescope Compact Array (ATCA) on the Broadband Backend (CABB). A total of three detections were made by \cite{Allison2012}, two of which were previously unknown. The sample explores the H{\sc i} content within nearby compact radio galaxies with young or newly triggered AGN. We also present our analysis made through consideration of infrared and X-ray properties. 

This work also involved the development of a data reduction pipeline for wide-band data, and an automated spectral-line finding method based on Bayesian inference. These tools are part of preparation for the First Large Absorption Survey in H{\sc i} (FLASH) with the Australian SKA Pathfinder \cite[ASKAP;][]{Deboer2009, Johnston2009, Schinckel2012}, and has been successfully employed in the ongoing ASKAP-FLASH commissioning survey \cite[e.g.][]{Allison2015}. FLASH's redshift regime of 0.4 $<$ z $<$ 1.0 will explore both associated and intervening (within another galaxy along the line of sight) H{\sc i} absorption, and allow us to extensively examine how the H{\sc i} content has evolved with redshift in a largely unexplored epoch. 

In this paper the standard cosmological model is used, with parameters $\Omega_\mathrm{M} = 0.27$, $\Omega_{\Lambda} = 0.73$ and $\mathrm{H}_{0}~=~71$\,km\,s$^{-1}$\,Mpc$^{-1}$. All uncertainties refer to the 68.3\,per\,cent confidence interval, unless otherwise stated.

\section{Sample Selection, Observations and Data Reduction}

\subsection{Observations}

\subsubsection{Sample Selection}

To study the H{\sc i} content of the inner circumnuclear regions of galaxies, we chose radio galaxies known to have a bright, compact core. Targets were selected at 20 GHz from the AT20G survey catalogue \citep{Murphy2010}, which were cross-matched with the 6dF Galaxy Survey \citep{Jones2009} to obtain redshift information \citep{Sadler2014}. In order to select compact sources, those with extended radio emission on scales greater than 15 arcseconds at 20 GHz were removed. Following this, the sample was restricted to 45 sources south of declination $\delta~=~-20\degree$, within the redshift range of 0.04 $<$ z $<$ 0.08. This range was chosen to search for the 21-cm H{\sc i} line within a convenient and available frequency range of the ATCA, and to avoid contaminating the spectra with H{\sc i} emission lines by probing sufficiently nearby galaxies. The instruments also imposed a minimum flux density limit of 50 mJy at 1.4~GHz (as absorption line strength depends on the brightness of the background source).

\begin{table*}
  \caption{Summary of observations, including those made by \citet{Allison2012}. In order of column, we list the observing group ID (obs. ID), observing dates, the ATCA array configuration, central frequency of the zoom channel, average integration time per source in hours, the number of observed sources, the average 1-$\sigma$ noise in the channels, and notes. }
  \begin{tabular}{lllccccl}
  \hline
 Obs. ID & Dates & Array config & $\rm{\nu}$ & t$_{\rm{int}}$ & n & $\sigma_\mathrm{chan}$ & Notes\\
& & & MHz & hr & & mJy/beam &\\
 \hline
A & 2011 Feb 03-06 & 6A & 1342 & 3.5 & 14 & 4.0 & Published by \citet{Allison2012}.\\
B & 2011 Apr 23-26 & 6A & 1342 & 3.5 & 15 & 4.0 & Published by \citet{Allison2012}.\\
C & 2013 Feb 13-16 & 6A & 1342 & 4.0 & 14 & 6.3 & Carried out in 3 contiguous 24 hour periods.\\
D & 2013 June 08-10 & 6C & 1310 & 1.5 & 9 & 7.4 & Made across three separate periods of 6--10 hours.\\
E & 2014 Sept 25-28 & 6A & 1340 & 4.0 & 14 & 3.4 & Carried out in 3 contiguous 24 hour periods.\\
\hline
\end{tabular}
\label{table:observing}
\end{table*}

The results from observations of 29 sources were reported by \cite{Allison2012}. Two sources within our sample (J062706--352916 and J164416--771548) were previously observed by \cite{Morganti2001} to optical depths upper limits of $\tau~<$~4.7\%, and so were not re-observed by us. These observations have since been extended by three separate observation runs made with the ATCA in order to expand to the sample:
\begin{itemize}  
\item{14 targets in February 2013 with the same selection criteria as that by \cite{Allison2012}. Of these targets, the source J091856--243829 was included for an intervening H{\sc i} search, but while we give its spectrum, it is not part of our sample.}
\item{9 targets in June 2013, in a higher redshift range (0.080~$<$~z~$<$~0.096). This redshift range was within the receiver limits and avoided specific radio frequency interference (RFI).}
\item{14 targets in September 2014. These were part of a separate sample of 68 compact GPS sources imaged with the Very Long Baseline Array (VLBA). They had also been selected from the AT20G survey with redshifts identified in the 6dF Galaxy Survey. These targets had additional constraints of declinations of $\delta >$ -40$\degree$, redshifts of 0.04~$<$~z~$<$~0.08, a minimum flux density of 50 mJy at 1.4 GHz, and an angular size of 0.2 arcseconds or less at 20 GHz.}
\end{itemize}

Several of the VLBA-imaged sources had been detected in the continuum with the Australian Long Baseline Array (LBA) by \cite{Hancock2009} at 4.8 GHz, where all targets were observed but unresolved on scales of $\sim$100 mas, implying linear sizes of less than 100 pc. The VLBA imaging spatial resolution is 0.5 mas, and hence will allow for a comprehensive study of the radio source morphology on parsec scales.

In total our sample consisted of 66 targets which are detailed in Table~2. A short description of each of the columns follows.

\medskip

\noindent  \textit{Column (1)}: The AT20G source name.

\noindent  \textit{Columns (2) \& (3)}: The right ascension and declination of the source at J2000.
  
\noindent  \textit{Column (4)}: The AT20G 20\,GHz flux density \citep{Murphy2010}.

\noindent  \textit{Column (5)}: $S_{1.4}$ is the NVSS 1.4\,GHz flux density \citep{Condon1998}.
 
\noindent  \textit{Column (6)}: $S_{0.843}$ is the SUMSS 843\,MHz flux density \citep{Mauch2003}.
 
\noindent  \textit{Columns (7) \& (8)}: The corresponding spectral indices between the two lower-frequencies and 20\,GHz are given by $\alpha_{1.4}^{20}$ and $\alpha_{0.843}^{20}$ respectively.
 
\noindent  \textit{Column (9)}: Observation ID. Refer to Table 1 for the details of each run.

\noindent  \textit{Column (10)}: $z_\mathrm{opt}$ is the optical spectroscopic redshift.

\noindent  \textit{Column (11)}: SC is the optical spectroscopic classification as defined by \cite{Mahony2011}. Aa means only absorption lines were detected, Ae corresponds to only emission lines, AeB broad emission lines, and Aae means both were observed.

\noindent   \textit{Column (12)}: $\mathrm{M}_{K}$ is the Two Micron All Sky Survey (2MASS) $K$-band absolute magnitude \citep{Skrutskie2006}.

\noindent   \textit{Column (13)}: The 20 GHz compactness, $6\mathrm{km}_\mathrm{vis}$, defined as the ratio of the measured visibility amplitude on the longest ATCA baselines to that on the shortest baselines \citep{Chhetri2013}. They find a lower limit of 0.86 for unresolved sources. 
  
\noindent   \textit{Column (14)}: FR is the Fanaroff-Riley type at low frequency (1.4 GHz) \citep{Fanaroff1974,Baldi2009,Ghisellini2011}.

\medskip

\begin{figure}
\includegraphics[width=1.0\linewidth]{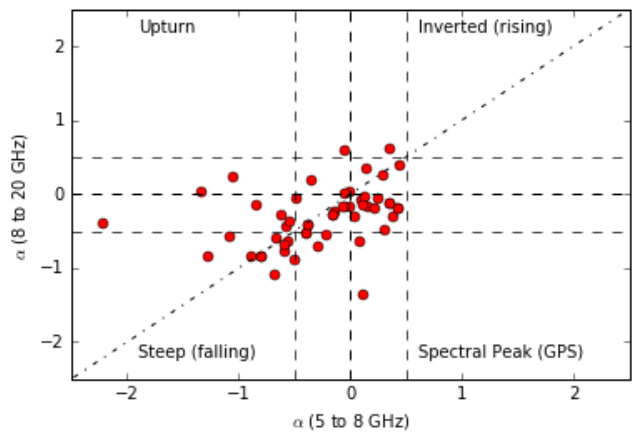}
\caption{Spectral indices for the 49 sources in our samples with measurements at 20 GHz, 8.4 GHz and 4.8 GHz. All but nine fall in a flat-spectrum regime in either axis (spectral indices of -0.5 $< \alpha <$ 0.5), with 26 flat in both indices. The majority are found to have either a decreasing spectral energy distribution with frequency (steep/falling) or with a spectral peak between 4.8 and 20 GHz (GPS).}
\label{figure:SED}
\end{figure}

\subsubsection{Spectral indices of radio sources}

Instead of selecting by spectral index, our survey targeted compact radio galaxies selected at 20 GHz, including both steep and flat-spectrum sources, and so our source selection differed from similar surveys, such as \cite{Gupta2006} and \cite{Chandola2011}, which targeted GPS and CSS radio sources. Of the full sample, 38 (58\%) are classified as flat-spectrum sources (-0.5 $< \alpha <$ 0.5 between 1.4/0.843 GHz and 20 GHz; Table 2).  

We examined additional flux density measurements to confirm how many of our compact core-selected sources are classified as flat-spectrum. Fig.~1 shows the spectral indices from the AT20G survey \citep{Murphy2010} for 49 of our sources with flux measurements at 20 GHz, 8.4 GHz and 4.8~GHz. Considering only these flux measurements, the majority have flat spectral indices. Most were found to have either a decreasing spectral energy distribution with frequency (steep/falling) or with a spectral peak between 4.8 and 20~GHz (GPS).

\subsubsection{ATCA Observations}

All observations were made using the CABB system \citep{Ferris2002}, with a 64 MHz zoom-band across 2048 channels. This band was centred at either 1310, 1340 or 1342 MHz to allow a simultaneous search for associated H{\sc i} absorption over the respective redshift ranges for each set of targets within a single zoom-band. The ATCA observations are detailed in Table 2. 

\afterpage{
\hbox{}

\begin{sidewaystable}
\begin{minipage}{\textwidth}
\caption{Properties of the full sample of AT20G sources searched for \mbox{H{\sc i}} absorption. See end of Section 2.1.1 for column descriptions.}
\begin{tabular}{@{}lccccccccccccccc@{}}
     \hline
      AT20G name & RA & Dec. & $S_\mathrm{20}$ & $S_\mathrm{1.4}$ & $S_\mathrm{0.843}$ & $\alpha_{1.4}^{20}$ & $\alpha_{0.843}^{20}$ & Obs. ID & $z_\mathrm{opt}$ & SC & $\mathrm{M}_{K}$ & $6\mathrm{km}_\mathrm{vis}$ & FR \\
      & (J2000) & (J2000)  & (mJy) & (mJy) & (mJy) & & & & & & \\
   (1) & (2) & (3) & (4) & (5) & (6) & (7) & (8) & (9) & (10) & (11) & (12) & (13) & (14)\\
      \hline
J001215--384618 $\star$ & 00 12 15.13 & -38 46 18.8 & 80 & 65.4 & 52.2 & 0.08 & 0.13 & A & 0.0755$^{a}$ & Aae & -25.63 & $\dotsm$ & 0 \\ 
J001605--234352 $\star$ & 00 16 05.81 & -23 43 52.1 & 68 & 275.8 & $\dotsm$ & -0.53 & $\dotsm$ & B & 0.0640$^{a}$ & Ae & -24.79 & 0.91 & 0 \\ 
J003908--222001 $\star$ & 00 39 08.17 & -22 20 01.3 & 49 & 113.5 & $\dotsm$ & -0.32 & $\dotsm$ & B  & 0.0644$^{b}$ & $\dotsm$ & -25.28 & 0.89 & 0 \\ 
J005734--012328 * & 00 57 34.88 & -01 23 28.0 & 55 & 5370.0$^{t}$ & $\dotsm$ & -1.72 & $\dotsm$ & E & 0.0447$^{d}$ & $\dotsm$ & -25.52 & 1.00 & {\sc i} \\ 
J010249--795604 $\star$ & 01 02 49.45 & -79 56 04.9 & 113 & $\dotsm$ & 837.9 & $\dotsm$ & -0.63 & B & 0.0570$^{c}$ & $\dotsm$ & -25.10 & 0.55 & 0 \\ 
J011132--730209 * & 01 11 32.25 & -73 02 09.9 & 74 & $\dotsm$ & 86.4 & $\dotsm$ & -0.05 & C & 0.0665$^{a}$ & Aa & -26.00 & 0.93 & 0 \\ 
J012820--564939 $\star$ & 01 28 20.38 & -56 49 39.7 & 189 & $\dotsm$ & 82.6 & $\dotsm$ & 0.26 & A & 0.0666$^{a}$ & Ae & -24.91 & 1.00 & 0 \\ 
J023749--823252 $\star$ & 02 37 49.54 & -82 32 52.8 & 50 & $\dotsm$ & 457.3 & $\dotsm$ & -0.70 & B & 0.0754$^{a}$ & Aae & -24.87 & 0.78 & 0 \\ 
J024326--561242 $\star$ & 02 43 26.57 & -56 12 42.3 & 114 & $\dotsm$ & 181.9 & $\dotsm$ & -0.15 & A & 0.0637$^{a}$ & Ae & -24.43 & 0.99 & 0 \\ 
J025926--394037 $\star$ & 02 59 26.51 & -39 40 37.7 & 71 & 984.8 & 1552.0 & -0.99 & -0.97 & B & 0.0662$^{a}$ & Aae & -25.66 & 0.47 & 0 \\ 
J025955--123635 $\star$ & 02 59 55.24 & -12 36 35.1 & 92 & 518.1 & $\dotsm$ & -0.65 & $\dotsm$ & D & 0.0867$^{a}$ & Aae & -25.77 & 0.79 & 0 \\ 
J031357--395403 $\star$ & 03 13 57.83 & -39 54 03.4 & 78 & 597.1 & 917.6 & -0.77 & -0.78 & D & 0.0803$^{a}$ & Aa & -25.49 & 0.20 & 0 \\ 
J031552--190644 *$\top$ & 03 15 52.16 & -19 06 44.5 & 108 & 100.1 & $\dotsm$ & 0.03 & $\dotsm$ & E & 0.0671$^{a}$ & Ae & -24.58 & 0.20 & {\sc i} \\ 
J031757--441416 $\star$ & 03 17 57.66 & -44 14 16.8 & 258 & $\dotsm$ & 1915.0 & $\dotsm$ & -0.63 & D & 0.0759$^{d}$ & $\dotsm$ & -26.88 & 0.33 & 0 \\ 
J033114--524148 * & 03 31 14.99 & -52 41 48.2 & 55 & $\dotsm$ & 51.0 & $\dotsm$ & 0.02 & C & 0.0666$^{e}$ & $\dotsm$ & -25.15 & 0.93 & 0 \\ 
J033913--173600 * & 03 39 13.73 & -17 36 00.9 & 87 & 170.0 & $\dotsm$ & -0.00 & $\dotsm$ & E & 0.0656$^{a}$ & Aa & -25.64 & 0.96 & 0 \\ 
J034630--342246 * & 03 46 30.56 & -34 22 46.1 & 102 & 1724.5 & 1789.4 & -1.06 & -0.90 & C & 0.0538$^{f}$ & $\dotsm$ & -24.60 & 0.92 & {\sc ii} \\ 
J035145--274311 * & 03 51 45.09 & -27 43 11.4 & 122 & 5340.2 & $\dotsm$ & -1.42 & $\dotsm$ & C & 0.0657$^{a}$ & Ae & -24.52 & 0.28 & {\sc ii}  \\ 
J035257--683117 * & 03 52 57.51 & -68 31 16.8 & 68 & $\dotsm$ & 409.4 & $\dotsm$ & -0.57 & D & 0.0657$^{l}$ & $\dotsm$ & -26.13 & 0.51 & 0 \\ 
J035410--265013 * & 03 54 10.07 & -26 50 13.7 & 98 & 86.5 & $\dotsm$ & 0.05 & $\dotsm$ & C & 0.0650$^{a}$ & Aa & -25.36 & 0.97 & 0 \\ 
J043022--613201 * & 04 30 22.00 & -61 32 01.0 & 148 & $\dotsm$ & 2790.0 & $\dotsm$ & -0.93 & C & 0.0555$^{a}$ & Aa & -25.77 & 0.85 & {\sc i} \\ 
J043754--425853 $\star$ & 04 37 54.73 & -42 58 53.9 & 119 & $\dotsm$ & 173.2 & $\dotsm$ & -0.12 & A & 0.0475$^{a}$ & Aae & -25.48 & 0.92 & 0 \\ 
J052200--285608 $\star$ & 05 22 00.78 & -28 56 08.4 & 43 & 574.1 & $\dotsm$ & -0.97 & $\dotsm$ & B & 0.0670$^{a}$ & Ae & -24.78 & 0.72 & 0 \\ 
J055712--372836 $\star$ & 05 57 12.45 & -37 28 36.3 & 82 & 301.8 & 457.5 & -0.49 & -0.54 & A & 0.0448$^{a}$ & Ae & -25.85 & 0.70 & 0 \\ 
J060555--392905 * & 06 05 55.98 & -39 29 05.0 & 79 & 108.8 & 84.3 & -0.12 & -0.02 & C & 0.0454$^{a}$ & Aa & -25.00 & 1.03 & 0 \\ 
J062706--352916 $\dagger$ & 06 27 06.73 & -35 29 16.0 & 688 & 4632.9 & 4592.0 & -0.72 & -0.60 & -- & 0.0549$^{a}$ & Aa & -26.24 & 0.69 & {\sc i}  \\ 
J065359--415144 * & 06 53 59.97 & -41 51 44.8 & 56 & $\dotsm$ & 727.5 & $\dotsm$ & -0.81 & D & 0.0910$^{a}$ & Aa & -26.08 & 0.78 & {\sc i} \\ 
J084452--100059 * & 08 44 52.73 & -10 00 59.0 & 46 & 293.6 & $\dotsm$ & -0.70 & $\dotsm$ & E & 0.0429$^{a}$ & Aa & -25.11 & 0.78 & {\sc i} \\ 
J090802--095937 * & 09 08 02.23 & -09 59 37.5 & 60 & 182.3 & $\dotsm$ & -0.42 & $\dotsm$ & E & 0.0534$^{a}$ & Aa & -25.26 & 0.81 & {\sc i} \\ 
J091856--243829 *$\bullet$ & 09 18 56.56 & -24 38 28.4 & 64 & 55.6 & $\dotsm$ & 0.05 & $\dotsm$ & C & 0.0561$^{a}$ & Aae & -23.43 & 1.05 & 0 \\ 
J092338--213544 $\star$ & 09 23 38.95 & -21 35 44.9 & 328 & 267.7 & $\dotsm$ & 0.08 & $\dotsm$ & B & 0.0520$^{g}$ & $\dotsm$ & -24.74 & 1.00 & {\sc i} \\ 
J114539--105350 * & 11 45 39.02 & -10 53 50.4 & 73 & 102.4 & $\dotsm$ & -0.13 & $\dotsm$ & E & 0.0774$^{a}$ & Aa & -26.08 & 0.90 & 0 \\ 
J121044--435437 $\star$ & 12 10 44.67 & -43 54 37.4 & 129 & $\dotsm$ & 122.7 & $\dotsm$ & 0.02 & A & 0.0693$^{a}$ & Aae & -25.07 & 1.04 & 0 \\ 
J123148--321314 * & 12 31 48.71 & -32 13 13.8 & 106 & 165.1 & 137.5 & -0.17 & -0.08 & D & 0.0884$^{a}$ & Aae & -25.27 & 0.99 & 0 \\ 
J125457--442456 $\star$ & 12 54 57.67 & -44 24 56.9 & 344 & $\dotsm$ & 368.1 & $\dotsm$ & -0.02 & A & 0.0411$^{a}$ & Aae & -25.47 & 0.98 & 0 \\ 
J125615--114635 * & 12 56 15.96 & -11 46 35.7 & 123 & 54.4 & $\dotsm$ & 0.31 & $\dotsm$ & E & 0.0579$^{a}$ & Aa & -25.06 & 0.98 & 0 \\ 
J125711--172434 * & 12 57 11.59 & -17 24 34.0 & 83 & 98.4 & $\dotsm$ & -0.06 & $\dotsm$ & E & 0.0470$^{a}$ & Aa & -26.83 & 0.88 & 0 \\ 
\hline
\end{tabular}
\end{minipage}
\label{table:sample_info}
\end{sidewaystable}
}
\afterpage{
\hbox{}

\begin{sidewaystable}
\begin{minipage}{\textwidth}
\begin{tabular}{@{}lccccccccccccccc@{}}
     \hline
      AT20G name & RA & Dec. & $S_\mathrm{20}$ & $S_\mathrm{1.4}$ & $S_\mathrm{0.843}$ & $\alpha_{1.4}^{20}$ & $\alpha_{0.843}^{20}$ & Obs. ID & $z_\mathrm{opt}$ & SC & $\mathrm{M}_{K}$ & $6\mathrm{km}_\mathrm{vis}$ & FR \\
      & (J2000) & (J2000)  & (mJy) & (mJy) & (mJy) & & & & & & \\
   (1) & (2) & (3) & (4) & (5) & (6) & (7) & (8) & (9) & (10) & (11) & (12) & (13) & (14)\\
      \hline
J131124--442240 $\star$ & 13 11 24.04 & -44 22 40.5 & 44 & $\dotsm$ & 208.3 & $\dotsm$ & -0.49 & B & 0.0506$^{a}$ & Aae & -26.39 & 0.86 & {\sc i}/{\sc ii}  \\ 
J132920--264022 $\star$ & 13 29 20.70 & -26 40 22.2 & 101 & 102.4 & $\dotsm$ & -0.01 & $\dotsm$ & B & 0.0502$^{a}$ & Aa & -25.14 & 1.03 & 0 \\ 
J135036--163449 * & 13 50 36.14 & -16 34 49.5 & 186 & 109.2 & $\dotsm$ & 0.20 & $\dotsm$ & D & 0.0977$^{a}$ & Ae & -25.70 & 0.99 & 0 \\ 
J135607--172433 * & 13 56 07.02 & -17 24 33.1 & 239 & 180.1 & $\dotsm$ & 0.11 & $\dotsm$ & E & 0.0747$^{a}$ & Aae & -24.31 & 0.99 & 0 \\ 
J140912--231550 * & 14 09 11.97 & -23 15 49.5 & 137 & 599.4 & $\dotsm$ & -0.55 & $\dotsm$ & D & 0.0864$^{a}$ & Aae & -25.13 & 0.99 & 0 \\ 
J151741--242220 *$\ddagger$ & 15 17 41.76 & -24 22 20.2 & 3449 & 2041.9 & $\dotsm$ & 0.20 & $\dotsm$ & E & 0.0490$^{a}$ & Aae & -25.71 & 1.00 & 0 \\ 
J164416--771548 *$\dagger$ & 16 44 16.03 & -77 15 48.5 & 399 & $\dotsm$ & 1165.9 & $\dotsm$ & -0.34 & -- & 0.0427$^{h}$ & Aae & -25.06 & 1.00 & {\sc ii}  \\ 
J165710--735544 * & 16 57 10.08 & -73 55 44.5 & 42 & $\dotsm$ & 99.3 & $\dotsm$ & -0.27 & C & 0.0712$^{a}$ & Aa & -25.03 & $\dotsm$ & 0 \\ 
J171522--652018 $\star$ & 17 15 22.19 & -65 20 18.6 & 53 & $\dotsm$ & 190.5 & $\dotsm$ & -0.40 & A & 0.0492$^{a}$ & Aae & -25.75 & 0.93 & 0 \\ 
J180957--455241 $\star$ & 18 09 57.79 & -45 52 41.2 & 1087 & $\dotsm$ & 1530.0 & $\dotsm$ & -0.11 & A & 0.0697$^{a}$ & AeB & -25.28 & $\dotsm$ & 0 \\ 
J181857--550815 * & 18 18 57.99 & -55 08 15.0 & 75 & $\dotsm$ & 185.6 & $\dotsm$ & -0.29 & C & 0.0726$^{a}$ & Aa & -25.93 & 0.92 & {\sc i}   \\ 
J181934--634548 $\star$ $\dagger$ & 18 19 34.99 & -63 45 48.2 & 1704 & $\dotsm$ & 19886.0 & $\dotsm$ & -0.78 & A,B & 0.0641$^{i}$ & Ae? & -25.00 & 0.80 & 0 \\ 
J191457--255202 * & 19 14 57.79 & -25 52 02.8 & 147 & 367.6 & $\dotsm$ & -0.34 & $\dotsm$ & E & 0.0631$^{i}$ & Aa & -26.25 & 0.83 & 0 \\ 
J192043--383107 $\star$ & 19 20 43.13 & -38 31 07.1 & 63 & 239.7 & 247.8 & -0.5 & -0.43 & B & 0.0452$^{a}$ & Aae & -25.01 & $\dotsm$ & 0 \\ 
J204552--510627 * & 20 45 52.29 & -51 06 27.7 & 54 & $\dotsm$ & 620.9 & $\dotsm$ & -0.77 & C & 0.0485$^{a}$ & Aa & -26.32 & 0.94 & {\sc i} \\
J205306--162007 * &  20 53 06.00 & -16 20 07.4 & 56 & 116.5 & $\dotsm$ & -0.27 & $\dotsm$ & E & 0.0430$^{a}$ & Aa & -25.56 & 0.86 & {\sc i} \\ 
J205401--424238 $\star$ & 20 54 01.79 & -42 42 38.7 & 86 & $\dotsm$ & 160.2 & $\dotsm$ & -0.20 & B & 0.0429$^{a}$ & Aae & -25.06 & 0.98 & 0 \\ 
J205754--662919 * & 20 57 54.01 & -66 29 19.6 & 49 & $\dotsm$ & 282.7 & $\dotsm$ & -0.55 & C & 0.0754$^{a}$ & Aa & -25.58 & 0.98 & {\sc i} \\ 
J205837--575636 $\star$ & 20 58 37.39 & -57 56 36.5 & 97 & $\dotsm$ & 846.2 & $\dotsm$ & -0.68 & A & 0.0524$^{a}$ & Aa & -25.33 & 0.85 & 0 \\ 
J210602--712218 $\star$ & 21 06 02.92 & -71 22 17.9 & 246 & $\dotsm$ & 1206.0 & $\dotsm$ & -0.50 & A & 0.0745$^{a}$ & Aa & -25.60 & 0.96 & 0 \\ 
J212222--560014 * & 21 22 22.81 & -56 00 14.6 & 58 & $\dotsm$ & 100.9 & $\dotsm$ & -0.17 & C & 0.0518$^{a}$ & Aae & -25.08 & 0.94 & {\sc i} \\ 
J214824--571351 * & 21 48 24.27 & -57 13 51.5 & 48 & $\dotsm$ & 292.5 & $\dotsm$ & -0.57 & D & 0.0806$^{d}$ & $\dotsm$ & -25.57 & 0.78 & 0 \\ 
J220253--563543 $\star$ & 22 02 53.31 & -56 35 43.0 & 69 & $\dotsm$ & 58.4 & $\dotsm$ & 0.05 & A & 0.0489$^{j}$ & $\dotsm$ & -24.42 & 1.03 & 0 \\ 
J220538--053531 * & 22 05 38.59 & -05 35 31.9 & 67 & 187.2 & $\dotsm$ & -0.39 & $\dotsm$ & E & 0.0573$^{m}$ & $\dotsm$ & -25.46 & 1.05 & 0 \\ 
J221220--251829 * & 22 12 20.77 & -25 18 29.0 & 71 & 304.4 & $\dotsm$ & -0.55 & $\dotsm$ & C & 0.0626$^{a}$ & Aa & -26.27 & 0.71 & 0 \\ 
J223931--360912 $\star$ & 22 39 31.26 & -36 09 12.5 & 64 & 661.0 & 842.5 & -0.88 & -0.81 & B & 0.0575$^{a}$ & Aa & -25.28 & $\dotsm$ & 0 \\ 
J231905--420648 $\star$ & 23 19 05.92 & -42 06 48.9 & 150 & $\dotsm$ & 911.6 & $\dotsm$ & -0.57 & A & 0.0543$^{a}$ & Aae & -26.16 & 0.79 & {\sc i}  \\ 
J233355--234340 $\star$ & 23 33 55.28 & -23 43 40.8 & 957 & 782.1 & $\dotsm$ & 0.08 & $\dotsm$ & B & 0.0477$^{k}$ & $\dotsm$  & -24.52 & 0.97 & 0 \\ 
J234129--291915 $\star$ & 23 41 29.72 & -29 19 15.3 & 120 & 239.6 & $\dotsm$ & -0.26 & $\dotsm$ & B & 0.0517$^{j}$ & $\dotsm$ & -26.23 & 0.86 & 0 \\ 
J234205--160840 * & 23 42 05.82 & -16 08 40.4 & 43 & 151.9 & $\dotsm$ & -0.47 & $\dotsm$ & E & 0.0649$^{a}$ & Aa & -24.17 & 0.80 & 0 \\ 
      \hline
    \end{tabular}
    \begin{tablenotes}
    \item[]* {Source observed in our 2013/2014 ATCA programmes (this paper)}
    \item[]$\star$ {Source searched in \citet{Allison2012}}
    \item[]$\top$ {Source searched for \mbox{H\,{\sc i}} absorption by \citet{Ledlow2001}}
    \item[]$\dagger$ {Source searched for \mbox{H\,{\sc i}} absorption by \citet{Morganti2001}}
    \item[]$\ddagger$ {Source searched for \mbox{H\,{\sc i}} absorption by \citet{VanGorkom1989}}
    \item[]$\bullet$ {Source searched for \mbox{H\,{\sc i}} absorption but not part of this survey sample (see Section 2.1.1)}
    \item[]Spectroscopic redshift reference:
      $^{a}$ {\citet{Jones2009}},
      $^{b}$ {\citet{Vettolani1989}},
      $^{c}$ {\citet{Jauncey1978}}, 
      $^{d}$ {\citet{Katgert1998}},
      $^{e}$ {\citet{Smith2004}}, \\
      $^{f}$ {\citet{Drinkwater2001}},
      $^{g}$ {\citet{Peterson1979}},
      $^{h}$ {\citet{Simpson1993}},
      $^{i}$ {\citet{Morganti2011}}, 
      $^{j}$ {\citet{Colless2003}}, 
      $^{k}$ {\citet{Wills1976}}, \\
      $^{l}$ {\citet{Mahony2011}},
      $^{m}$ {\citet{Cava2009}}
    \end{tablenotes}
\end{minipage}

\end{sidewaystable}


}

Like the survey of \cite{Allison2012}, sources were observed in 20 or 30 minute scans and interleaved with 1.5 or 2 minute observations of a nearby bright calibrator for gain and phase calibration. PKS~1934--638 \citep{Reynolds1994}\footnote{http://www.narrabri.atnf.csiro.au/calibrators/} was observed regularly as the primary bandpass and flux calibrator, and for some nearby targets as a secondary calibrator. All 2048 zoom channels were used corresponding to a bandwidth of 64 MHz, although channels affected by RFI were removed during data reduction. 

\subsection{Data Reduction}

All flagging, calibration and imaging of the data were performed using tasks from the {\sc miriad}\footnote{www.atnf.csiro.au/computing/software/miriad/} package \citep*{Sault1995} and implemented using a purpose-built H{\sc i} data reduction pipeline written in Python \citep{Allison2012}. Channels that are known to contain significant RFI (e.g. birdies, CABB-generated self-interference due to clock harmonics) were flagged out on the fly by the correlator and applied when data was loaded using {\sc atlod}. {\sc pgflag} and {\sc blflag} were used both before and after calibration for each target and calibrator to identify and flag out remaining RFI. Care was taken to avoid excessive flagging of data and hence the removal of any putative spectral lines, and comparisons were made between flagged and unflagged data. Typically less than 0.1\% of data was flagged by the correlator, and less than 2\% in total, although the spectra for the June 2013 observations were far more corrupted by RFI (upward of 10\% removed in a few cases).

The absolute flux scale and spectral index corrections based on the flux model of PKS 1934-638 \citep{Reynolds1994} were applied to each of the secondary calibrators. We then calculated the time-varying bandpass and gain corrections for the amplitude and phase for each of these, and applied their calibration solutions to their paired target sources. We repeated the remaining methods and process described by \cite{Allison2012} for obtaining the source spectra. Fig.~1 displays the extracted spectrum for each of the targets sources, where the arrow below the spectra indicates the expected position for any associated 21-cm H{\sc i} absorption lines, calculated using the known optical spectroscopic redshift of the target.

We analysed the spectra through a Bayesian linefinder and parameterisation pipeline \citep{Allison2012b}. Improvements made to the pipeline made since \cite{Allison2012b} have allowed for higher-order polynomial model fits to be made for the continuum, to account for any spectral artefacts that could not be successfully removed, e.g. bandpass calibrator ripples. As a result, there are now fewer false positive detections.

\section{Results}
\subsection{Detections of HI}

We obtained four detections of H{\sc i} absorption (three previously unknown) in addition to those obtained by \cite{Allison2012}. Fig.~4 shows the optical SuperCosmos Sky Survey \citep{Hambly2001} and our observed radio continuum images for the targets with H{\sc i} absorption, and results of spectral model fitting. Table 3 summarises the properties of the individual Gaussian components which were fitted using Bayesian model comparisons of the new detections. 

These four detections add to the previous three \citep{Allison2012}, giving a total of 7/66 detections, or a detection rate of 11\%. Of these seven detections, four have been categorised as a flat-spectrum source (-0.5 $< \alpha <$ 0.5 between 1.4/0.843 GHz and 20 GHz; Table~2; also seen in Fig. 1 for 5 to 8 GHz and 8 to 20 GHz). We have a H{\sc i} absorption detection rate of $\sim$10\% within both the flat-spectrum and steep-spectrum sources (indicated in Table~2). 

\begin{figure*}
\captionsetup[subfigure]{aboveskip=-2.5pt,belowskip=4pt}
\centering
\small
\begin{minipage}{\textwidth}
\begin{subfigure}[b]{0.5\textwidth}
  \vskip 0pt
  \centering
  \subcaption{AT20G J005734-012328}
  \includegraphics[width=1.0\linewidth]{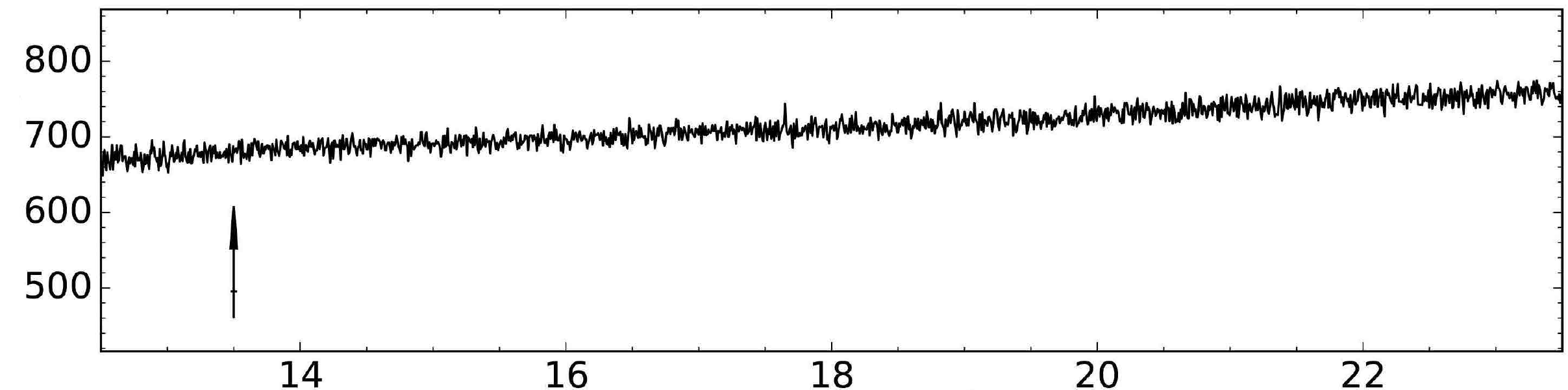}  
\end{subfigure}%
\begin{subfigure}[b]{0.5\textwidth}
  \vskip 0pt
  \centering
  \subcaption{AT20G J011132-730209}
  \includegraphics[width=1.00\linewidth]{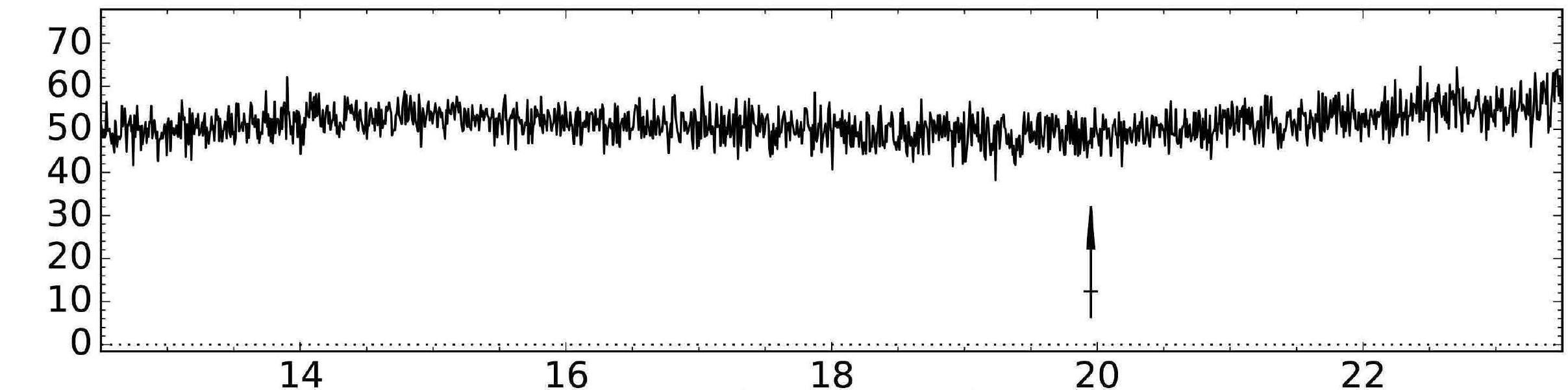}  
\end{subfigure}
\begin{subfigure}[b]{0.5\textwidth}
  \vskip 0pt
  \centering
  \subcaption{AT20G J025955-123635}
  \includegraphics[width=1.00\linewidth]{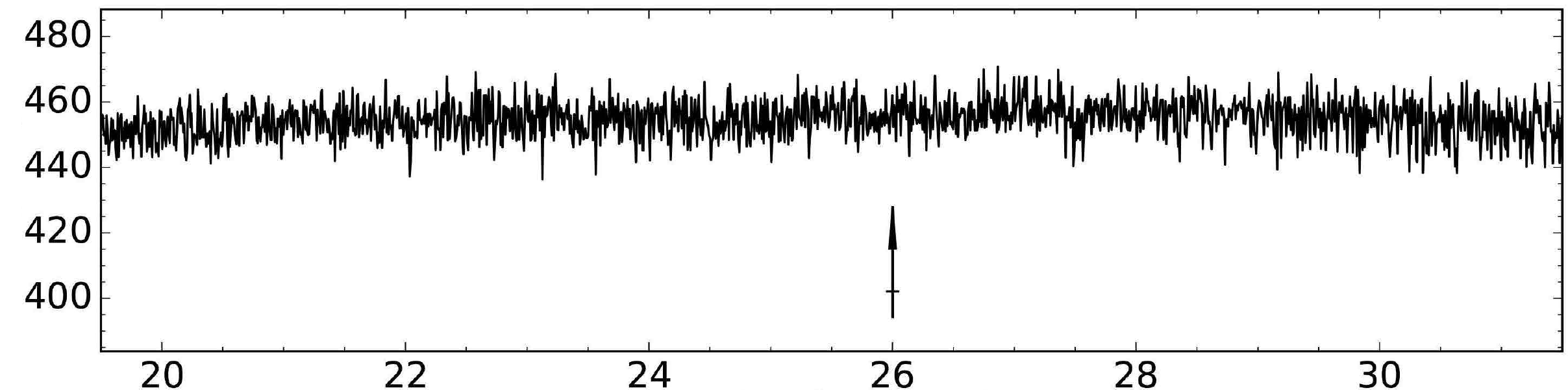}  
\end{subfigure}%
\begin{subfigure}[b]{0.5\textwidth}
  \vskip 0pt
  \centering
  \subcaption{AT20G J031357-395403}
  \includegraphics[width=1.0\linewidth]{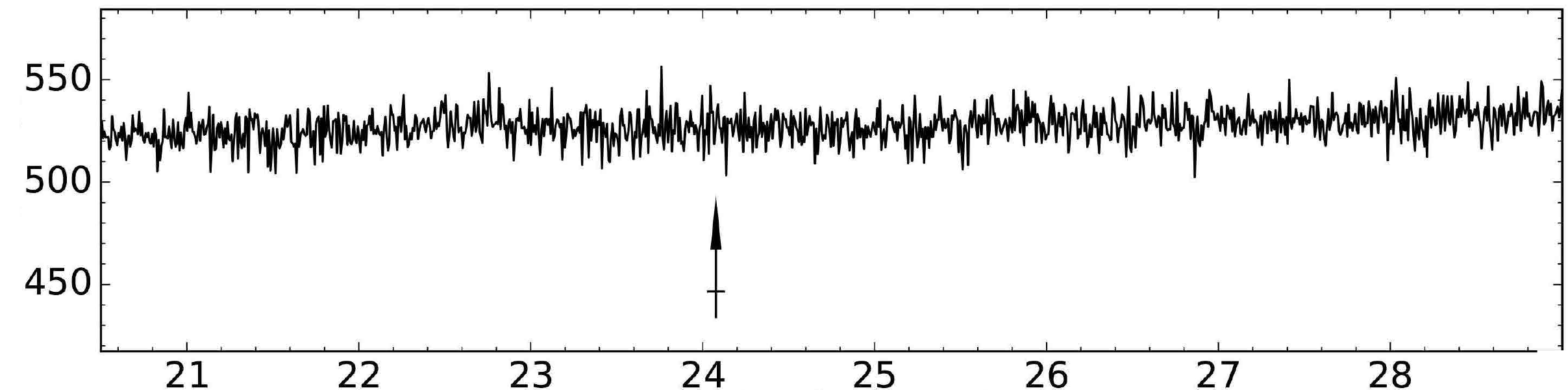}  
\end{subfigure}
\begin{subfigure}[b]{0.5\textwidth}
  \vskip 0pt
  \centering
  \subcaption{AT20G J031552-190644}
  \includegraphics[width=1.0\linewidth]{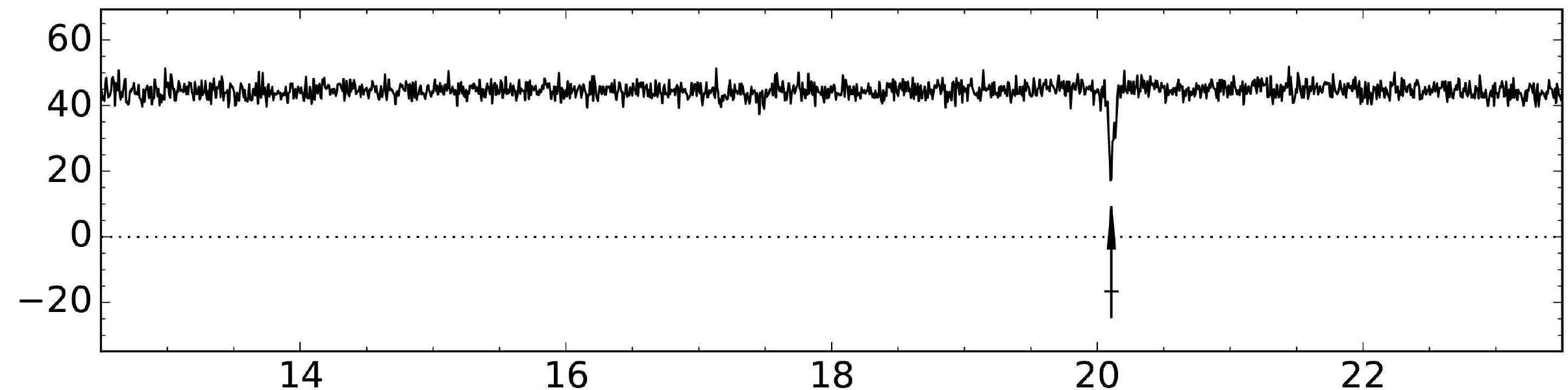}  
\end{subfigure}%
\begin{subfigure}[b]{0.5\textwidth}
  \vskip 0pt
  \centering
  \subcaption{AT20G J033114-524148}
  \includegraphics[width=1.0\linewidth]{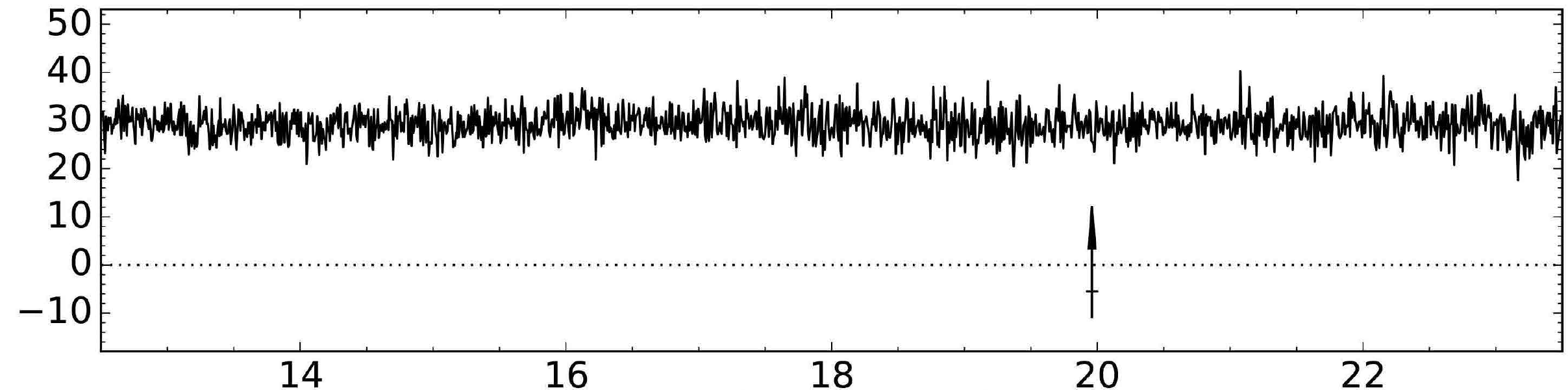}  
\end{subfigure}
\begin{subfigure}[b]{0.5\textwidth}
  \vskip 0pt
  \centering
  \subcaption{AT20G J033913-173600}
  \includegraphics[width=1.0\linewidth]{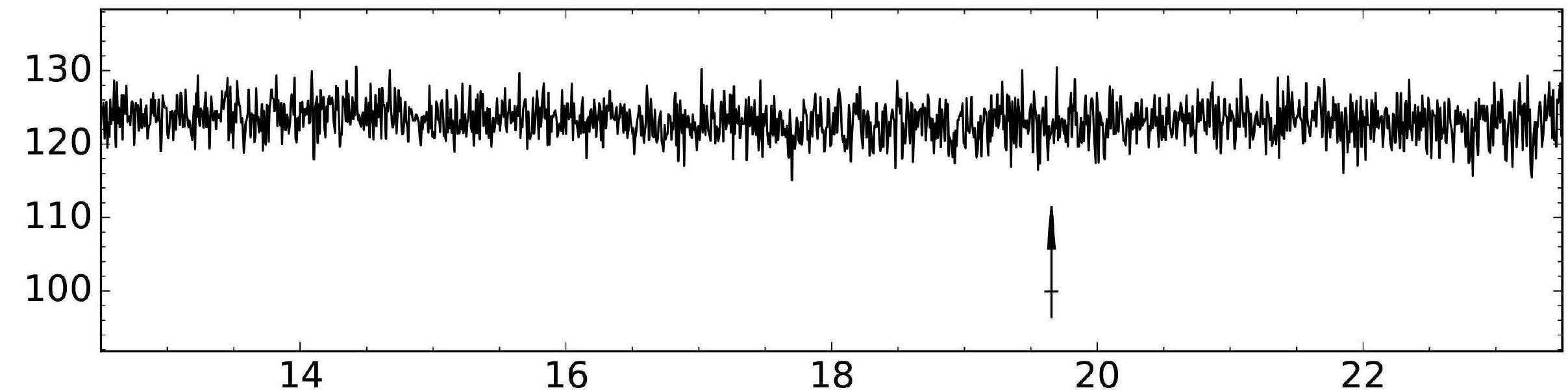}  
\end{subfigure}%
\begin{subfigure}[b]{0.5\textwidth}
  \vskip 0pt
  \centering
  \subcaption{AT20G J034630-342246}
  \includegraphics[width=1.0\linewidth]{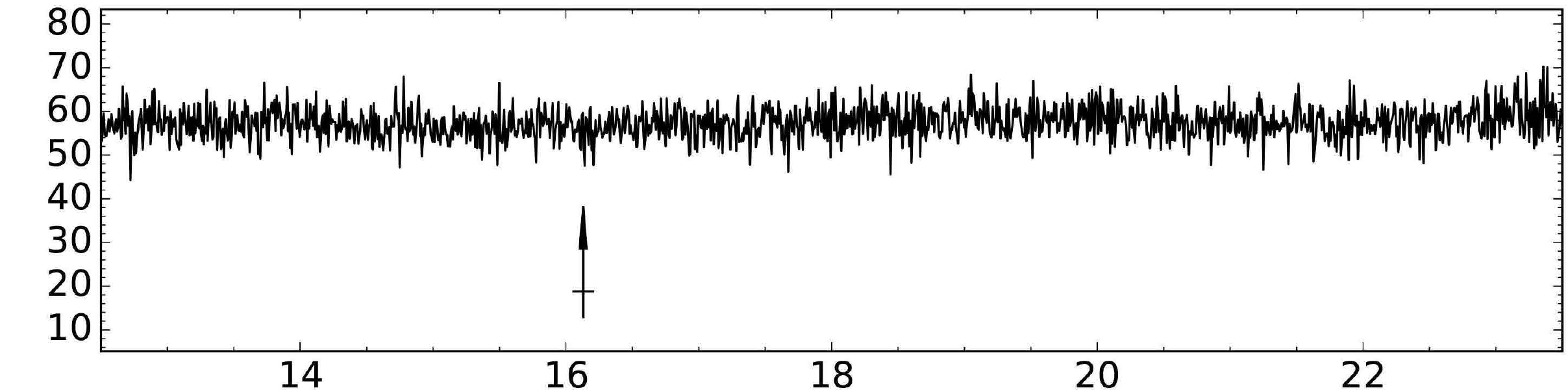}  
\end{subfigure}
\begin{subfigure}[b]{0.5\textwidth}
  \vskip 0pt
  \centering
  \subcaption{AT20G J035145-274311}
  \includegraphics[width=1.0\linewidth]{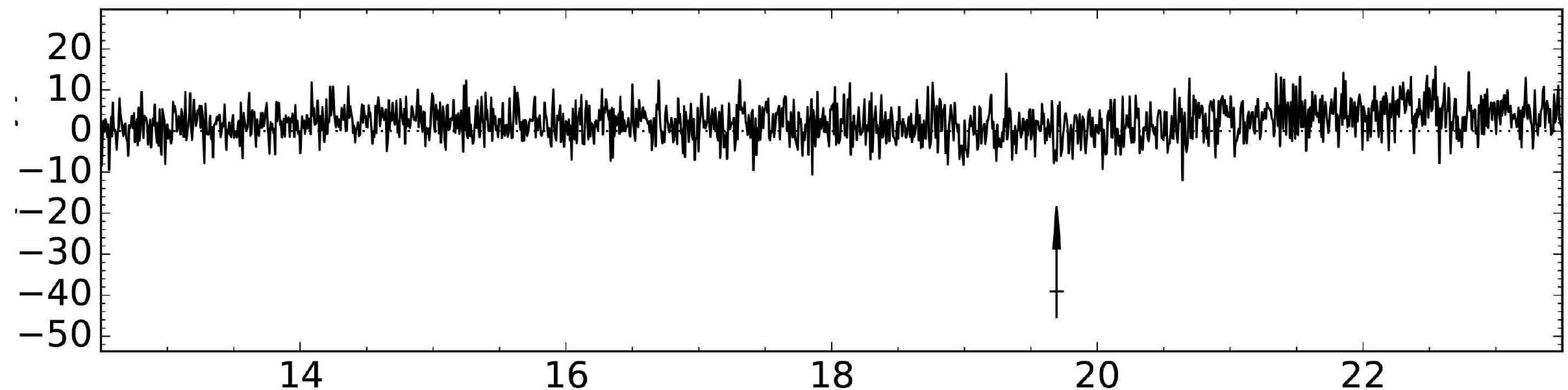}  
\end{subfigure}%
\begin{subfigure}[b]{0.5\textwidth}
  \vskip 0pt
  \centering
  \subcaption{AT20G J035257-683117}
  \includegraphics[width=1.0\linewidth]{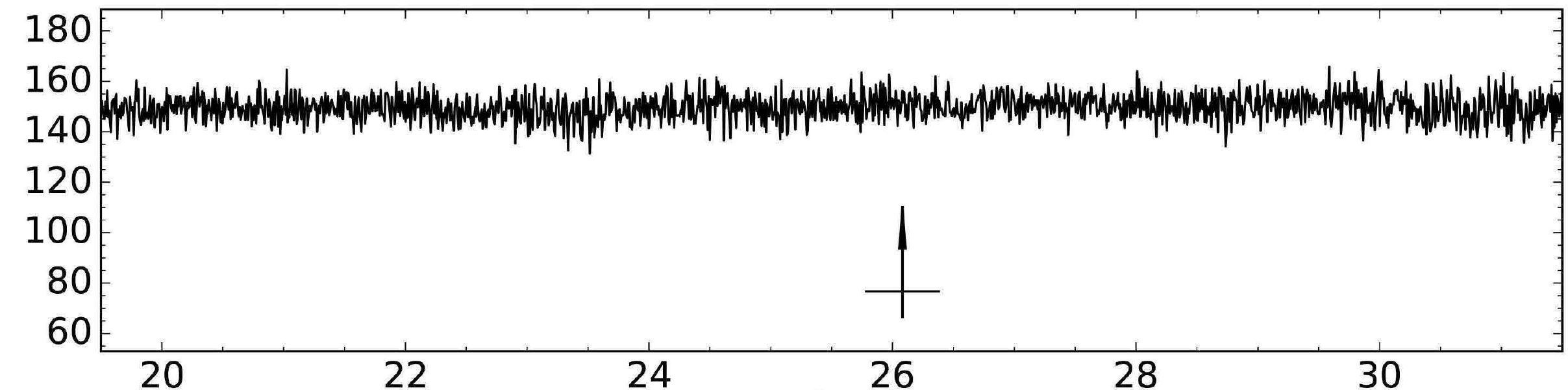}  
\end{subfigure}
\begin{subfigure}[b]{0.5\textwidth}
  \vskip 0pt
  \centering
  \subcaption{AT20G J035410-265013}
  \includegraphics[width=1.0\linewidth]{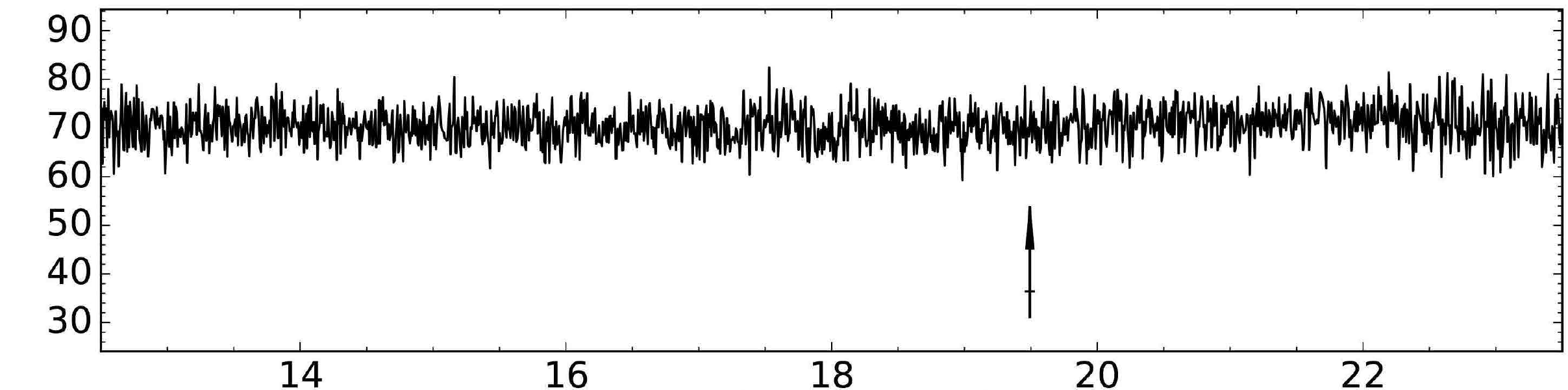}  
\end{subfigure}%
\begin{subfigure}[b]{0.5\textwidth}
  \vskip 0pt
  \centering
  \subcaption{AT20G J043022-613201}
  \includegraphics[width=1.0\linewidth]{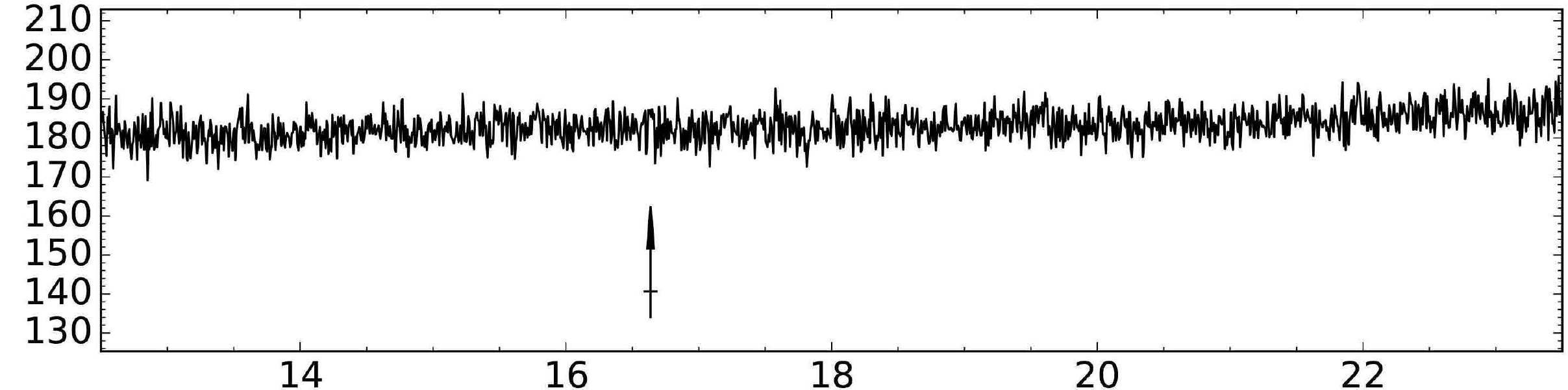}  
\end{subfigure}
\begin{subfigure}[b]{0.5\textwidth}
  \vskip 0pt
  \centering
  \subcaption{AT20G J060555-392905}
  \includegraphics[width=1.0\linewidth]{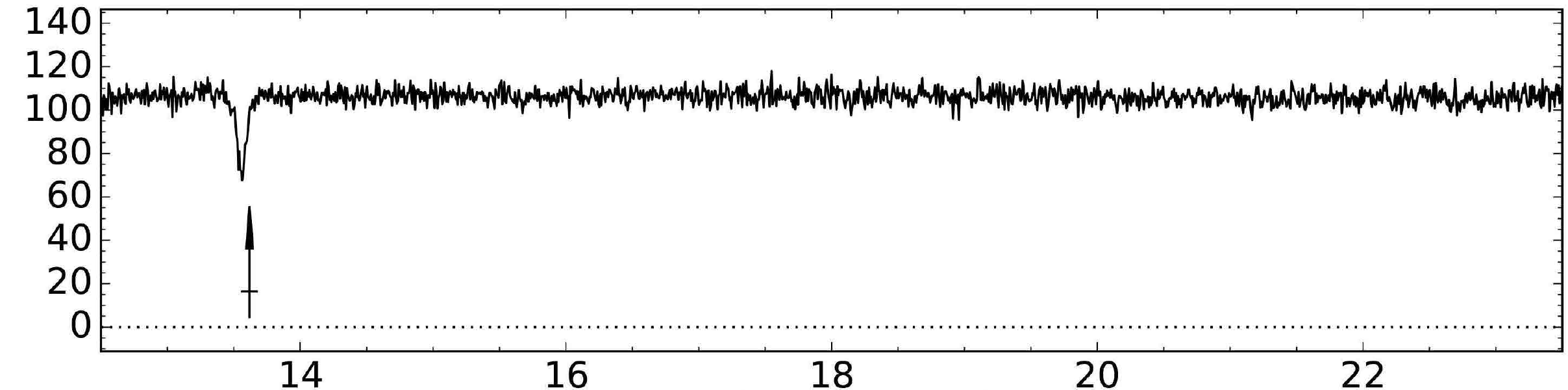}  
\end{subfigure}%
\begin{subfigure}[b]{0.5\textwidth}
  \vskip 0pt
  \centering
  \subcaption{AT20G J065359-415144}
  \includegraphics[width=1.0\linewidth]{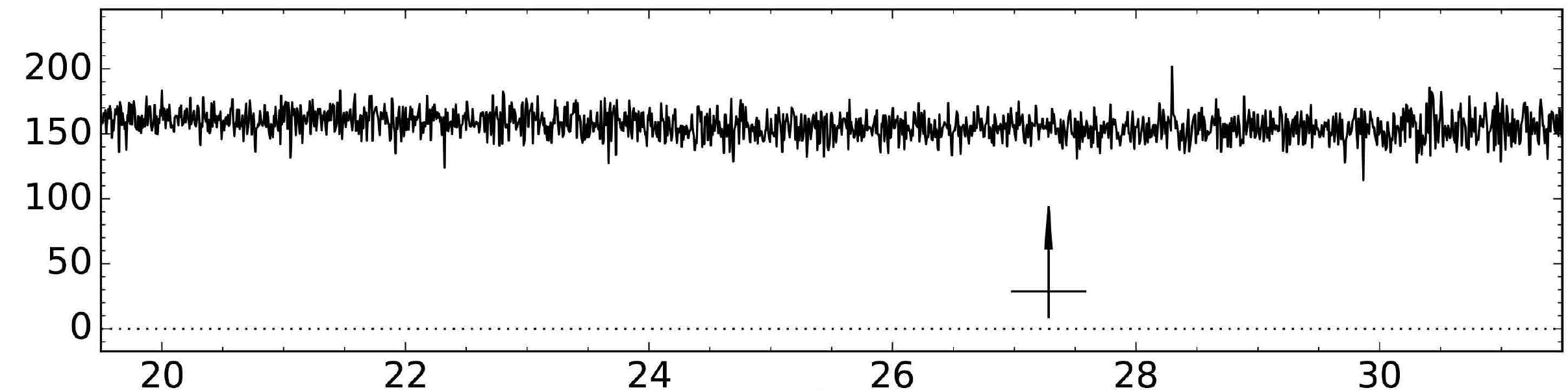}  
\end{subfigure}
\begin{subfigure}[b]{0.5\textwidth}
  \vskip 0pt
  \centering
  \subcaption{AT20G J084452-100059}
  \includegraphics[width=1.0\linewidth]{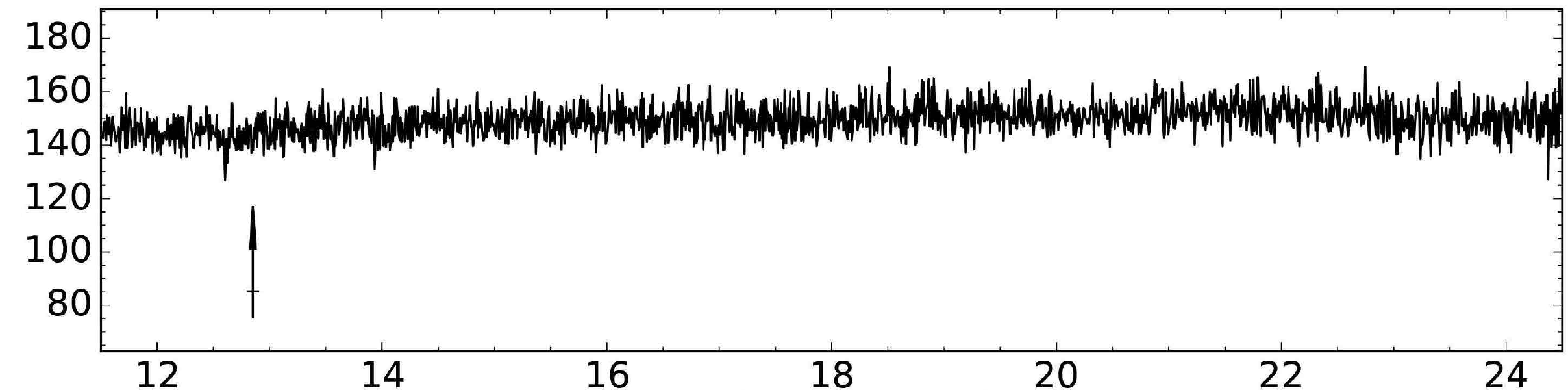}  
\end{subfigure}%
\begin{subfigure}[b]{0.5\textwidth}
  \vskip 0pt
  \centering
  \subcaption{AT20G J090802-095937}
  \includegraphics[width=1.0\linewidth]{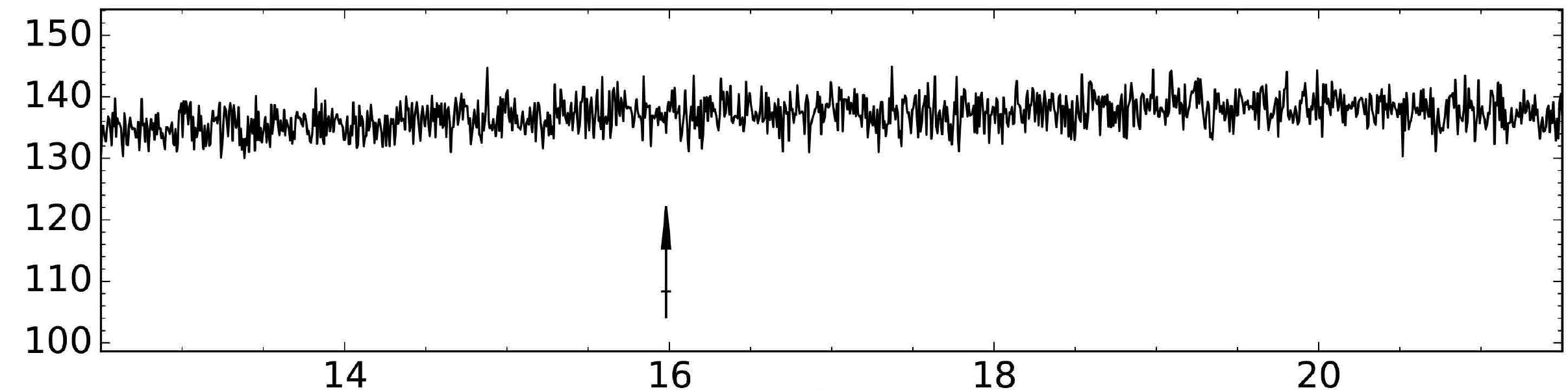}  
\end{subfigure}\\
\caption{CABB 21-cm spectra for newly observed sources in our sample (see Table 2), with the amplitude in flux (mJy) plotted against the Doppler corrected barycentric velocity (\textit{cz}, in 10$^{3}$ km~s$^{-1}$). The arrow indicates the optical redshift of the host galaxy, where any associated H{\sc i} absorption feature would be expected. The 1$\sigma$ redshift error is indicated by the horizontal line on the arrow.}%
\end{minipage}
\end{figure*}
\begin{figure*}
\captionsetup[subfigure]{aboveskip=-2pt,belowskip=4pt}
\ContinuedFloat
\begin{minipage}{\textwidth}
\centering
\small
\begin{subfigure}[b]{0.5\textwidth}
  \vskip 0pt
  \centering
  \subcaption{AT20G J091856-243829}
  \includegraphics[width=1.0\linewidth]{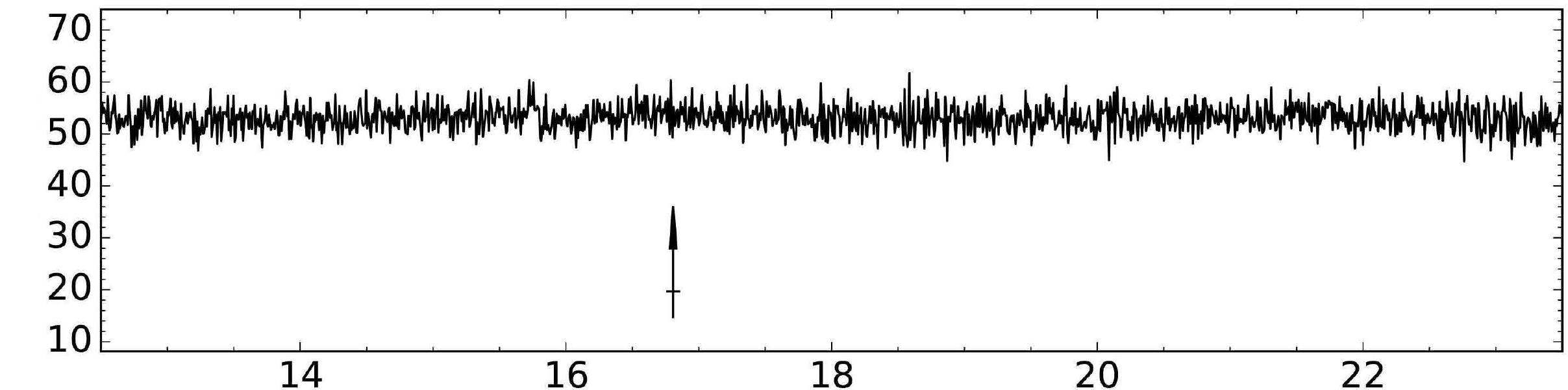}  
\end{subfigure}%
\begin{subfigure}[b]{0.5\textwidth}
  \vskip 0pt
  \centering
  \subcaption{AT20G J114539-105350}
  \includegraphics[width=1.0\linewidth]{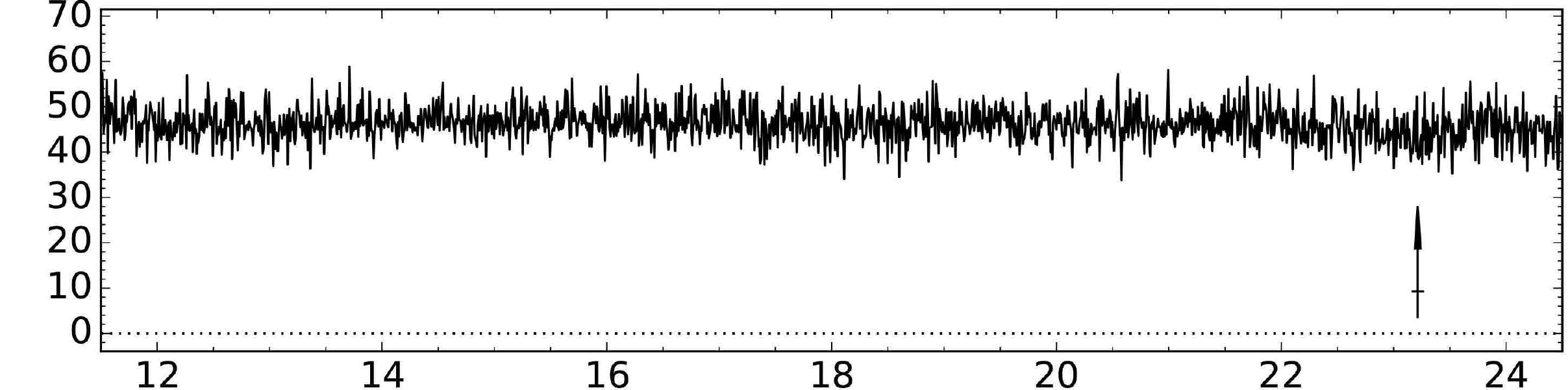}  
\end{subfigure}\\
\begin{subfigure}[b]{0.5\textwidth}
  \vskip 0pt
  \centering
  \subcaption{AT20G J123148-321314}
  \includegraphics[width=1.0\linewidth]{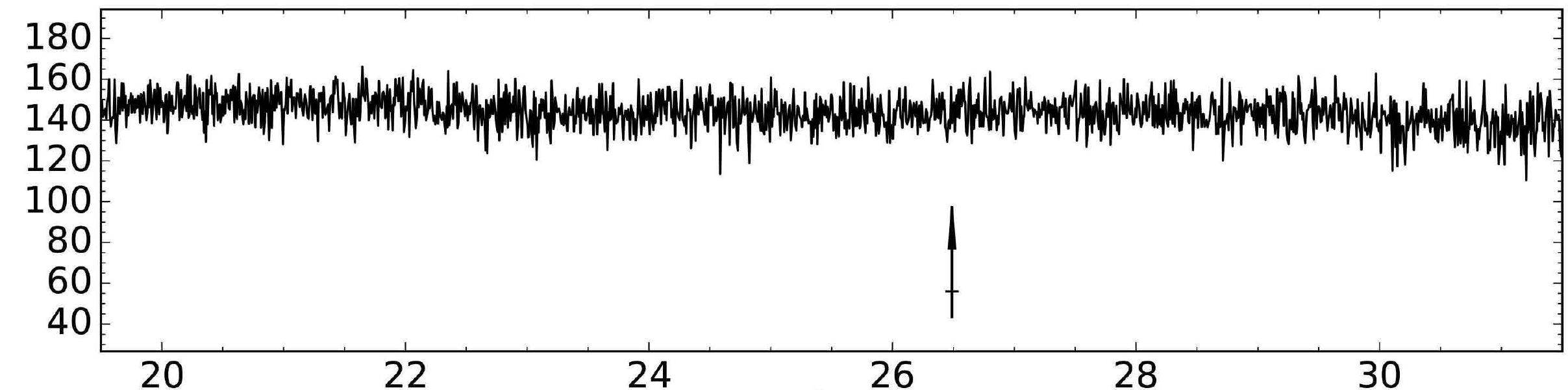}  
\end{subfigure}%
\begin{subfigure}[b]{0.5\textwidth}
  \vskip 0pt
  \centering
  \subcaption{AT20G J125615-114635}
  \includegraphics[width=1.0\linewidth]{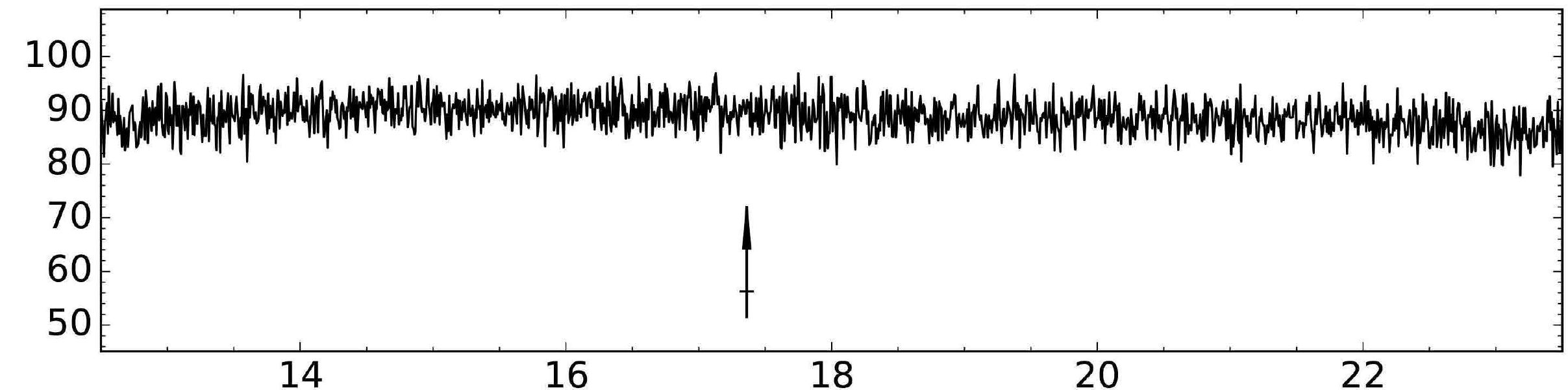}  
\end{subfigure}
\begin{subfigure}[b]{0.5\textwidth}
  \vskip 0pt
  \centering
  \subcaption{AT20G J125711-172434}
  \includegraphics[width=1.0\linewidth]{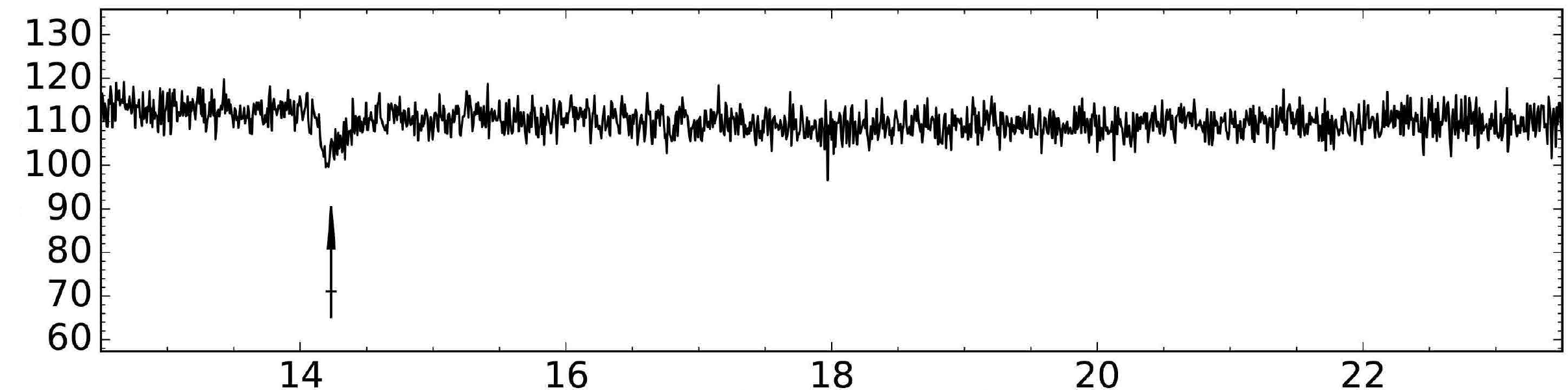}  
\end{subfigure}%
\begin{subfigure}[b]{0.5\textwidth}
  \vskip 0pt
  \centering
  \subcaption{AT20G J135036-163449}
  \includegraphics[width=1.0\linewidth]{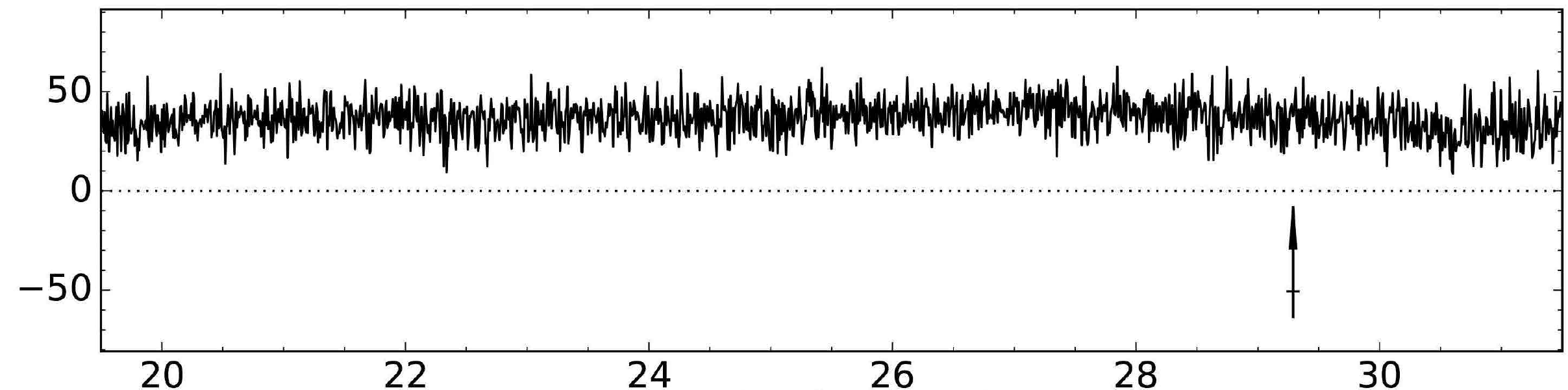}  
\end{subfigure}
\begin{subfigure}[b]{0.5\textwidth}
  \vskip 0pt
  \centering
  \subcaption{AT20G J135607-172433}
  \includegraphics[width=1.0\linewidth]{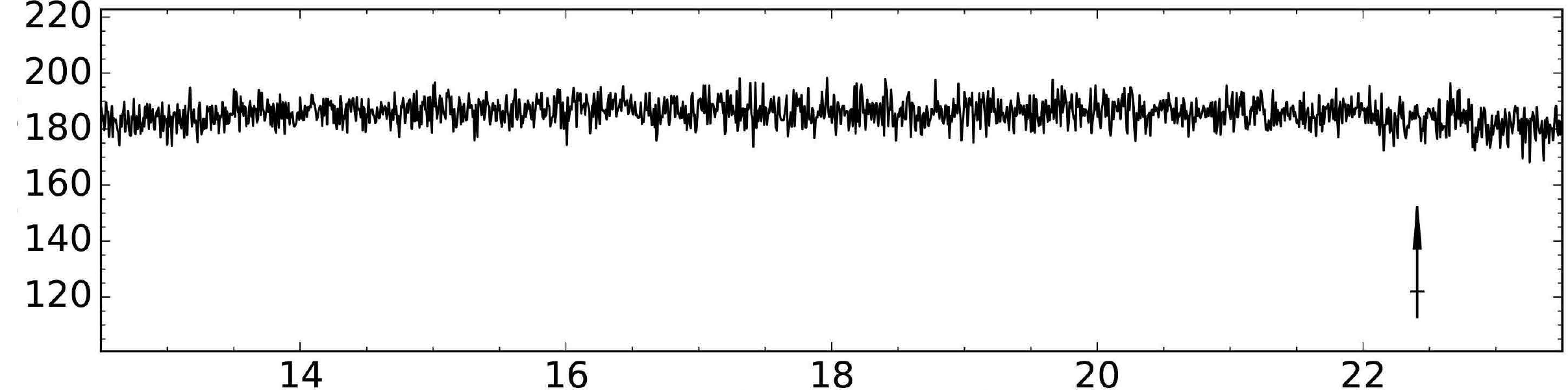}  
\end{subfigure}%
\begin{subfigure}[b]{0.5\textwidth}
  \vskip 0pt
  \centering
  \subcaption{AT20G J140912-231550}
  \includegraphics[width=1.0\linewidth]{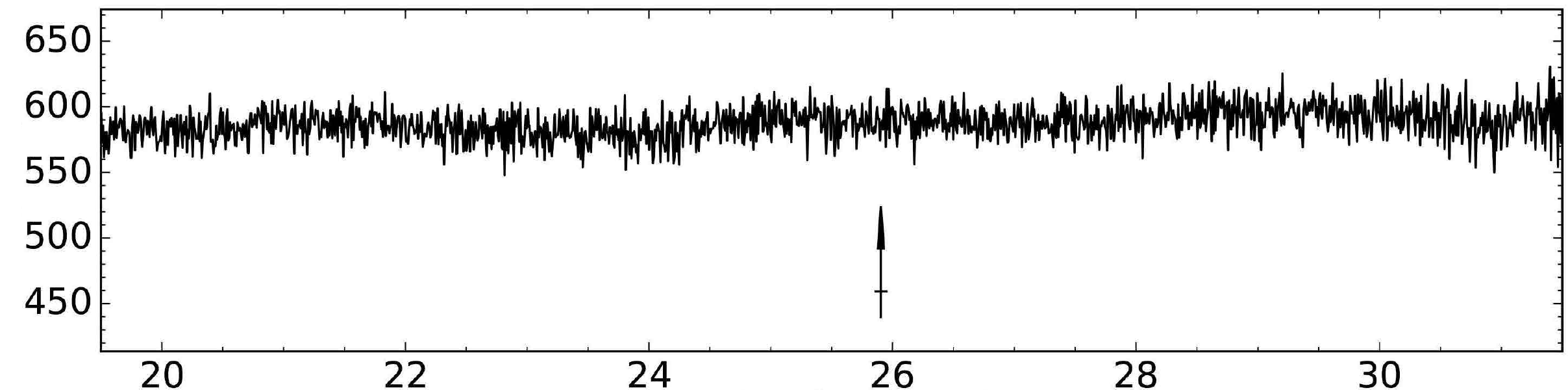}  
\end{subfigure}
\begin{subfigure}[b]{0.5\textwidth}
  \vskip 0pt
  \centering
  \subcaption{AT20G J151741-242220}
  \includegraphics[width=1.0\linewidth]{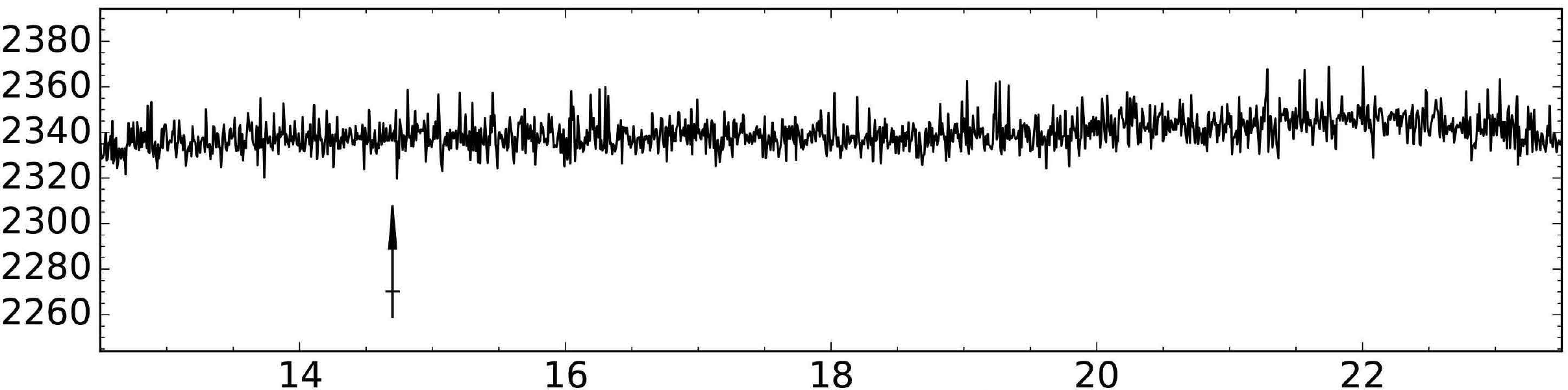}  
\end{subfigure}%
\begin{subfigure}[b]{0.5\textwidth}
  \vskip 0pt
  \centering
  \subcaption{AT20G J165710-735544}
  \includegraphics[width=1.0\linewidth]{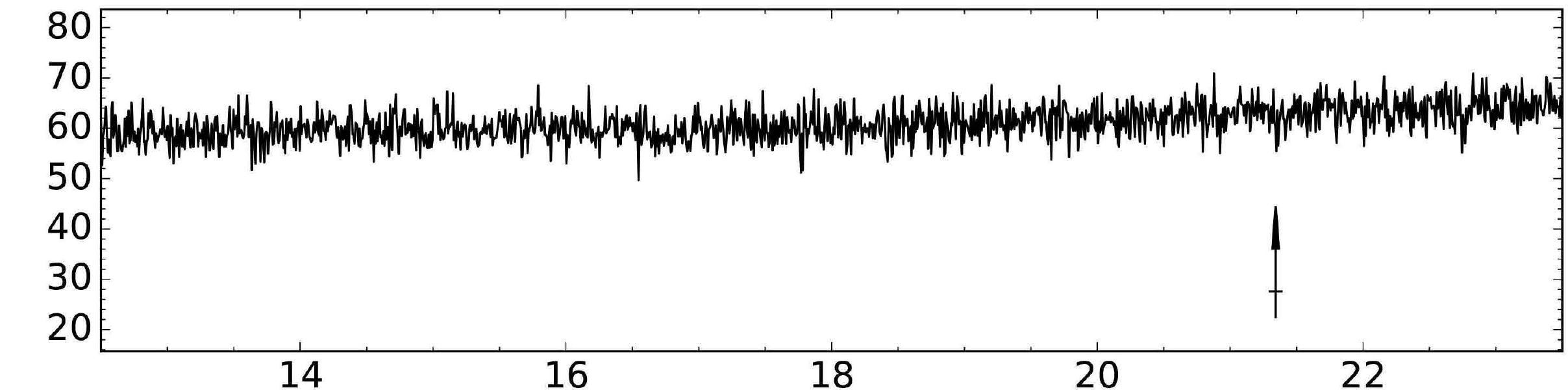}  
\end{subfigure}
\begin{subfigure}[b]{0.5\textwidth}
  \vskip 0pt
  \centering
  \subcaption{AT20G J181857-550815}
  \includegraphics[width=1.0\linewidth]{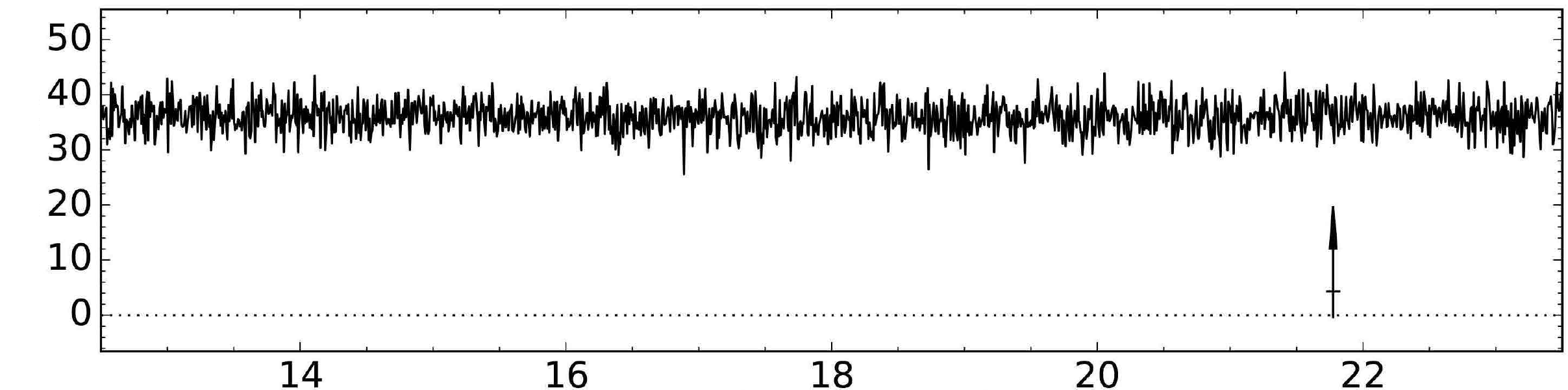}  
\end{subfigure}%
\begin{subfigure}[b]{0.5\textwidth}
  \vskip 0pt
  \centering
  \subcaption{AT20G J191457-255202}
  \includegraphics[width=1.0\linewidth]{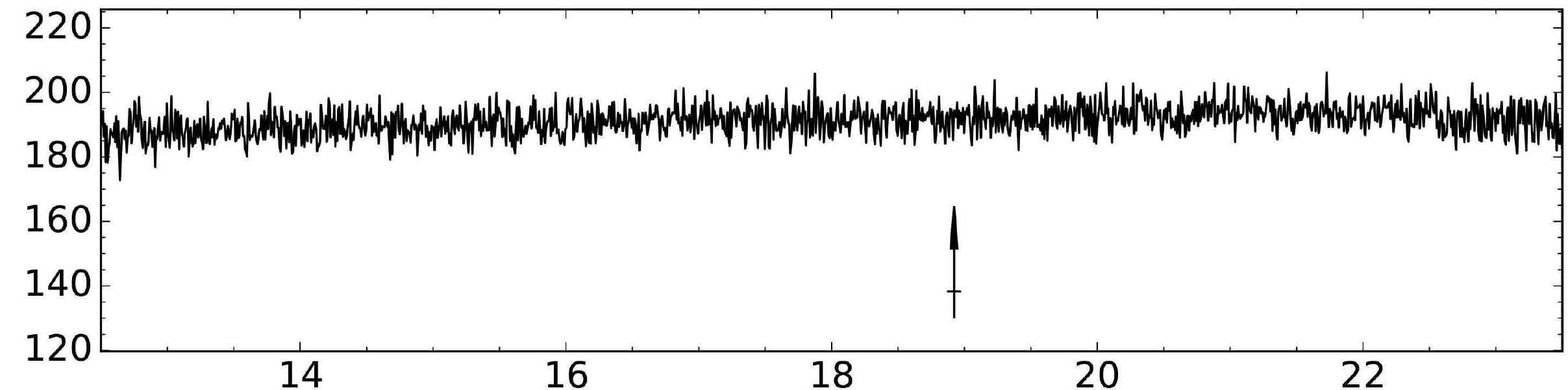}  
\end{subfigure}
\begin{subfigure}[b]{0.5\textwidth}
  \vskip 0pt
  \centering
  \subcaption{AT20G J204552-510627}
  \includegraphics[width=1.0\linewidth]{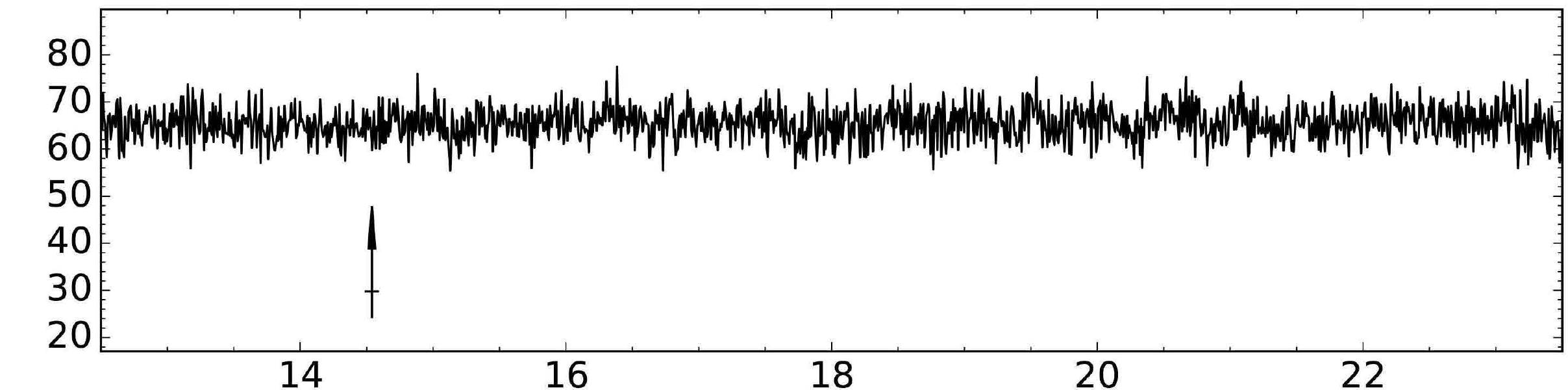}  
\end{subfigure}%
\begin{subfigure}[b]{0.5\textwidth}
  \vskip 0pt
  \centering
  \subcaption{AT20G J205306-162007}
  \includegraphics[width=1.0\linewidth]{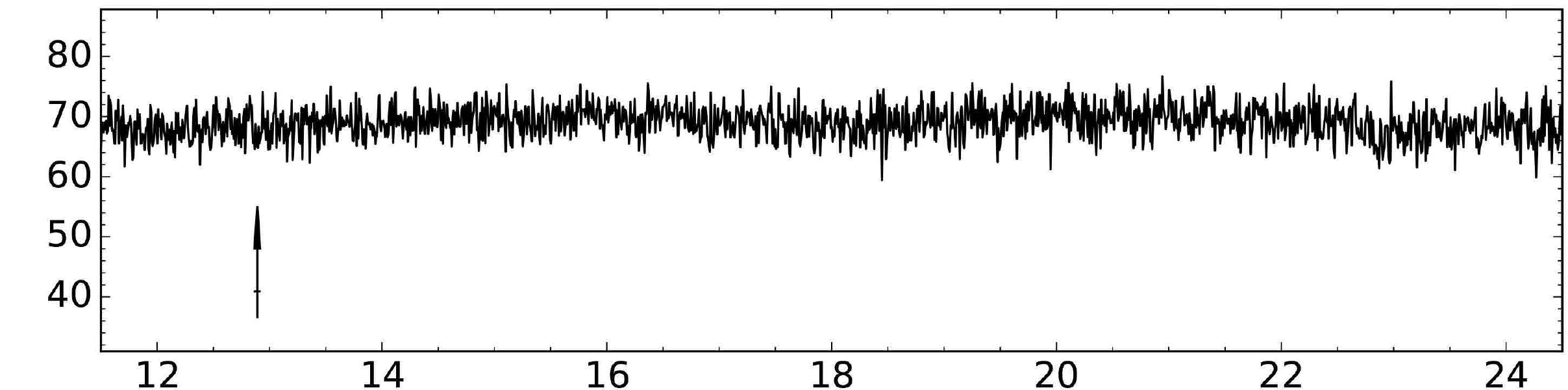}  
\end{subfigure}
\begin{subfigure}[b]{0.5\textwidth}
  \vskip 0pt
  \centering
  \subcaption{AT20G J205754-662919}
  \includegraphics[width=1.0\linewidth]{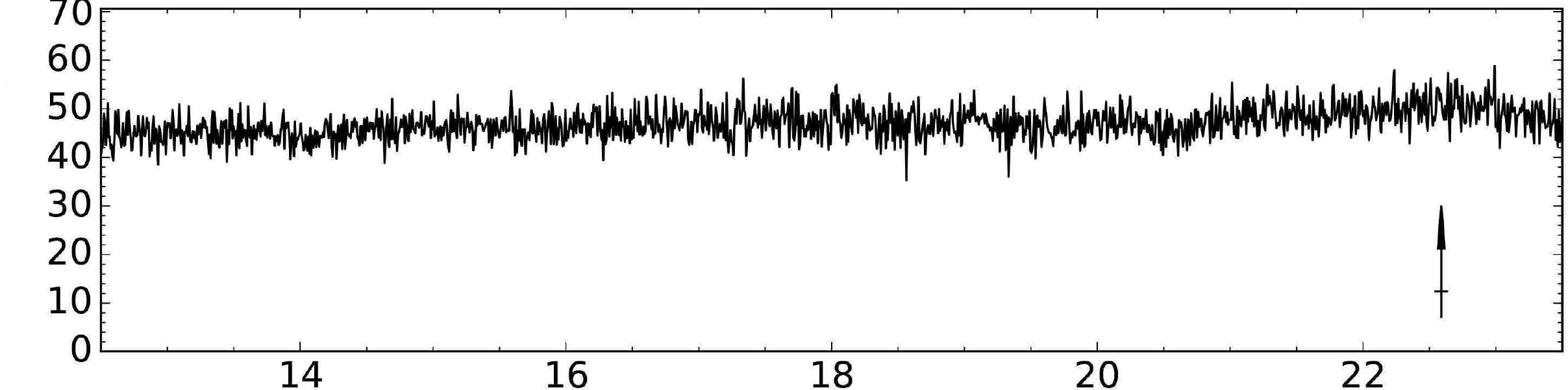}  
\end{subfigure}%
\begin{subfigure}[b]{0.5\textwidth}
  \vskip 0pt
  \centering
  \subcaption{AT20G J212222-560014}
  \includegraphics[width=1.0\linewidth]{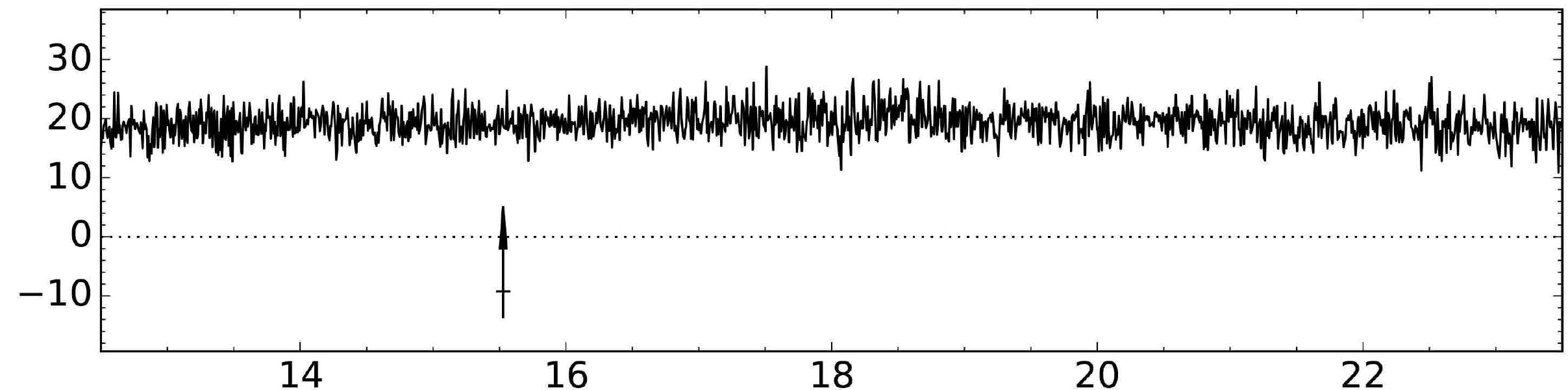}  
\end{subfigure}
\caption{\it{Continued.}}
\end{minipage}
\end{figure*}
\begin{figure*}
\captionsetup[subfigure]{aboveskip=-2pt,belowskip=4pt}
\ContinuedFloat
\begin{minipage}{\textwidth}
\centering
\small
\begin{subfigure}[b]{0.5\textwidth}
  \vskip 0pt
  \centering
  \subcaption{AT20G J214824-571351}
  \includegraphics[width=1.0\linewidth]{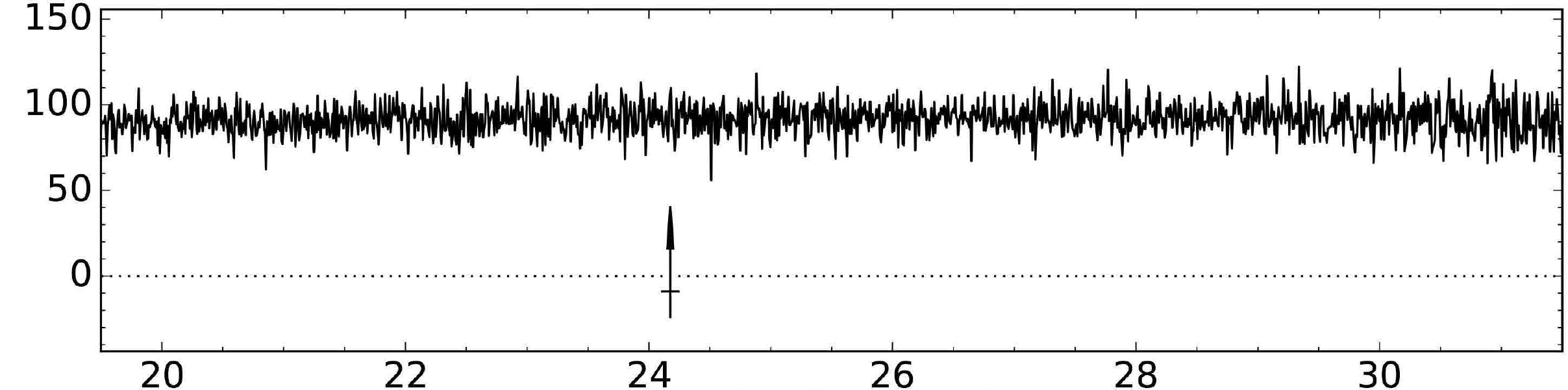}  
\end{subfigure}%
\begin{subfigure}[b]{0.5\textwidth}
  \vskip 0pt
  \centering
  \subcaption{AT20G J220538-053531}
  \includegraphics[width=1.0\linewidth]{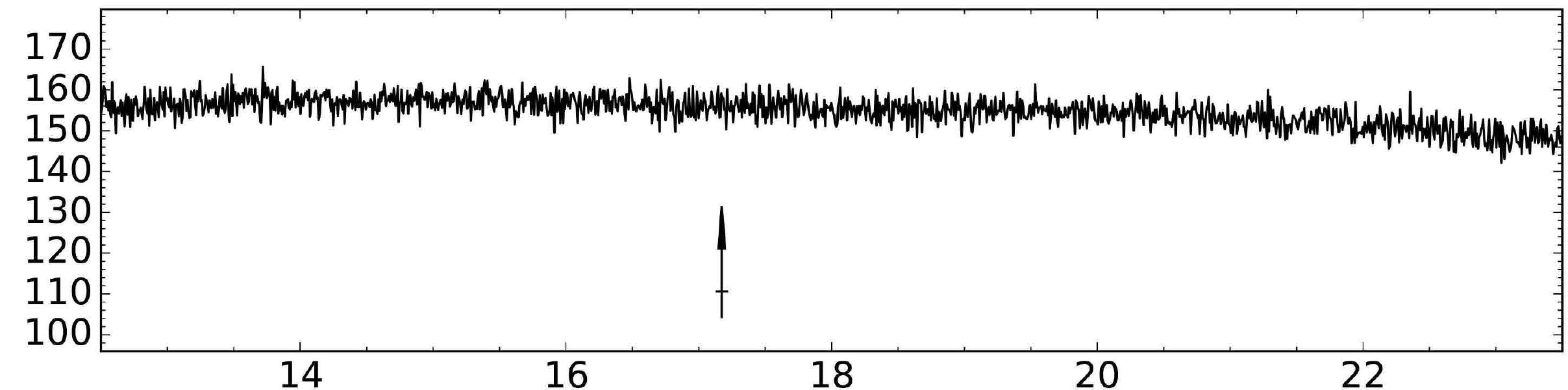}  
\end{subfigure}
\begin{subfigure}[b]{0.5\textwidth}
  \vskip 0pt
  \centering
  \subcaption{AT20G J221220-251829}
  \includegraphics[width=1.0\linewidth]{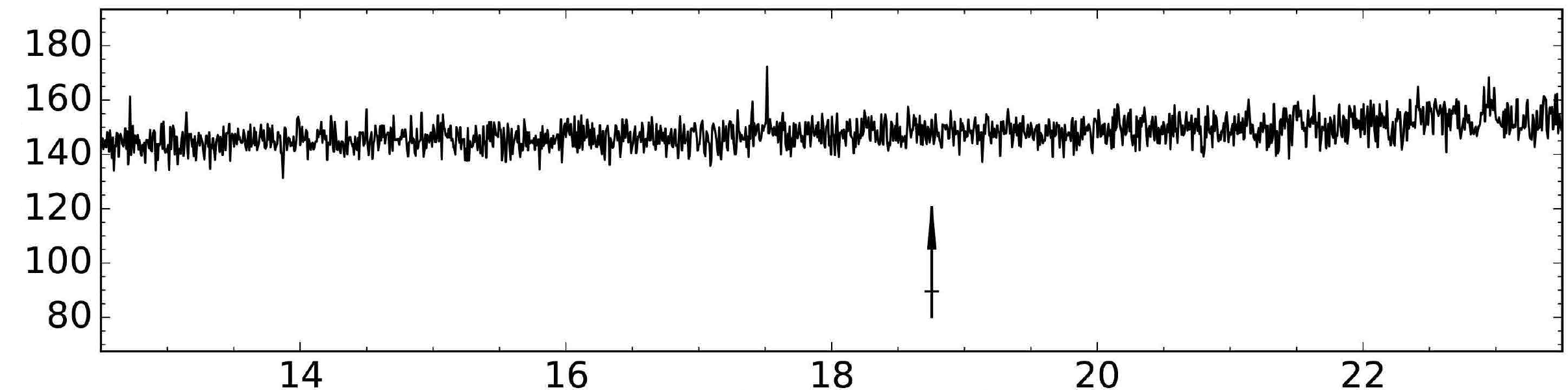}  
\end{subfigure}%
\begin{subfigure}[b]{0.5\textwidth}
  \vskip 0pt
  \centering
  \subcaption{AT20G J234205-160840}
  \includegraphics[width=1.0\linewidth]{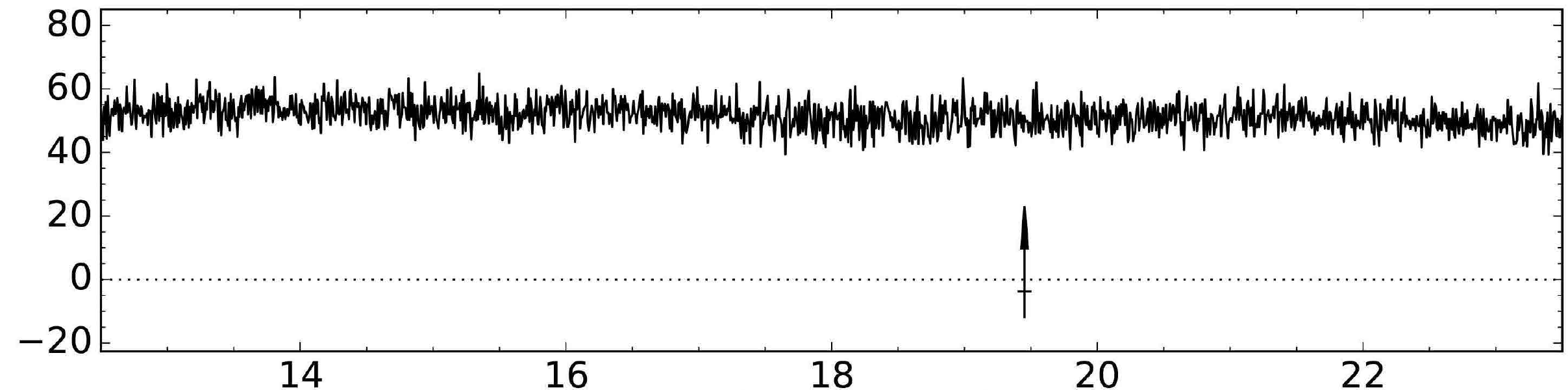}  
\end{subfigure}
\caption{\it{Continued.}}
\end{minipage}
\label{fig:all_spectra}
\end{figure*}

Table 4 gives the optical depth and H{\sc i} column density for each of the sources observed in 2013-2014. The 21-cm optical depth is given by
\begin{equation}
  \tau(\nu) = -\ln{\left(1 - {\Delta{S_\mathrm{line}(\nu)} \over f S_\mathrm{C}(\nu)}\right)},
\end{equation}
where $S_\mathrm{C}$ is the continuum flux density at the optical spectroscopic redshift, $\Delta{S_\mathrm{line}}$ is the spectral-line depth (detection) or 3-$\sigma$ rms noise (non-detection), and $f$ is the covering factor of the background source. In the optically thin regime ($\Delta{S_\mathrm{line}}/S_\mathrm{C} \lesssim$ 0.3), the above expression reduces to:
\begin{equation}
  \tau(\nu) \approx {\Delta{S_\mathrm{line}(\nu)} \over f S_\mathrm{C}} \approx {\tau_\mathrm{obs}(\nu) \over f},
\end{equation}
where $\tau_\mathrm{obs}(v) \equiv {\Delta{S_\mathrm{line}} \over S_\mathrm{C}}$ is the observed optical depth as a function of velocity. The H{\sc i} column density (in cm$^{-2}$) is related to the integrated optical depth (km~s$^{-1}$) as follows \cite[e.g.][]{Wolfe1975}:
\begin{eqnarray}
  N_\mathrm{HI} & = & 1.823 \times 10^{18}\,T_\mathrm{spin} \int{\tau(v)\mathrm{d}v} \nonumber \\
  & \approx & 1.823 \times 10^{18}\,{T_\mathrm{spin} \over f} \int{\tau_\mathrm{obs}(v)\mathrm{d}v},
\end{eqnarray}
where $T_\mathrm{spin}$ is the harmonic mean spin temperature of the gas (in K). 

\subsection{Notes on detections}

\begin{figure}
\includegraphics[scale=0.6, angle=-90]{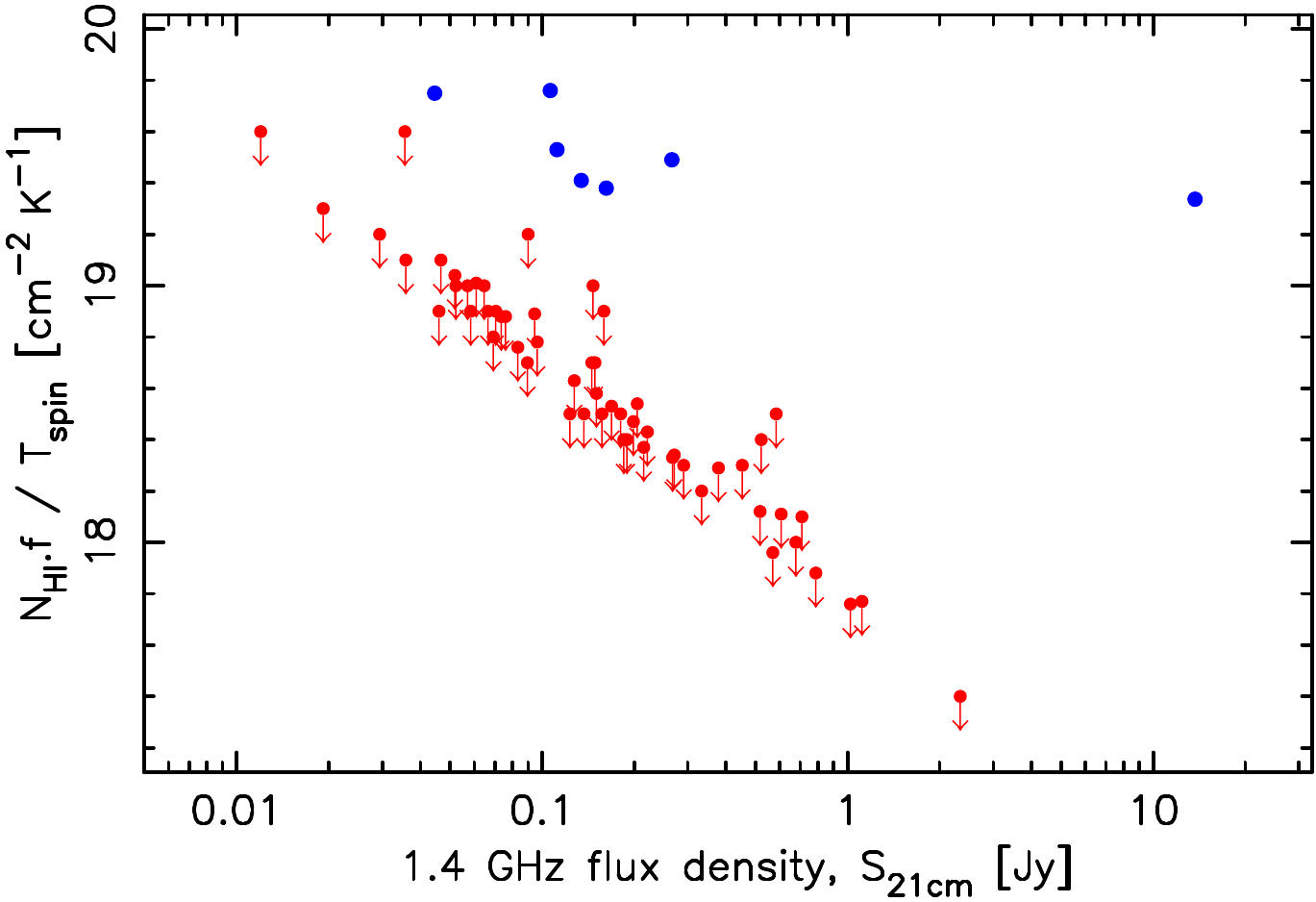}
\caption{The 21-cm line strength versus the measured flux density of the source. Blue points give our detections, and red the upper limits of our non-detections. A few sources had shorter integration times and so lie above the range for our detections (see Table 2).}
\label{figure:strength_flux}
\end{figure}

Our four new H{\sc i} detections have a large range of peak optical depths (from as high as $\tau_{\rm{obs}} \sim$ 87\% to a tentative detection at $\tau_{\rm{obs}} \sim$ 3\%). Our non-detections show a similar range of upper limits on the peak optical depths (Table~4), an expected result given the range of flux densities and sensitivity of our survey (Fig.~3). Some sources had higher upper limits found for their line strengths due to shorter integration times in the June 2013 observation run (Table~2). 

\textbf{J031552-190644 (z = 0.0671)} - A previously known strong H{\sc i} absorption feature \citep{Ledlow2001} is detected against the compact flat spectrum core of this well-studied FR-I radio source. The AGN is hosted by a disc-dominated spiral galaxy \citep{Ledlow1998}. The host galaxy is located close to the centre of the galaxy cluster A428. 

\cite{Ledlow2001} found that the very narrow peak absorption in the host galaxy gave a large apparent optical depth of 0.98 $\pm$ 0.06 that is only seen against the radio core and not against either of the two observed radio jets, hence ruling out a large, diffuse H{\sc i} halo. They measure a column density of N$_{\rm{HI}}$ $\simeq$ 1.82 $\times$ 10$^{18} \rm{\frac{T_{S}}{f}\tau} \Delta \nu_{\rm{FWHM}}$ cm$^{-2}$, with a FWHM velocity width of $\Delta \nu_{\rm{FWHM}}$~=~34~km~s$^{-1}$ giving a value of N$_{\rm{HI}}$ $\simeq$ 6.19 $\times$ 10$^{19} \rm{\frac{T_{S}}{f}\tau}$ cm$^{-2}$. \cite{Ledlow2001} suggests that, given this high optical depth, the absorbing gas is either of a high density, or the sightline goes through the disc in such a way to maximise the amount of absorbing gas (e.g. through a spiral arm or very dense cloud, possibly originating in the ISM of the host galaxy due to the narrow linewidth they observe). 

Our measurement of the peak optical depth is similar to that of \cite{Ledlow2001}; we find the H{\sc i} column density to be N$_{\rm{HI}}$ = 5.62 ($\pm$ 0.27) $\times$ 10$^{19}  \frac{T_{S}}{f}$ cm$^{-2}$ (Table 4). The absorption line feature in our spectrum is best fitted by two narrow Gaussians, both located around the spectroscopic optical redshift, and matches the narrow, deep feature that \cite{Ledlow2001} observed. 

\textbf{J060555-392905 (z = 0.0452)} - This is a flat-spectrum radio source with a new H{\sc i} detection. It was spectroscopically classified as an AGN source with only absorption features (Aa) within its optical spectrum by \cite{Mahony2011}. The presence of a foreground star affects both the spectrum and the visual classification of the morphology of the host galaxy (Fig.~4).

\begin{table*}
\normalsize
 \centering
  \caption{The inferred values for parameters from fitting a multiple Gaussian spectral-line model for the new sources we detect \mbox{H{\sc i}} absorption. See \citet{Allison2012} for parameters on the other H{\sc i} detections. $z_{\mathrm{centre}}$ is the redshift of the centre of the absorption component; $\Delta{v}_{\mathrm{FWHM},i}$ is the FWHM; $\Delta{S}_{i}$ is the depth and $R$ is the natural logarithm of the ratio of probability for this model versus the no spectral-line model (the Bayes odds ratio).}
  
  \begin{tabular}{lrrrrrrrr}
  \hline
AT20G name & $z_{\mathrm{centre}}$ & $\Delta{v}_{\mathrm{FWHM},i}$ & 1$\sigma$ error & $\Delta{S}_{i}$ & 1$\sigma$ error & $R$ \\
  & & (km\,s$^{-1}$) & (km\,s$^{-1}$)& (mJy) & (mJy)& & \\
 \hline
J031552-190644 & 0.0671 & 29 & $^{+4}_{-6}$ & 26 & $^{+2}_{-5}$ & 284.6$\pm$0.3\\ & 0.0672 & 19 & $^{+22}_{-~5}$ & 12 & $^{+3}_{-2}$ &  \\
J060555-392905 & 0.0448 & 50 & $^{+1}_{-1}$ & 51 & $^{+1}_{-1}$ & 471.4$\pm$0.3\\
J084452-100059 & 0.0423 & 450 & $^{+40}_{-80}$ & 4 & $^{+1}_{-1}$ & 13.8$\pm$0.3\\
J125711-172434 & 0.0474 & 100 & $^{+110}_{-~30}$ & 8 & $^{+1}_{-2}$ & 91.1$\pm$0.4\\ & 0.0477 & 210 & $^{+~10}_{-120}$ & 5 & $^{+2}_{-1}$ &  \\

\hline
\end{tabular}
\label{table:analysis_det}
\end{table*}

\begin{table*}
\normalsize
 \centering
  \caption{A summary of derived \mbox{H{\sc i}} absorption properties from model fitting for our 2013 and 2014 observing runs. See \citet{Allison2012} for information on the other observations. $\sigma_\mathrm{chan}$ is the estimated uncertainty per channel; $S_\mathrm{C}$ is the spectral flux density of the continuum model at either the position of peak absorption or the optical redshift estimate; 1$\sigma$ error is the continuum model's flux density error, $\Delta{S}_\mathrm{line}$ is the peak spectral-line depth; $\tau_\mathrm{obs,peak}$ is the observed peak optical depth and $\int{\tau_\mathrm{obs}\mathrm{d}v}$ is the observed velocity integrated optical depth. Upper limits assumed a single Gaussian spectral line of FWHM = 30 kms$^{-1}$, and peak depth = 3$\sigma_\mathrm{chan}$.}
\begin{tabular}{lcrrrrrrrr}
  \hline
      AT20G name & $\sigma_\mathrm{chan}$ & $S_\mathrm{C}$ & 1$\sigma$ error & $\Delta{S}_\mathrm{line}$ & $\tau_\mathrm{obs,peak}$ & $\int{\tau_\mathrm{obs}\mathrm{d}v}$ & $\log_{10}{\left(\mathrm{N}_\mathrm{HI}f\over T_\mathrm{spin}\right)}$ \\
      & (mJy) & (mJy) & (mJy) & (mJy) & & (km\,s$^{-1}$) & cm$^{-2}$ K$^{-1}$\\

 \hline
J005734-012328 & 3.70 & 677.2 & $^{+0.5}_{-0.5}$ & \textless11.1 & \textless0.02 & \textless0.52 & \textless18.0\\

J011132-730209 & 2.96 & 66.5 & $^{+0.1}_{-0.1}$ & \textless8.9 & \textless0.13 & \textless4.26 & \textless18.9\\

J025955-123635 & 4.93 & 452.1 & $^{+0.2}_{-0.2}$ & \textless14.8 & \textless0.03 & \textless1.04 & \textless18.3\\

J031357-395403 & 7.79 & 521.4 & $^{+0.2}_{-0.2}$ & \textless23.4 & \textless0.04 & \textless1.43 & \textless18.4\\

J031552-190644 & 2.00 & 44.5 & $^{+0.1}_{-0.1}$ & 25.9$^{+1.5}_{-1.6}$ & 0.87$^{+0.09}_{-0.08}$ & 31.16$^{+1.73}_{-1.76}$ & 19.75$^{+0.02}_{-0.02}$\\

J033114-524148 & 2.66 & 29.4 & $^{+0.1}_{-0.1}$ & \textless8.0 & \textless0.27 & \textless8.66 & \textless19.2\\

J033913-173600 & 2.40 & 123.4 & $^{+0.1}_{-0.1}$ & \textless7.2 & \textless0.06 & \textless1.86 & \textless18.5\\

J034630-342246 & 2.99 & 57.1 & $^{+0.1}_{-0.1}$ & \textless9.0 & \textless0.16 & \textless5.01 & \textless19.0\\

J035145-274311 & 2.84 & 12.0 & $^{+0.1}_{-0.2}$ & \textless8.5 & \textless0.71 & \textless22.76 & \textless19.6\\

J035257-683117 & 4.33 & 149.0 & $^{+0.2}_{-0.2}$ & \textless13.0 & \textless0.09 & \textless2.78 & \textless18.7\\

J035410-265013 & 3.26 & 70.5 & $^{+0.1}_{-0.1}$ & \textless9.8 & \textless0.14 & \textless4.43 & \textless18.9\\

J043022-613201 & 3.29 & 181.0 & $^{+0.1}_{-0.1}$ & \textless9.9 & \textless0.05 & \textless1.74 & \textless18.5\\

J060555-392905 & 2.96 & 106.4 & $^{+0.1}_{-0.1}$ & 34.8$^{+1.4}_{-1.5}$ & 0.40$^{+0.02}_{-0.02}$ & 31.86$^{+1.41}_{-1.44}$ & 19.76$^{+0.02}_{-0.02}$\\

J065359-415144 & 7.49 & 159.4 & $^{+0.3}_{-0.3}$ & \textless22.5 & \textless0.14 & \textless4.50 & \textless18.9\\

J084452-100059 & 2.65 & 134.3 & $^{+0.2}_{-0.2}$ & 4.1$^{+0.6}_{-0.7}$ & 0.03$^{+0.01}_{-0.01}$ & 14.20$^{+2.64}_{-2.71}$ & 19.41$^{+0.08}_{-0.08}$\\

J090802-095937 & 2.57 & 137.0 & $^{+0.1}_{-0.1}$ & \textless7.7 & \textless0.06 & \textless1.80 & \textless18.5\\

J091856-243829 & 2.04 & 53.1 & $^{+0.1}_{-0.1}$ & \textless6.1 & \textless0.12 & \textless3.68 & \textless18.8\\

J114539-105350 & 3.75 & 46.7 & $^{+0.1}_{-0.1}$ & \textless11.2 & \textless0.24 & \textless7.69 & \textless19.1\\

J123148-321314 & 8.71 & 146.9 & $^{+0.3}_{-0.3}$ & \textless26.1 & \textless0.18 & \textless5.68 & \textless19.0\\

J125615-114635 & 2.82 & 89.5 & $^{+0.1}_{-0.1}$ & \textless8.5 & \textless0.09 & \textless3.02 & \textless18.7\\

J125711-172434 & 2.83 & 111.9 & $^{+0.1}_{-0.1}$ & 10.8$^{+1.2}_{-1.3}$ & 0.10$^{+0.01}_{-0.01}$ & 18.46$^{+1.76}_{-1.65}$ & 19.53$^{+0.04}_{-0.04}$\\

J135036-163449 & 8.44 & 35.6 & $^{+0.3}_{-0.3}$ & \textless25.3 & \textless0.71 & \textless22.72 & \textless19.6\\

J135607-172433 & 2.70 & 185.2 & $^{+0.1}_{-0.1}$ & \textless8.1 & \textless0.04 & \textless1.40 & \textless18.4\\

J140912-231550 & 11.40 & 583.4 & $^{+0.4}_{-0.4}$ & \textless34.2 & \textless0.06 & \textless1.87 & \textless18.5\\

J151741-242220 & 3.22 & 2336.3 & $^{+0.2}_{-0.2}$ & \textless9.7 & \textless0.01 & \textless0.13 & \textless17.4\\

J165710-735544 & 2.48 & 58.4 & $^{+0.1}_{-0.1}$ & \textless7.4 & \textless0.13 & \textless4.07 & \textless18.9\\

J181857-550815 & 2.37 & 35.8 & $^{+0.1}_{-0.1}$ & \textless7.1 & \textless0.20 & \textless6.33 & \textless19.1\\

J191457-255202 & 2.99 & 189.4 & $^{+0.1}_{-0.1}$ & \textless9.0 & \textless0.05 & \textless1.51 & \textless18.4\\

J204552-510627 & 3.52 & 64.7 & $^{+0.1}_{-0.1}$ & \textless10.6 & \textless0.16 & \textless5.21 & \textless19.0\\

J205306-162007 & 2.41 & 69.2 & $^{+0.1}_{-0.1}$ & \textless7.2 & \textless0.10 & \textless3.34 & \textless18.8\\

J205754-662919 & 2.29 & 46.0 & $^{+0.1}_{-0.1}$ & \textless6.9 & \textless0.15 & \textless4.77 & \textless18.9\\

J212222-560014 & 2.43 & 19.2 & $^{+0.1}_{-0.1}$ & \textless7.3 & \textless0.38 & \textless12.15 & \textless19.3\\

J214824-571351 & 8.84 & 90.1 & $^{+0.3}_{-0.3}$ & \textless26.5 & \textless0.29 & \textless9.40 & \textless19.2\\

J220538-053531 & 2.94 & 157.1 & $^{+0.1}_{-0.1}$ & \textless8.8 & \textless0.06 & \textless1.79 & \textless18.5\\

J221220-251829 & 4.57 & 145.6 & $^{+0.1}_{-0.1}$ & \textless13.7 & \textless0.09 & \textless3.01 & \textless18.7\\

J234205-160840 & 2.78 & 52.2 & $^{+0.1}_{-0.1}$ & \textless8.3 & \textless0.16 & \textless5.10 & \textless19.0\\

\hline

\end{tabular}
\label{table:analysis_all}
\end{table*}

\begin{figure*}
\begin{minipage}{\textwidth}
\centering
\captionsetup[subfigure]{aboveskip=-2pt,belowskip=0pt}
\subcaption{J031552-190644}
\begin{subfigure}[b]{0.45\textwidth}
  \vskip 0pt
  \centering
  \includegraphics[width=0.69\linewidth]{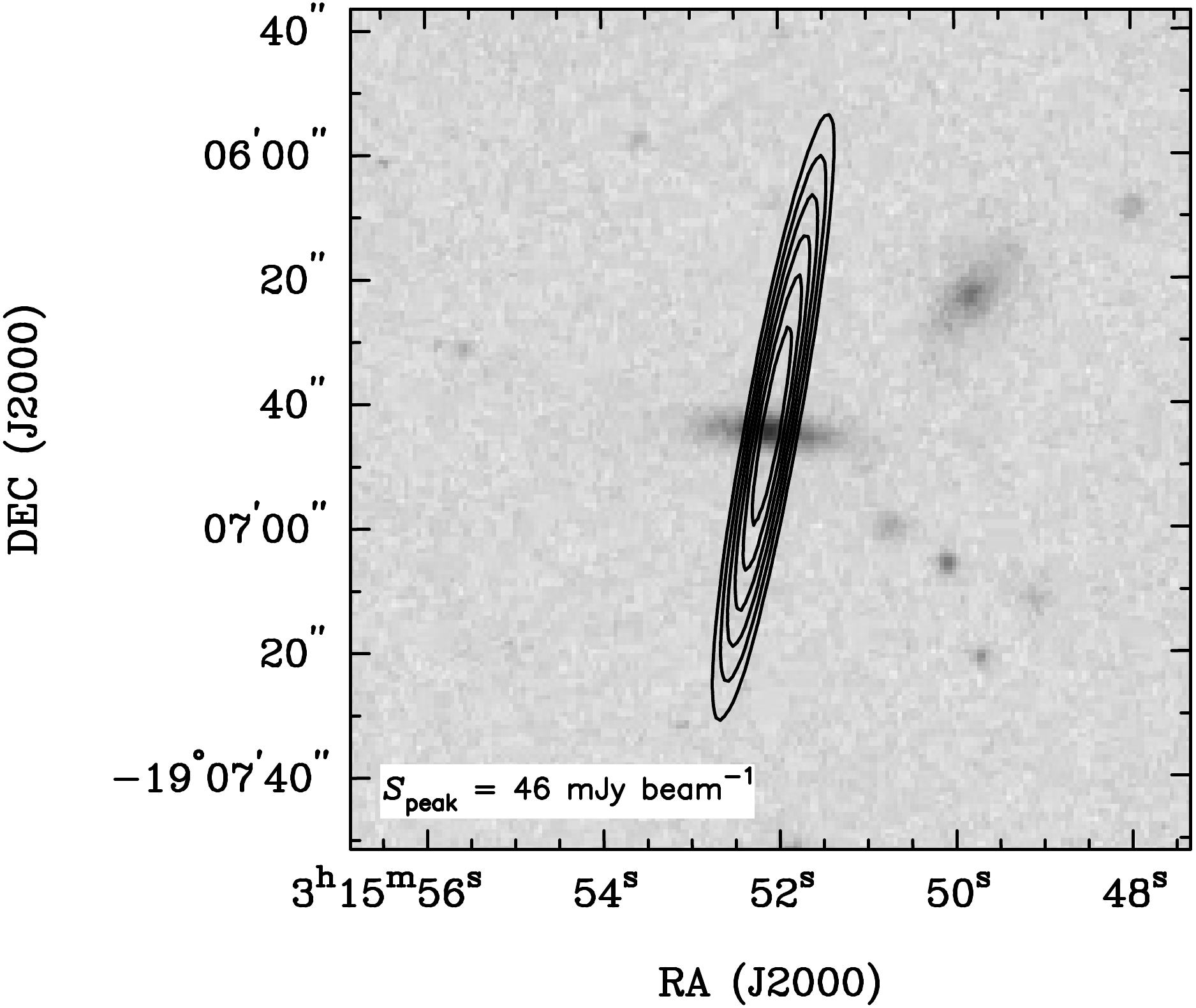}
\end{subfigure}\hspace{-5em}%
\begin{subfigure}[b]{0.45\textwidth}
  \vskip 0pt
  \centering
  \includegraphics[width=0.85\linewidth]{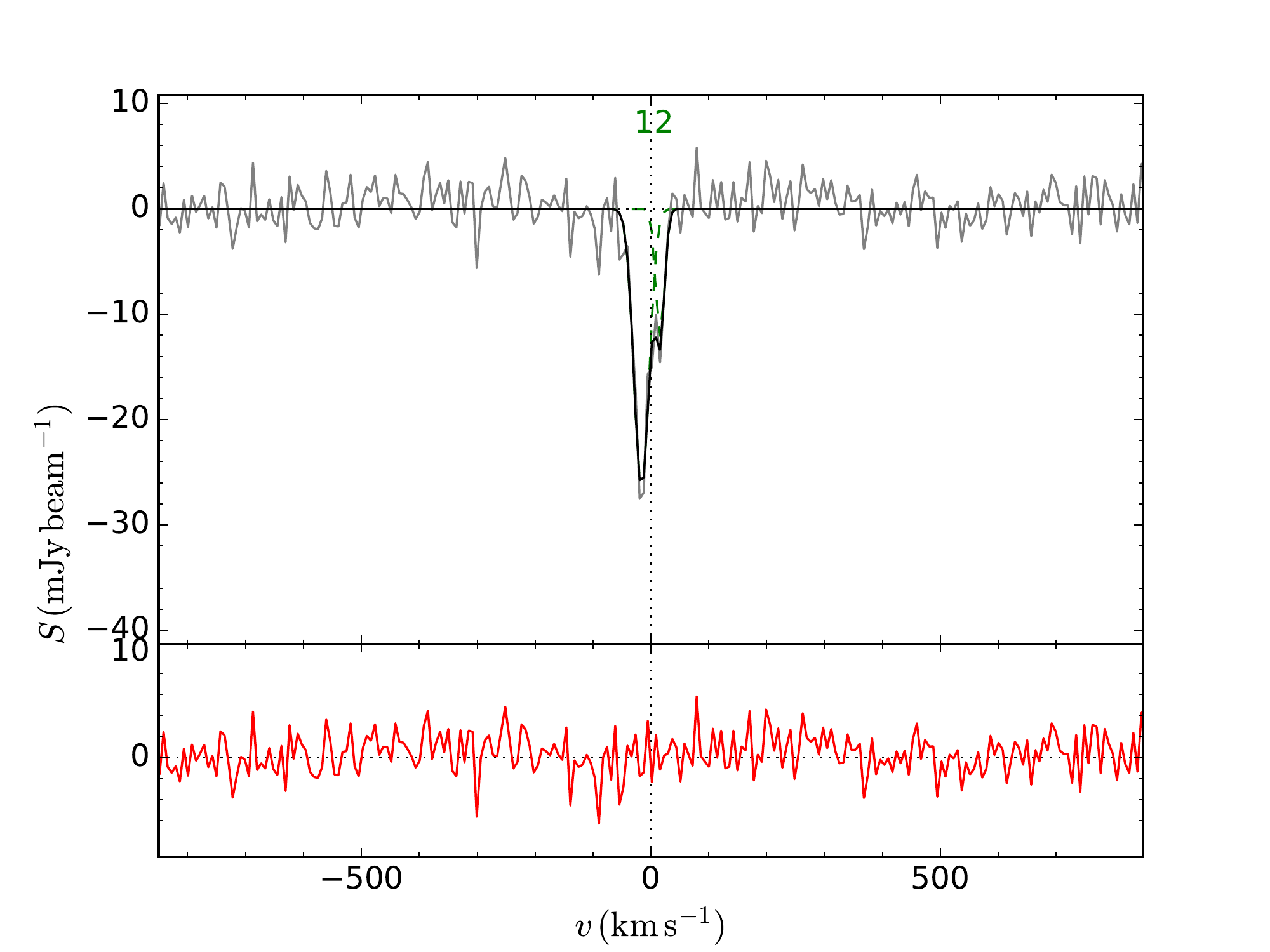}
\end{subfigure}
\subcaption{J060555-392905}
\begin{subfigure}[b]{0.45\textwidth}
  \vskip 0pt
  \centering
  \includegraphics[width=0.69\linewidth]{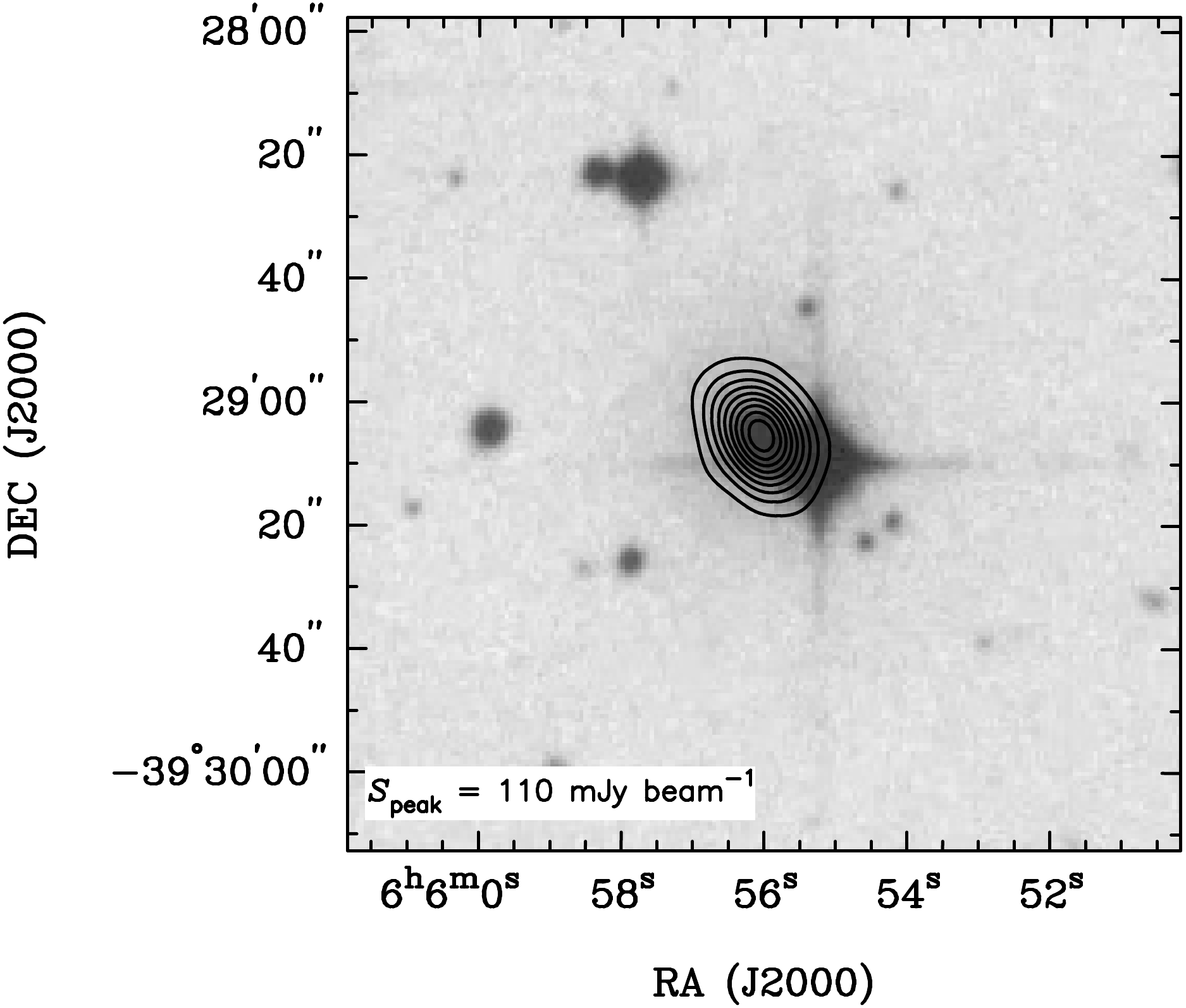}
\end{subfigure}\hspace{-5em}%
\begin{subfigure}[b]{0.45\textwidth}
  \vskip 0pt
  \centering
  \includegraphics[width=0.85\linewidth]{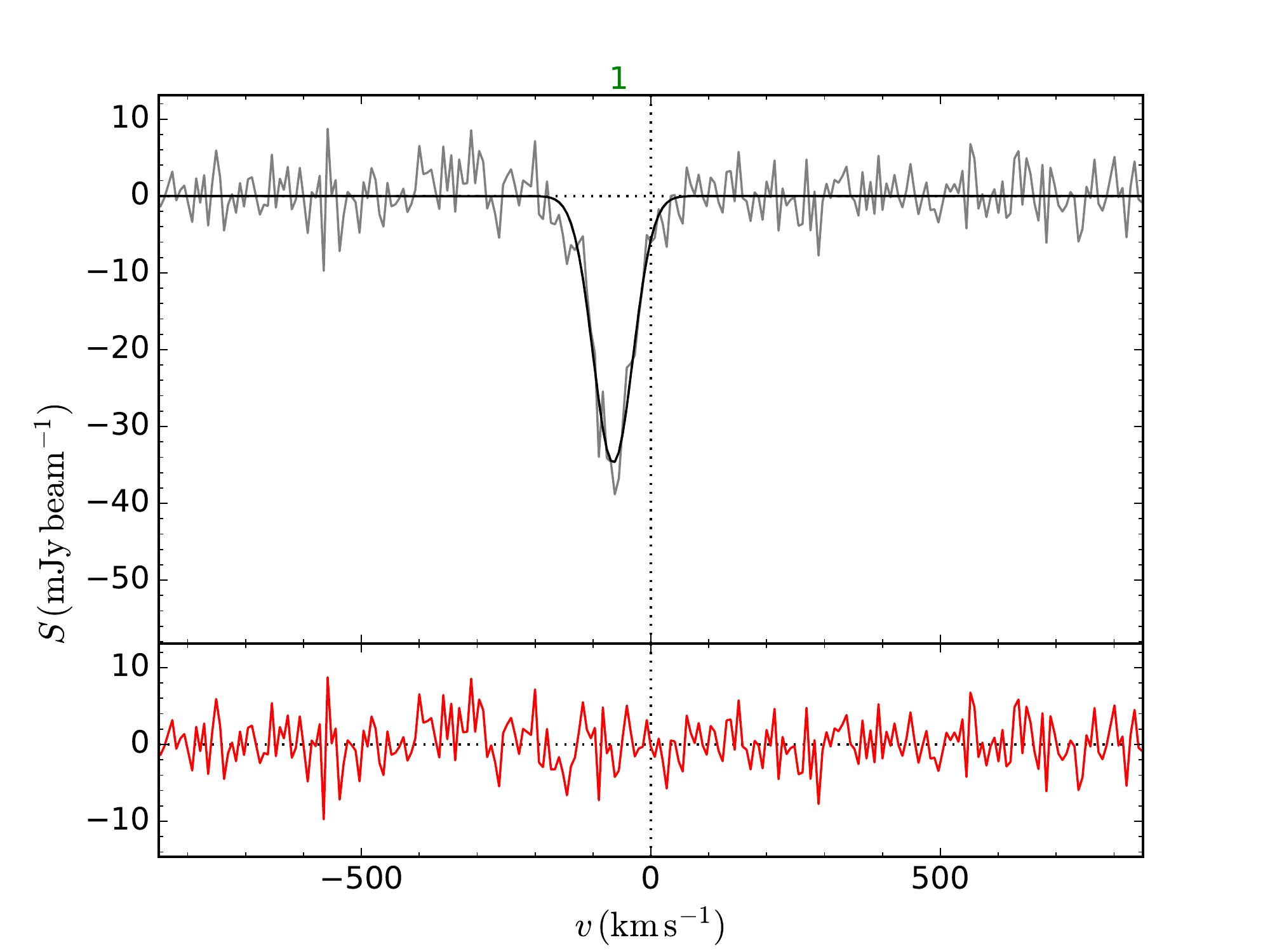}
\end{subfigure}
\subcaption{J084452-100059}
\begin{subfigure}[b]{0.45\textwidth}
  \vskip 0pt
  \centering
  \includegraphics[width=0.69\linewidth]{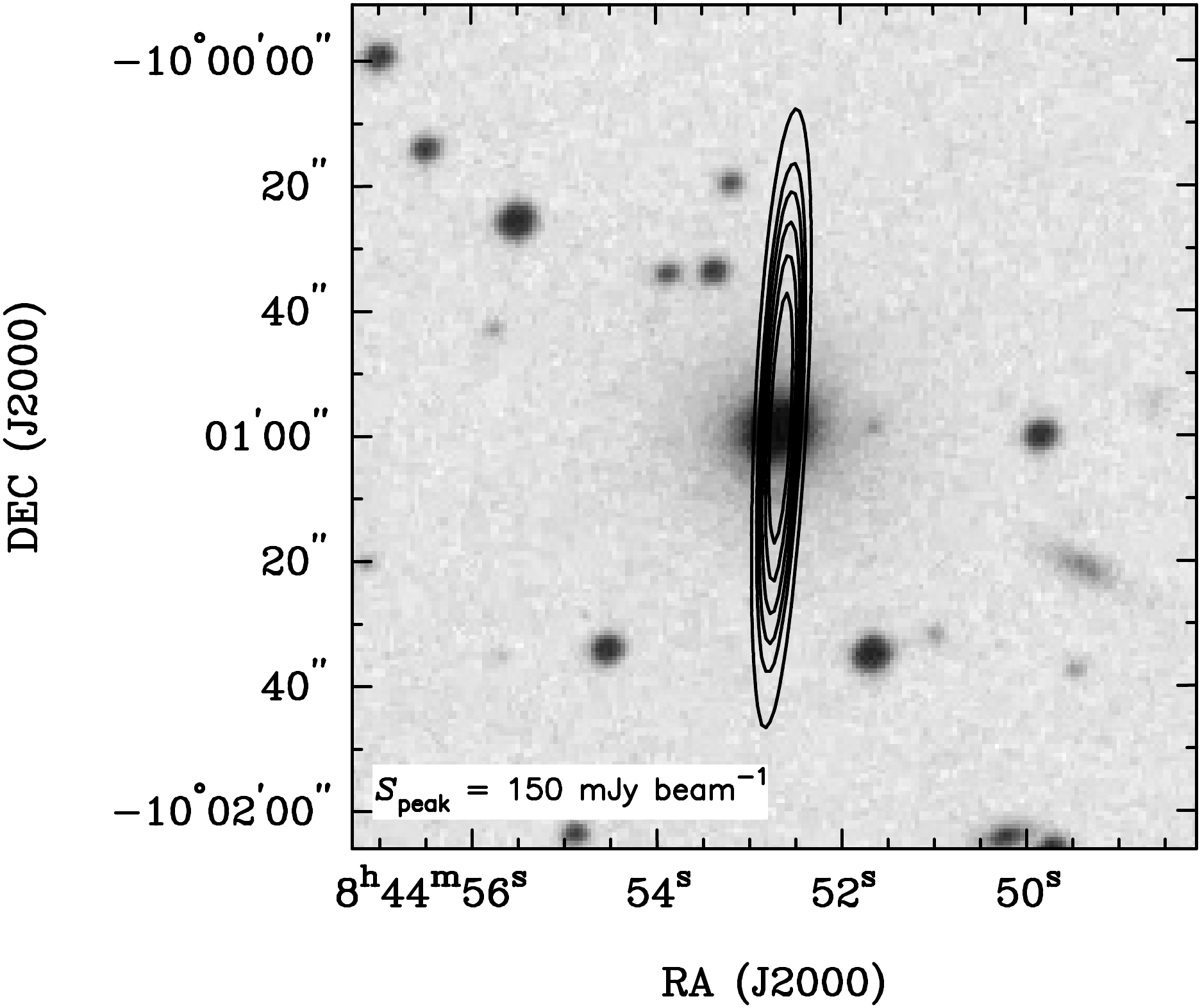}
\end{subfigure}\hspace{-5em}%
\begin{subfigure}[b]{0.45\textwidth}
  \vskip 0pt
  \centering
  \includegraphics[width=0.85\linewidth]{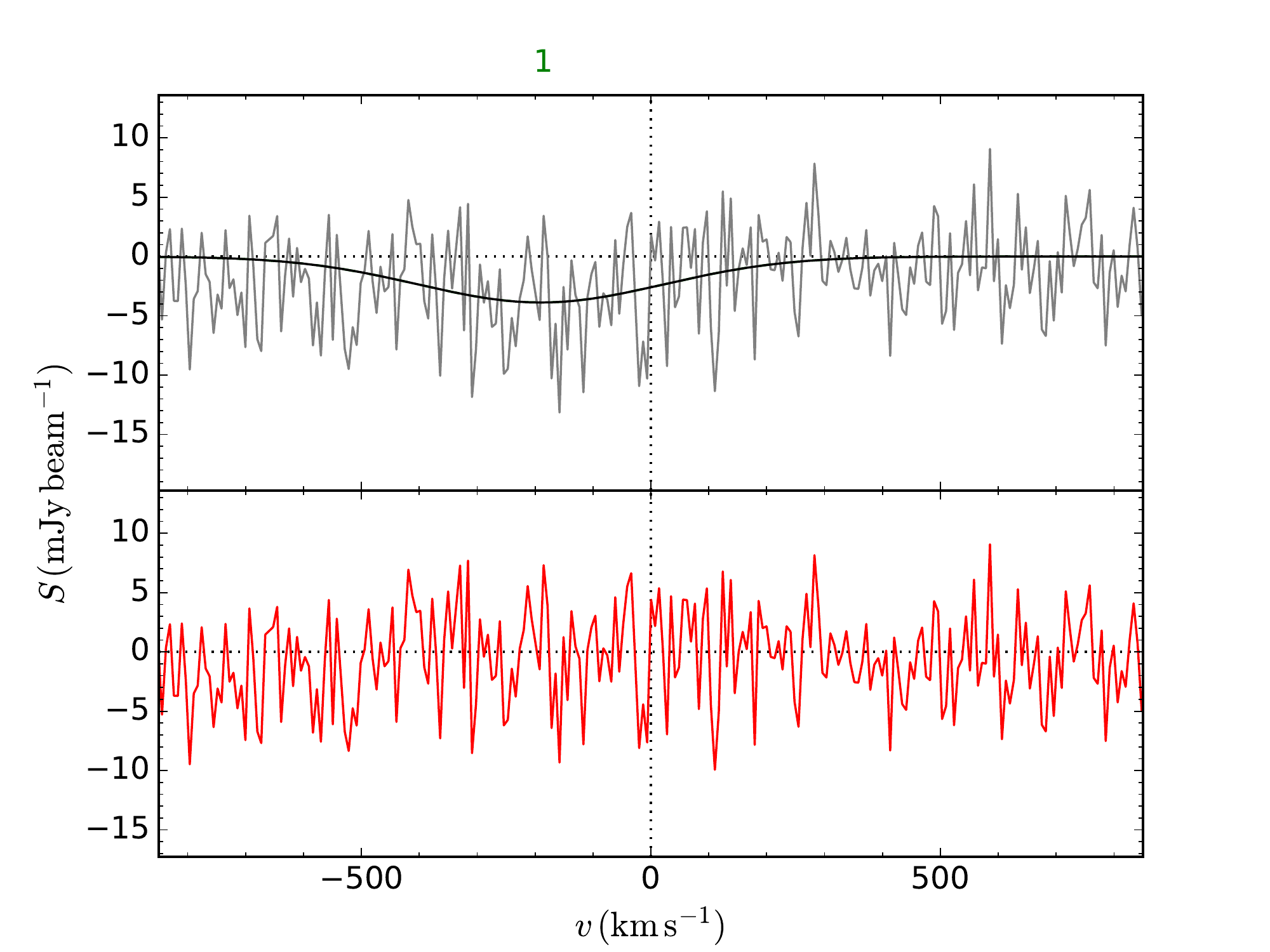}
\end{subfigure}
\subcaption{J125711-172434}
\begin{subfigure}[b]{0.45\textwidth}
  \vskip 0pt
  \centering
  \includegraphics[width=0.69\linewidth]{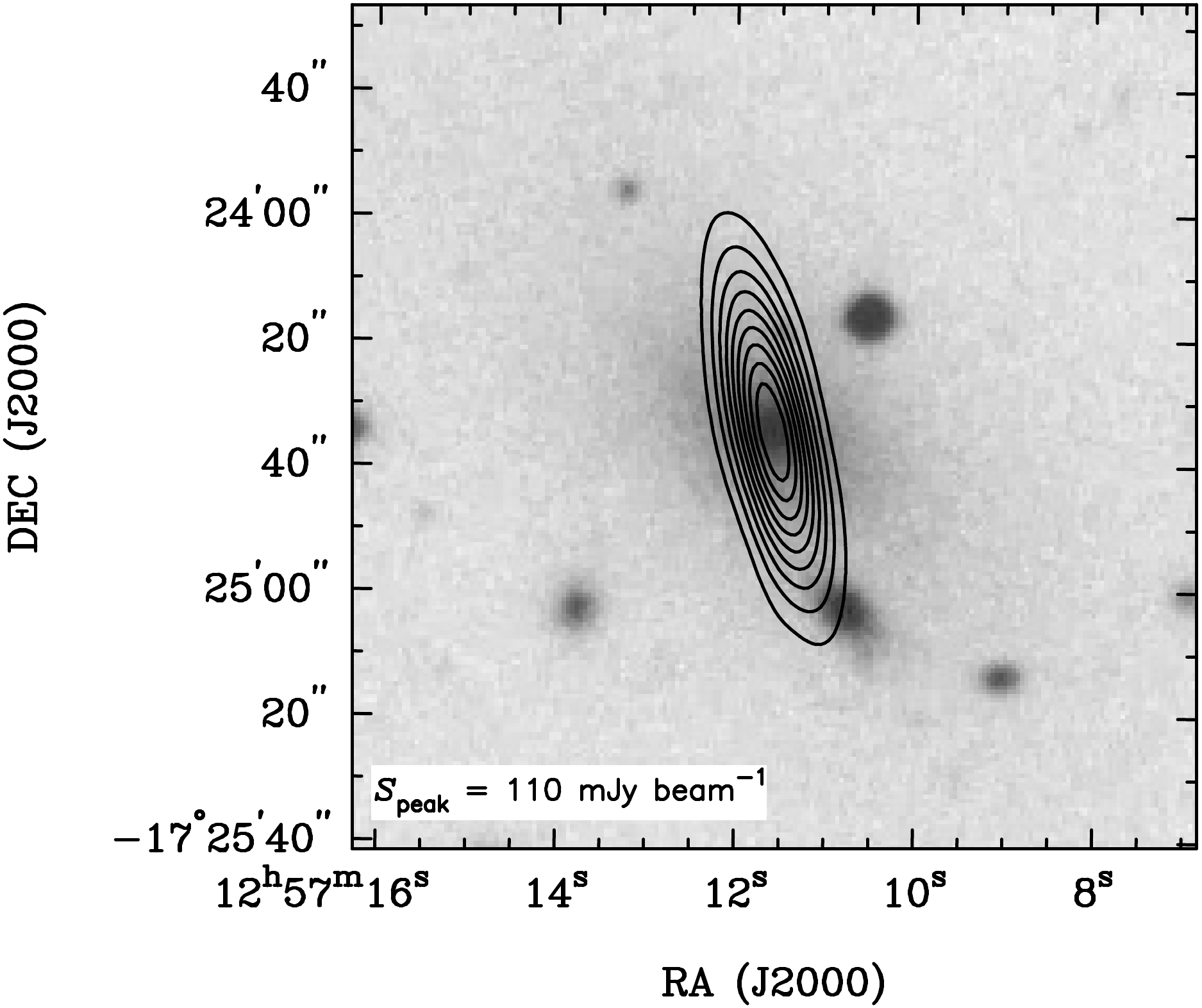}
\end{subfigure}\hspace{-5em}%
\begin{subfigure}[b]{0.45\textwidth}
  \vskip 0pt
  \centering
  \includegraphics[width=0.85\linewidth]{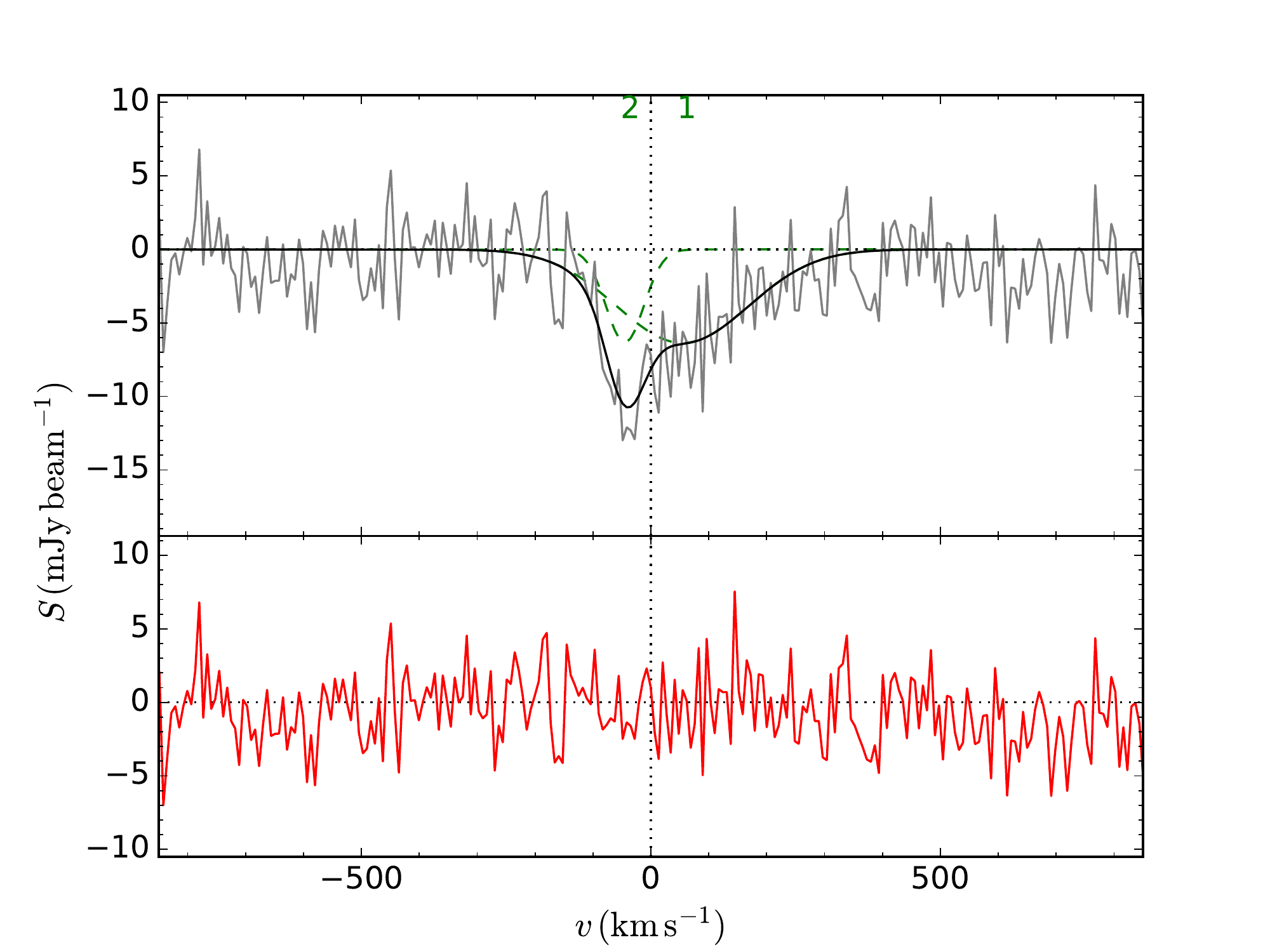}
\end{subfigure}
\end{minipage}
\caption{Detections of H{\sc i} 21-cm absorption. Left: radio image taken during observations superimposed onto a grey scale SuperCosmos Sky Survey image \citep{Hambly2001}. Contours are set to 10\% intervals of the peak flux density of each source. Right: the corresponding spectra from the ATCA observations centred at the optical spectroscopic redshift. The top panel shows the data following subtraction of the best-fitting continuum model (grey), and the best-fitting multiple Gaussian spectral-line model (black). The individual components are displayed in green. The red line in the bottom half represents the data after subtraction of this model fit.}
\label{figure:spectra_detection}
\end{figure*}
\begin{figure*}
\ContinuedFloat
\begin{minipage}{\textwidth}
\centering
\captionsetup[subfigure]{aboveskip=-2pt,belowskip=0pt}
\subcaption{J001605-234352}
\begin{subfigure}[b]{0.45\textwidth}
  \vskip 0pt
  \centering
  \includegraphics[width=0.69\linewidth]{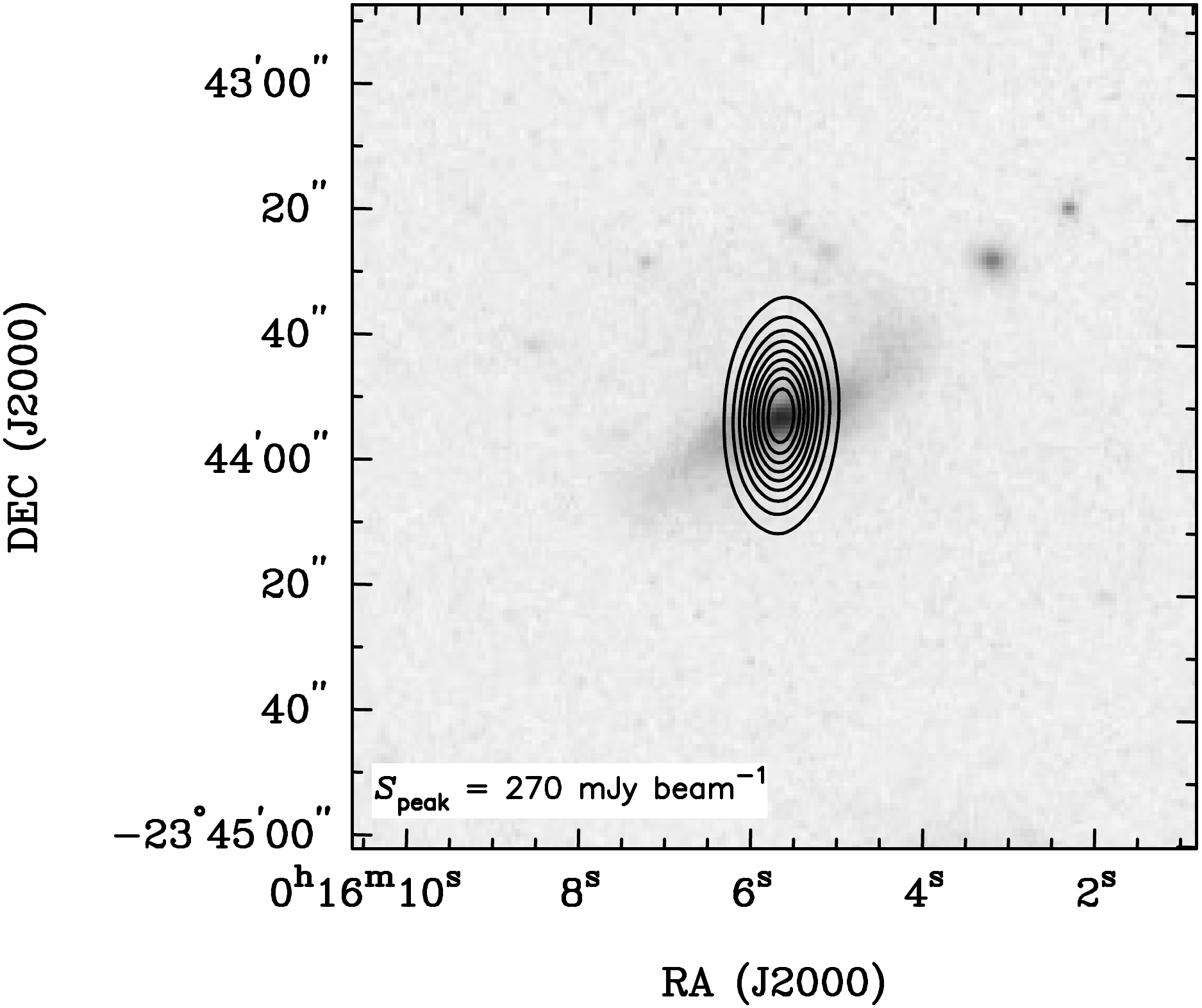}
\end{subfigure}\hspace{-5em}%
\begin{subfigure}[b]{0.45\textwidth}
  \vskip 0pt
  \centering
  \includegraphics[width=0.85\linewidth]{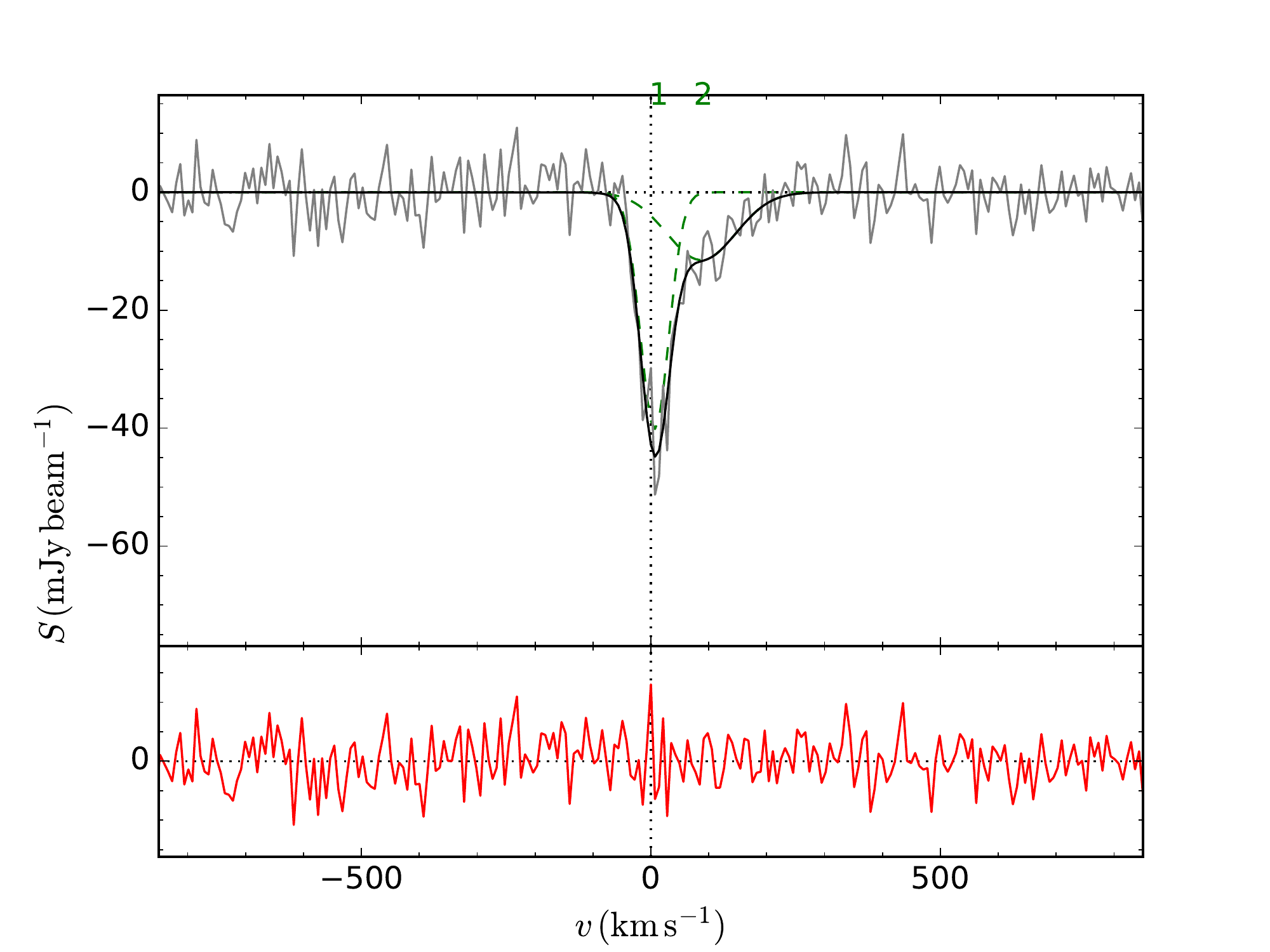}
\end{subfigure}
\subcaption{J181934-634548}
\begin{subfigure}[b]{0.45\textwidth}
  \vskip 0pt
  \centering
  \includegraphics[width=0.69\linewidth]{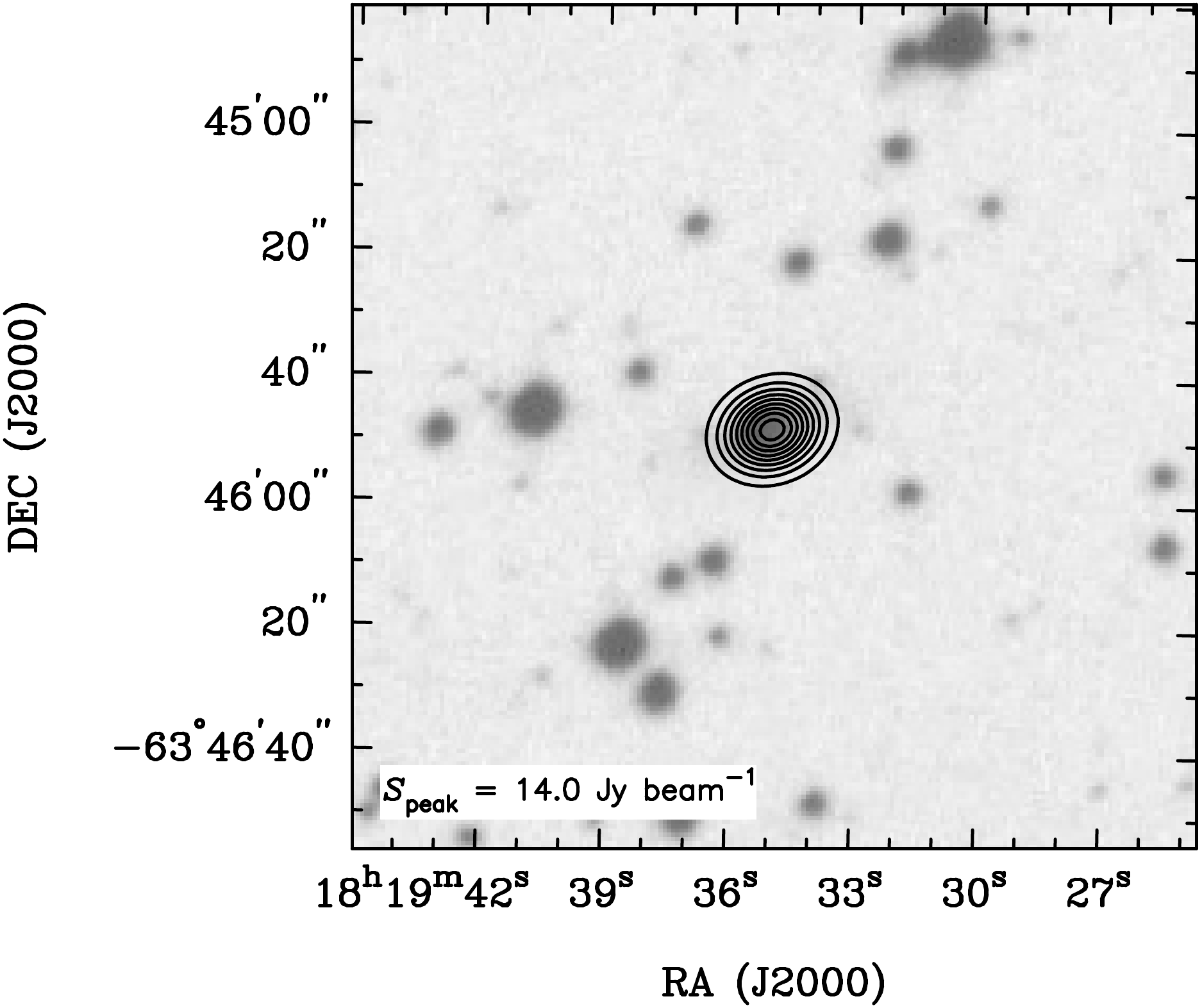}
\end{subfigure}\hspace{-5em}%
\begin{subfigure}[b]{0.45\textwidth}
  \vskip 0pt
  \centering
  \includegraphics[width=0.85\linewidth]{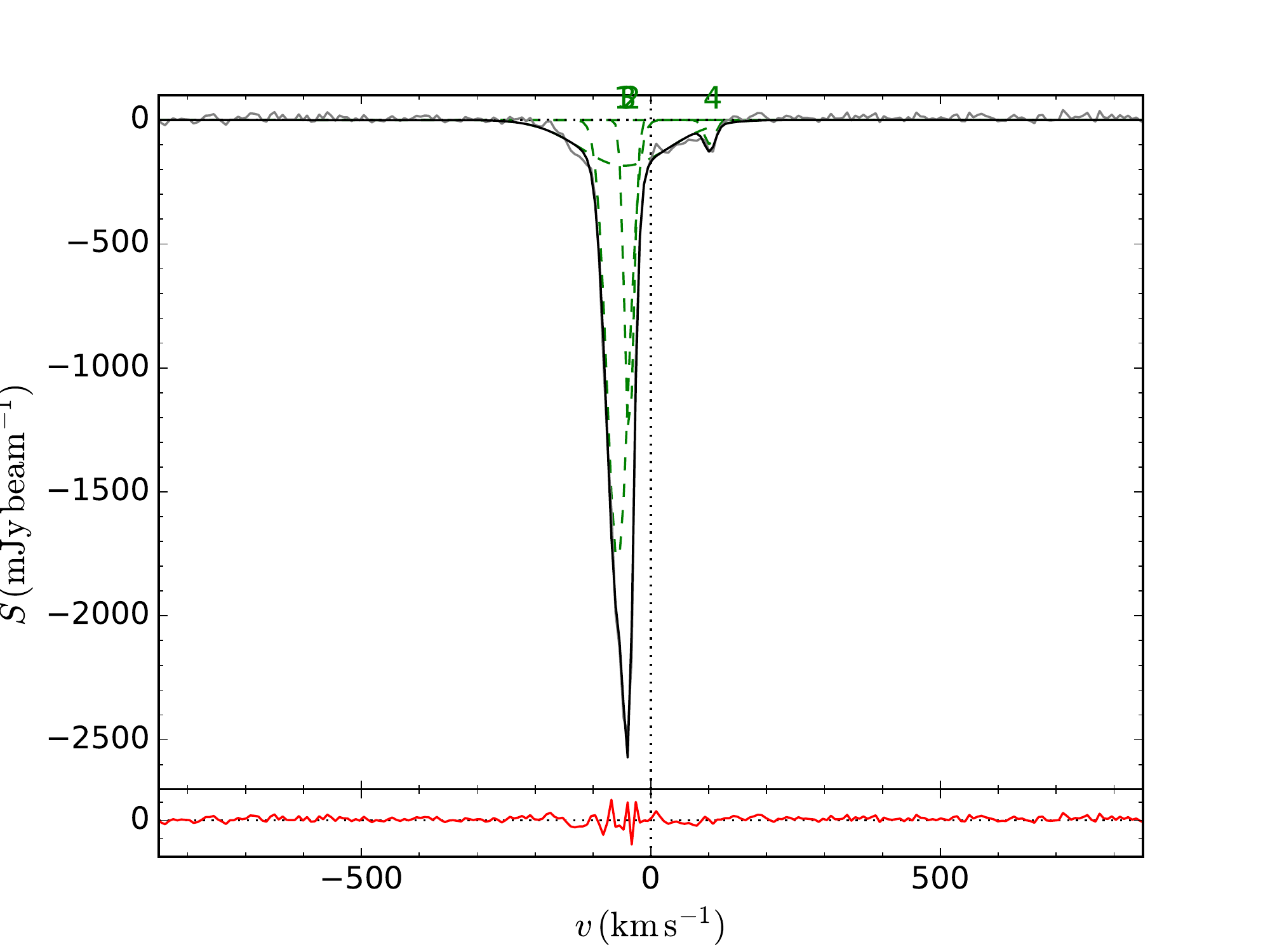}
\end{subfigure}
\subcaption{J205401-424238}
\begin{subfigure}[b]{0.45\textwidth}
  \vskip 0pt
  \centering
  \includegraphics[width=0.69\linewidth]{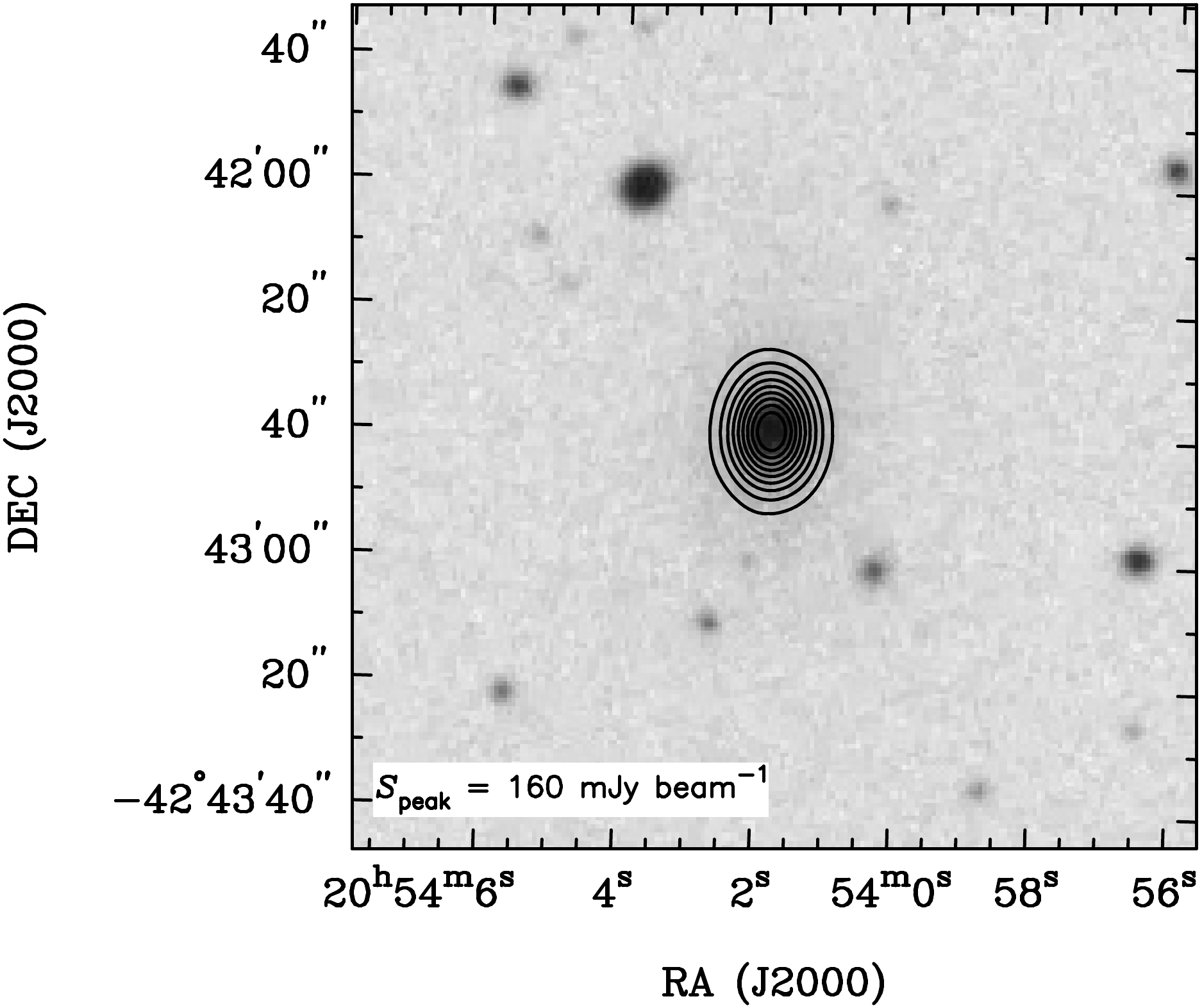}
\end{subfigure}\hspace{-5em}%
\begin{subfigure}[b]{0.45\textwidth}
  \vskip 0pt
  \centering
  \includegraphics[width=0.85\linewidth]{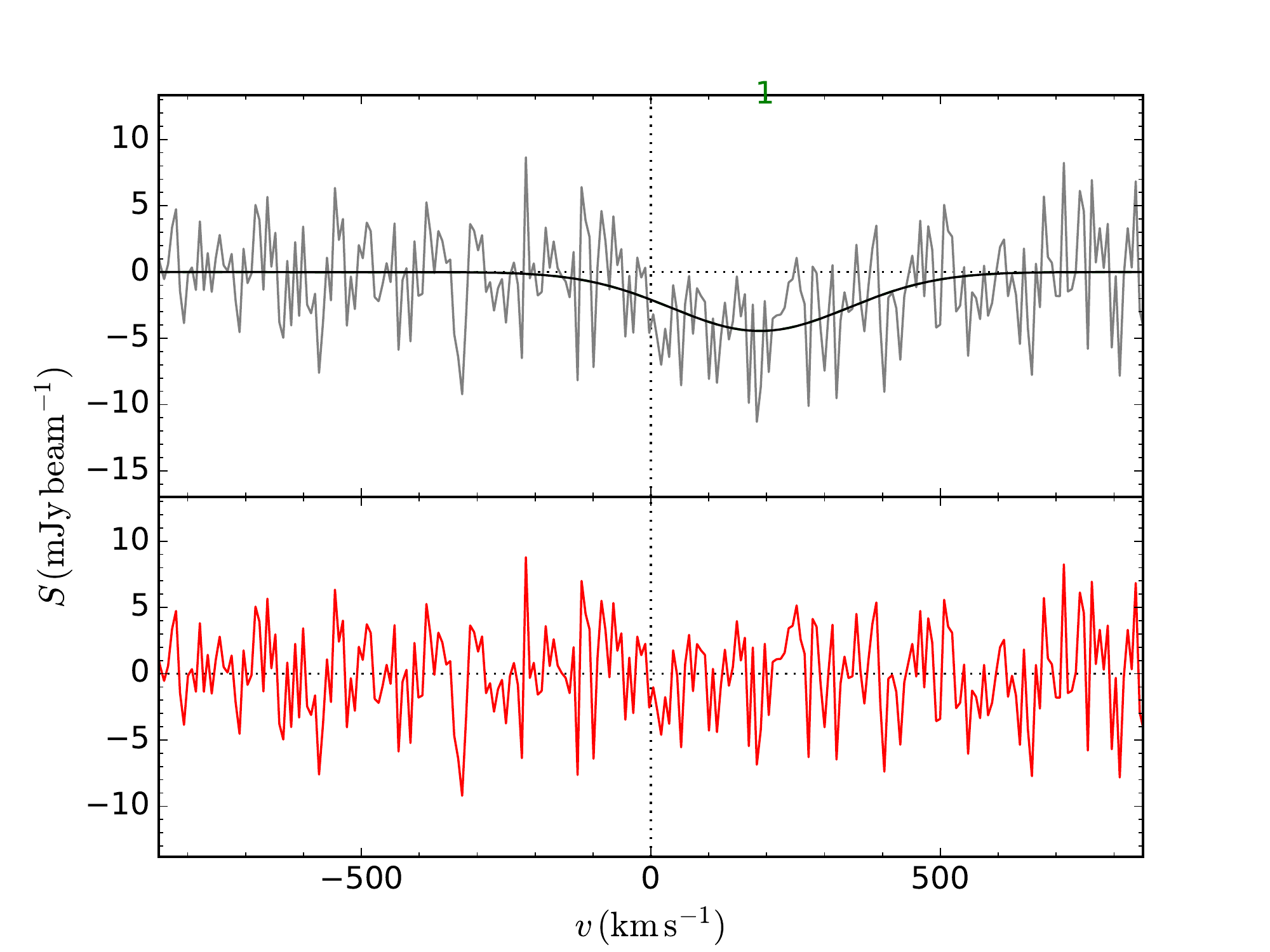}
\end{subfigure}
\end{minipage}
\caption{Continued. These three H{\sc i} detections and observations were published by \citet{Allison2012} and updated here. The tentative detection in J205401-424238 was verified by \citet{Allison2013}.}
\end{figure*}

A one-component, narrow Gaussian fit slightly blueshifted ($\sim$ 68 $\pm$ 4 km~s$^{-1}$) from the spectroscopic optical redshift was the most statistically likely in modelling (Table 3). Given the narrow absorption line and its slightly blueshifted location, the cold H{\sc i} gas we observe could be located close to the host galaxy's AGN, and perhaps be an outflow of a clump of gas. However, the galaxy's mid-infrared colours suggest at a star-forming host galaxy with a low AGN accretion rate (Section 5.1.1), and hence the absorption could be due to a gas-rich galactic disc. More information about the morphology of this host galaxy is required to aid this interpretation. 

\textbf{J084452-100059 (z = 0.0423)} - We find a tentative new detection of a weak broad line against this steep-spectrum source (spectral index of $\alpha = -0.7$). As with J060555-392905, it was spectroscopically identified as an Aa galaxy in \cite{Mahony2011}. It is classified as a FR-I type galaxy due to its extended low-frequency emission in NVSS. Beyond that there is a lack of information regarding the morphology of the galaxy within the literature. Its infrared colours are similar to those of elliptical galaxies, and its optical image suggests this morphology as well (Fig.~2). 

The feature is best modelled by a wide, weak absorption line centred slightly blueward of the source galaxy's optical redshift (shifted by $\sim$ 180 km s$^{-1}$). This detection has low significance compared to the other detections (Table~3). However, the logarithmic odds ratio of R = 13.9~$\pm$~0.3  means the spectral line model is favoured over the continuum fit (where there is a higher probability for a model to be true relative to another if it has a (log) Bayes odds ratio R $>$ 1). 

It is a rare broad feature without an associated narrow, deeper line. Like the previously reported H{\sc i} detection in J205401-424238 \citep{Allison2012}, this line is apparent throughout the observation in both polarization feeds, and does not appear in spectra that are extracted from spatial positions significantly off the source. If real, this is the second such feature within our sample \cite[Fig. 4, and][]{Allison2013} that may have been dismissed by visual inspection if not for our spectral line detection technique. To fully verify this detection however, we require further more sensitive 21-cm observations. 

\begin{figure}
\includegraphics[width=1.0\linewidth]{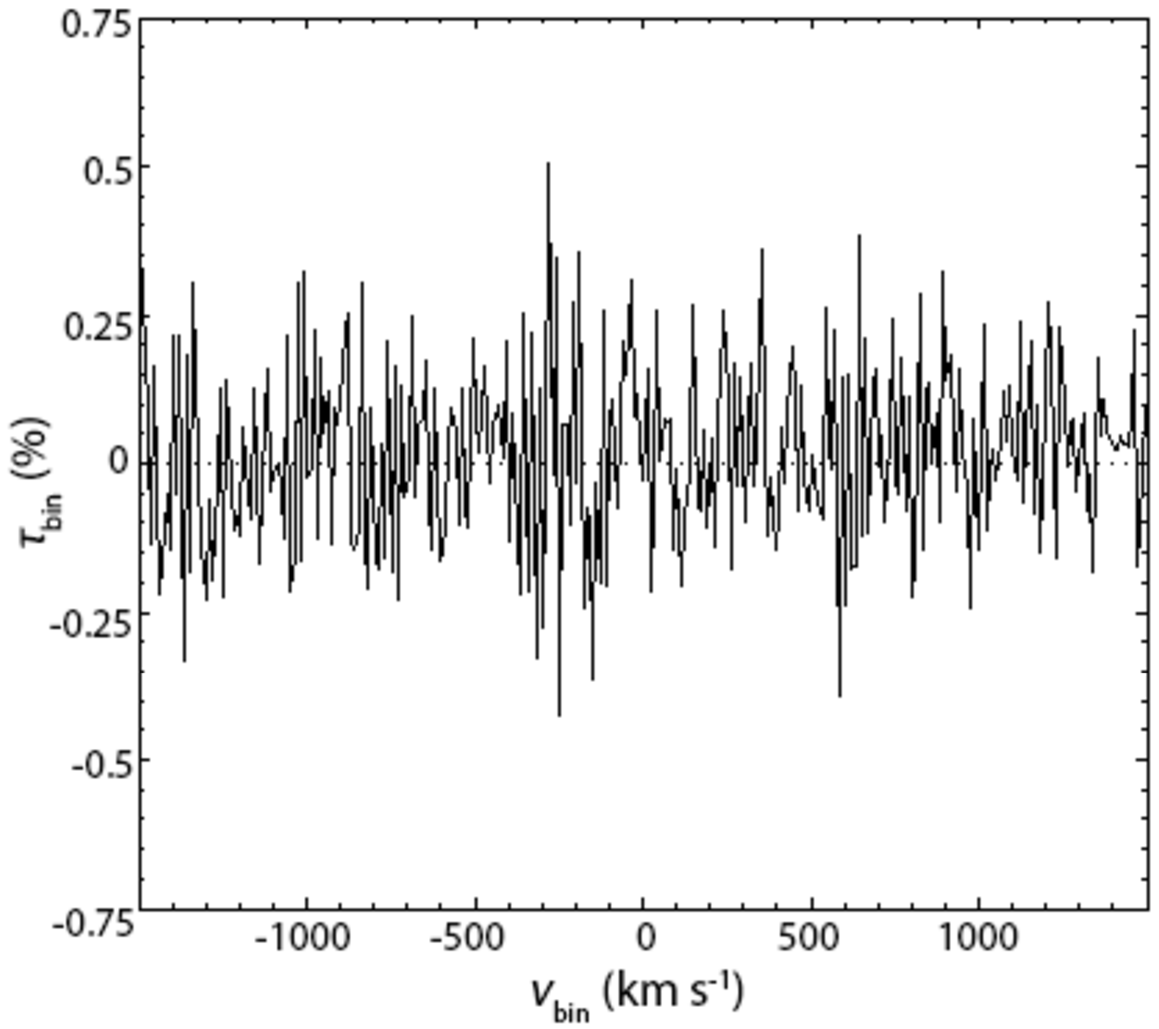}
\caption{The binned optical depth spectra for the 57 sources with non-detections of H{\sc i} absorption in our sample. The binned velocity is given relative to the optical redshifts (see Table 2). The data have been binned to 10 km~s$^{-1}$, similar to the spectral resolution of the individual spectra. A velocity range of ±1500~km~s$^{-1}$ is shown for clarity. We find no evidence for a statistical detection of H{\sc i} absorption.}
\label{figure:stacked}
\end{figure}

As the WISE mid-infrared colours indicate this galaxy has a low star formation rate, the absorbing gas we observe may only lie close to the AGN rather than being distributed throughout the galaxy, possibly fuelling the radio AGN \citep{Maccagni2014, Maccagni2016}. It may also represent an outflow of gas created by the AGN feedback, given the blueward offset of the line peak. 

\textbf{J125711-172434 (z = 0.0475)} - We find a new detection against this flat spectrum radio source, which is classified as a cD4 elliptical galaxy with a very faint yet significantly extended corona \citep{deVaucouleurs1991}. It is also the brightest source in the cluster A1644 \citep{Hoessel1980, Postman1995}. This cluster is the host of a few  X-ray sources, including 3XMM J125712.0-172431 which has a X-ray flux of 1.39 $\times$ 10$^{-13}$ ergs cm$^{-2}$ s$^{-1}$ \citep{Rosen2015}. This X-ray activity could be attributed to the AGN in the host galaxy, or with the cluster's hot intergalactic gas. The galaxy was found to only have absorption line features within its optical spectrum \citep{Mahony2011}, which contrasts with the X-ray emission detected.

We find two separate components in our spectral line modelling. The first is a narrow feature centred slightly blueward ($\sim$ 35 km~s$^{-1}$) of the optical spectroscopic redshift with $\Delta v_{\rm{FWHM}}$ = 133.5$^{+100.4}_{-64.5}$, and a broader component slightly redshifted by $\sim$ 24 km~s$^{-1}$ with $\Delta v_{\rm{FWHM}}$ = 154.1$^{+84.3}_{-85.1}$ km~s$^{-1}$ (Table 3). As this is an elliptical host galaxy whose mid-infrared WISE colours indicate a low star formation rate (Section 5.1.1), the features here may be tracing gas close to the AGN, and as we find a broad line, the gas may be unsettled to be distributed across such scales.

\subsection{Testing for a statistical detection in the stacked non-detected galaxies}

\begin{figure*}
\captionsetup[subfigure]{aboveskip=-1pt,belowskip=5pt}
\begin{minipage}{\textwidth}
\centering
\begin{subfigure}[b]{0.5\textwidth}
  \vskip 0pt
  \centering
  \subcaption{J011132-730209}
  \includegraphics[width=1.0\linewidth]{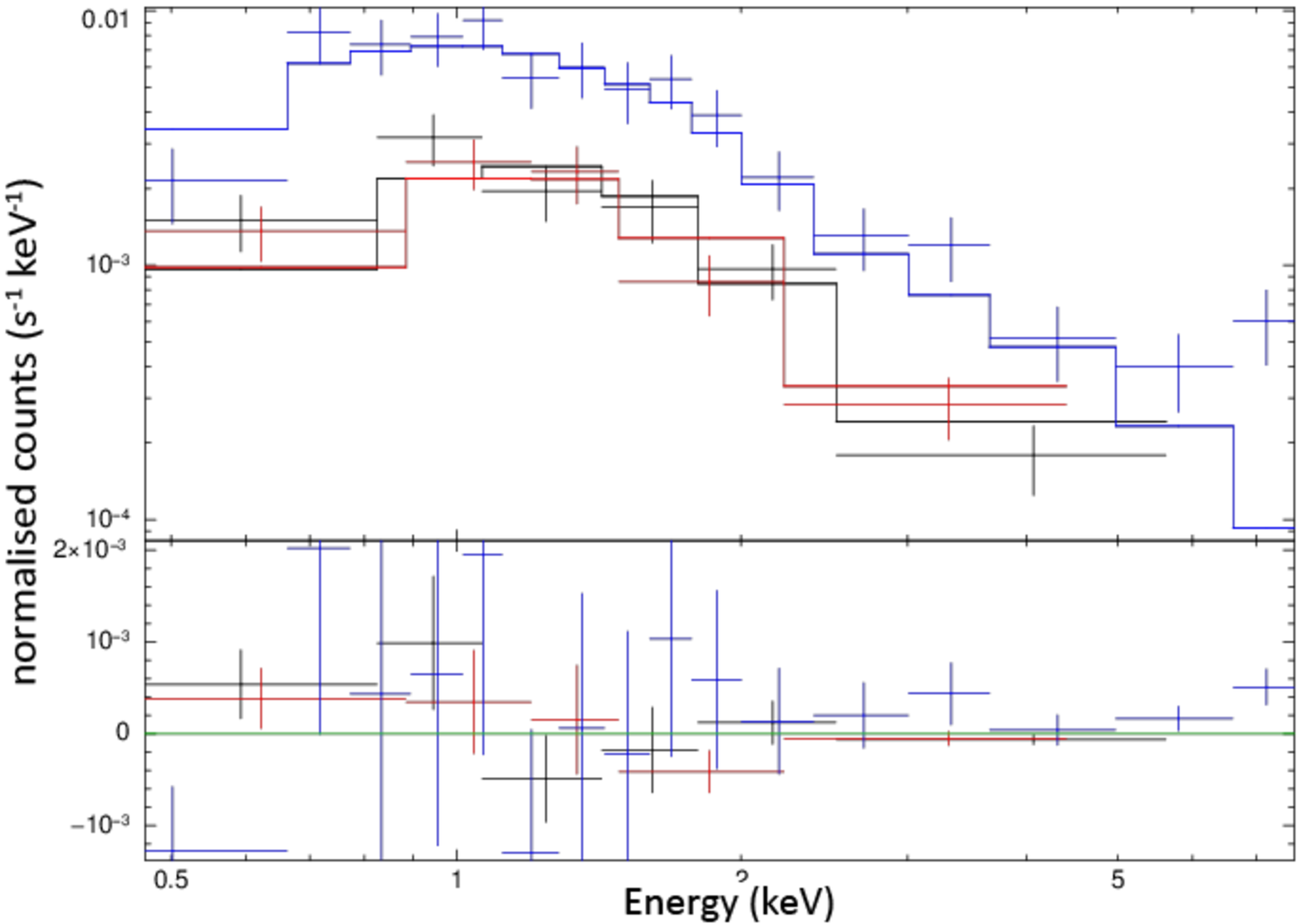}  
\end{subfigure}%
\begin{subfigure}[b]{0.5\textwidth}
  \vskip 0pt
  \centering
  \subcaption{J125711-172439}
  \includegraphics[width=1.0\linewidth]{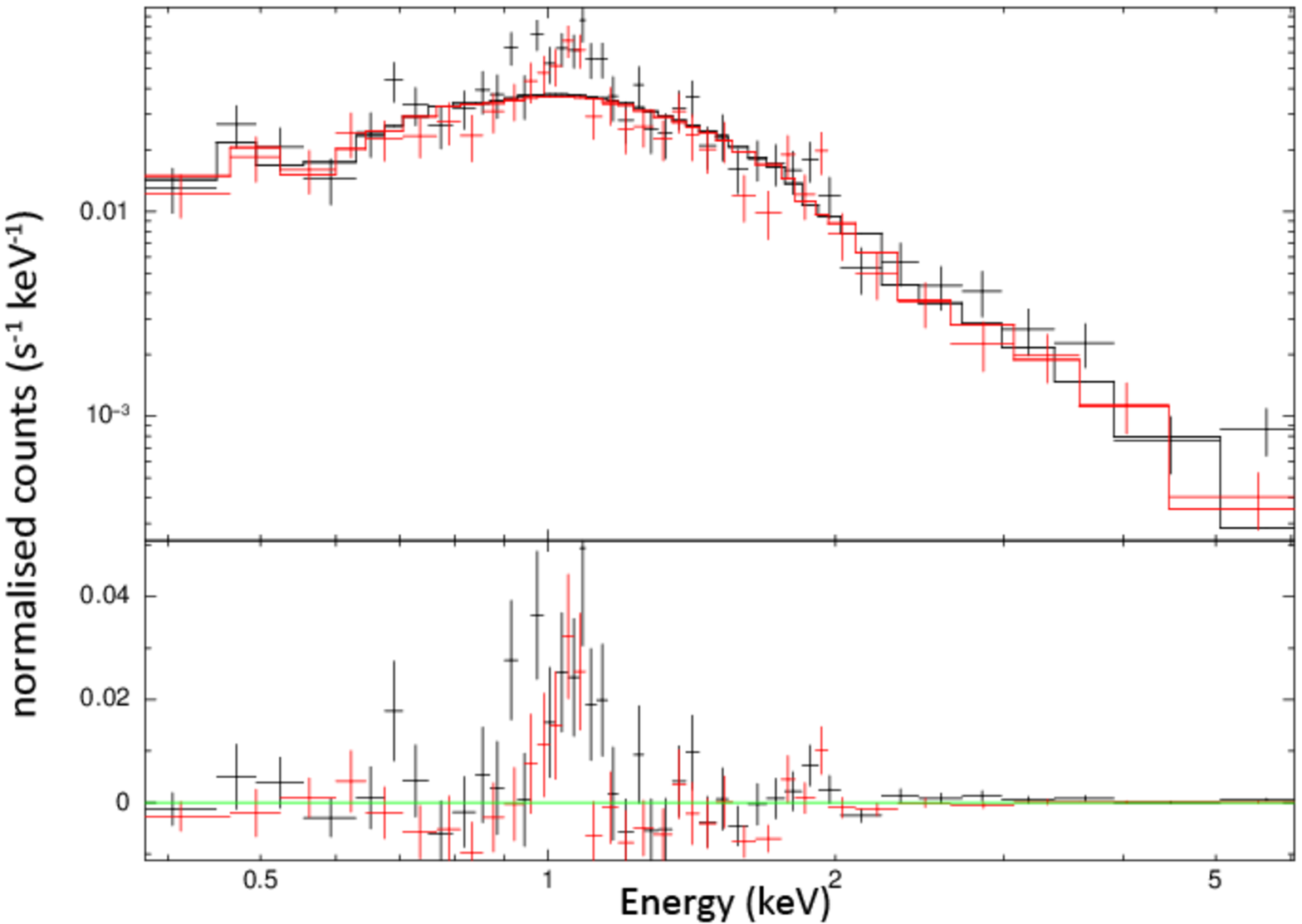}  
\end{subfigure}
\begin{subfigure}[b]{0.5\textwidth}
  \vskip 0pt
  \centering
  \subcaption{J181934-634548}
  \includegraphics[width=1.0\linewidth]{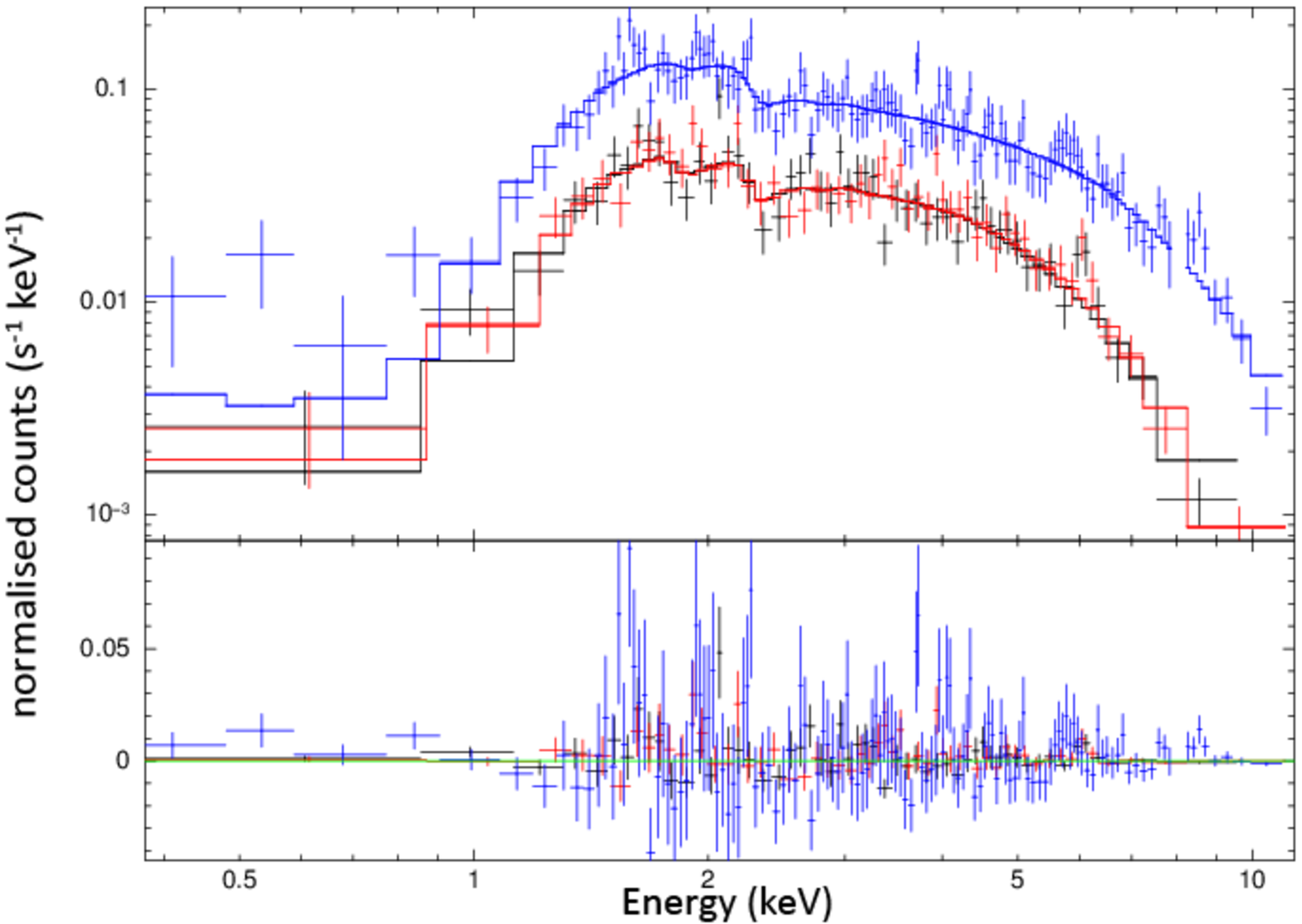}  
\end{subfigure}%
\begin{subfigure}[b]{0.5\textwidth}
  \vskip 0pt
  \centering
  \subcaption{J220538-053531}
  \includegraphics[width=1.0\linewidth]{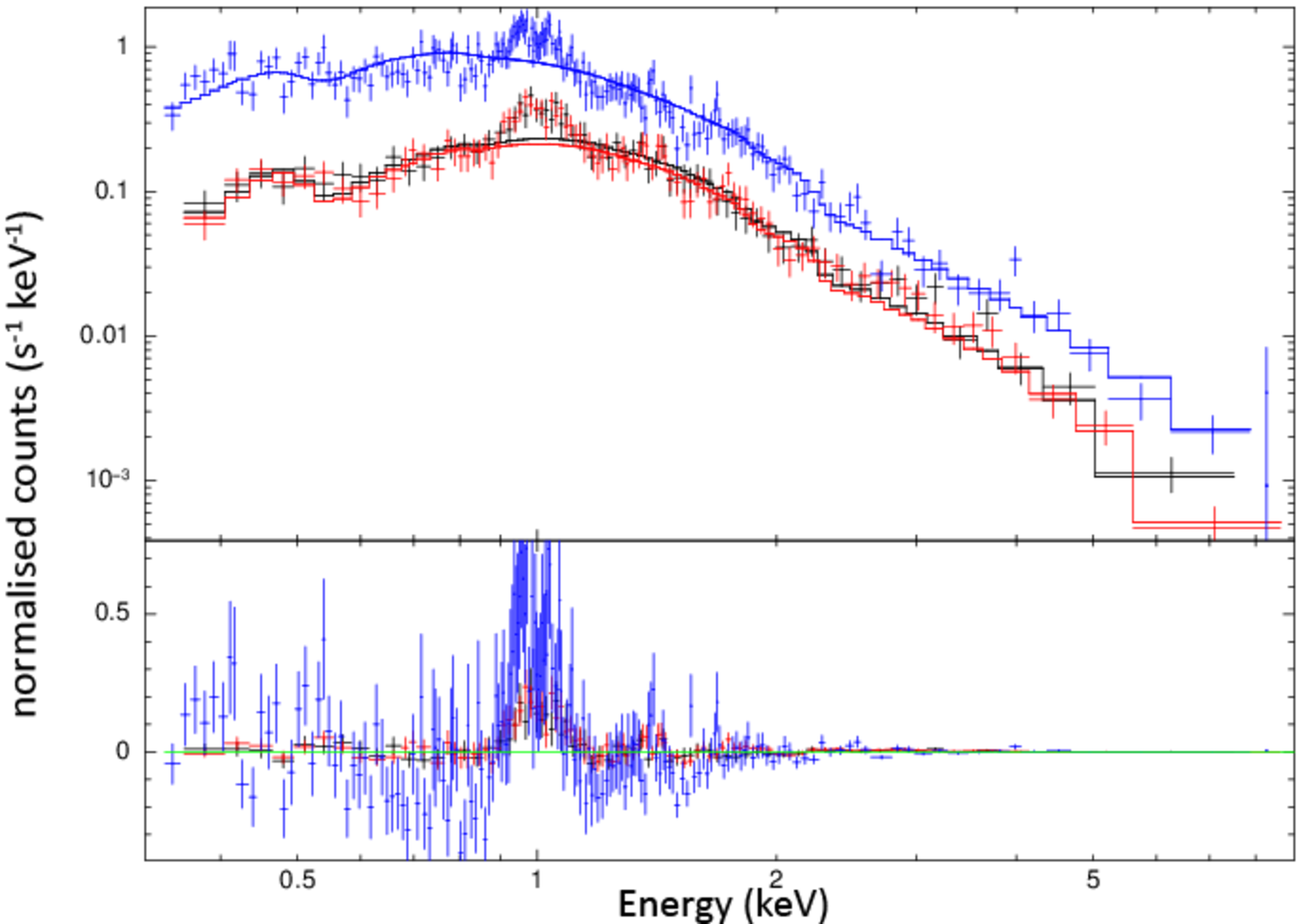} 
\end{subfigure}
\caption{X-ray spectra from the 3XMM-DR5 catalogue \citep{Rosen2015} for the 11 X-ray detected sources within our sample. The remaining 55 sources were either not observed by XMM or not detected as X-ray sources. For each image, the top panel gives the normalised counts for the EMOS1, EMOS2 and EPN detectors (black, red and blue respectively) versus energy, and the bottom panel gives the residual to the absorbed power law model we fitted. The sources on this page, including two with H{\sc i} absorption features in their radio spectra (J125711-172434 and J181934-634548), exhibit a decrease in normalised counts toward the low energy end (`soft X-rays') of a factor of two or greater. This is characteristic of photoelectric absorption and Compton scattering X-rays, and an indicator of a high column density of hydrogen (for these sources, N$_{\rm{H}} >$ 2 $\times$ 10$^{21}$ cm$^{-2}$). The remaining seven X-ray emitting sources do not have a large amount of hydrogen to diminish the observed emission in the soft X-ray end of the spectra, and have lower hydrogen column densities.}
\end{minipage}
\end{figure*}
\begin{figure*}
\captionsetup[subfigure]{aboveskip=-1pt,belowskip=5pt}
\ContinuedFloat
\begin{minipage}{\textwidth}
\centering
\small
\begin{subfigure}[b]{0.5\textwidth}
  \vskip 0pt
  \centering
  \subcaption{J031757-441416}
  \includegraphics[width=1.0\linewidth]{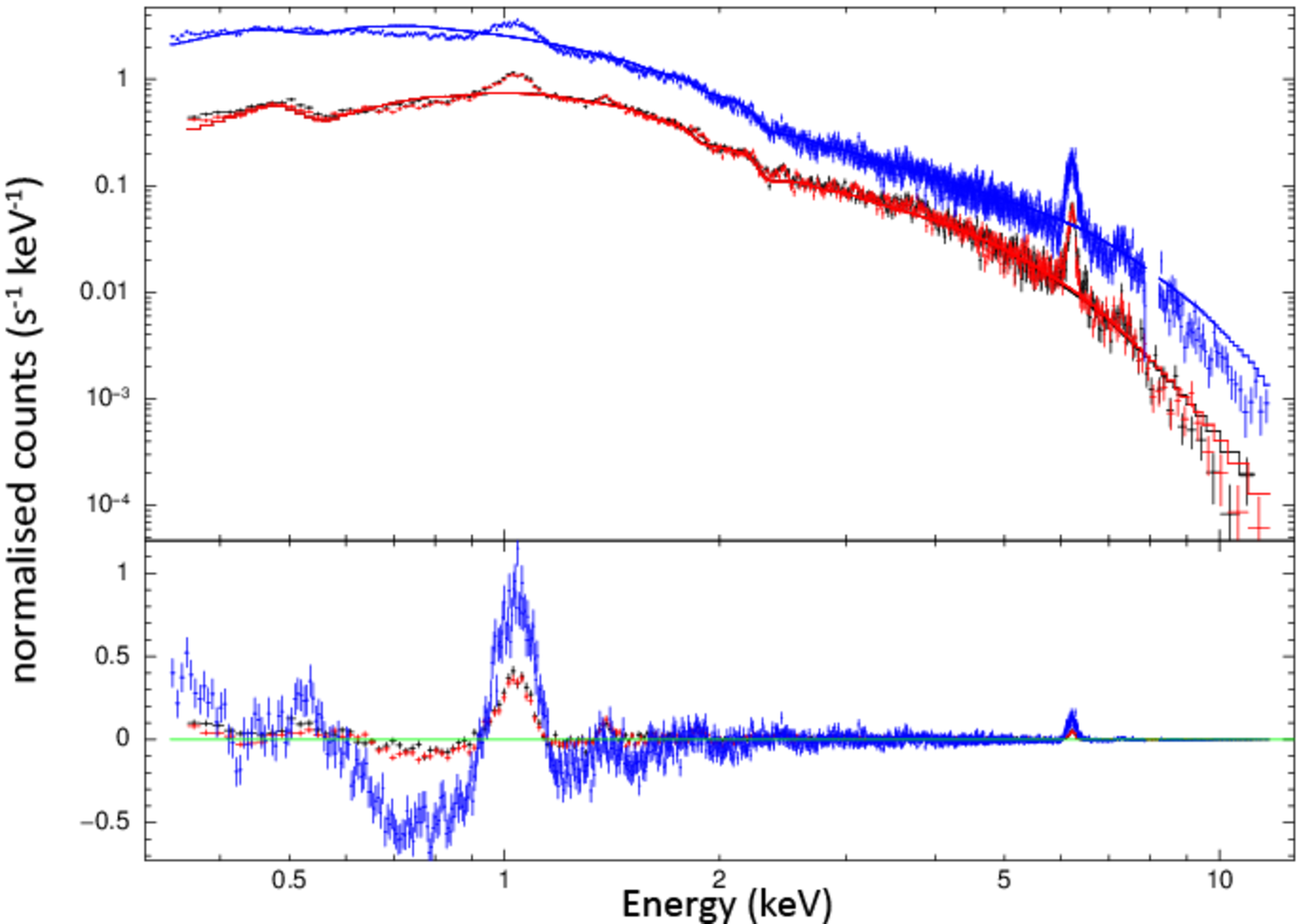}  
\end{subfigure}%
\begin{subfigure}[b]{0.5\textwidth}
  \vskip 0pt
  \centering
  \subcaption{J033114-524148}
  \includegraphics[width=1.0\linewidth]{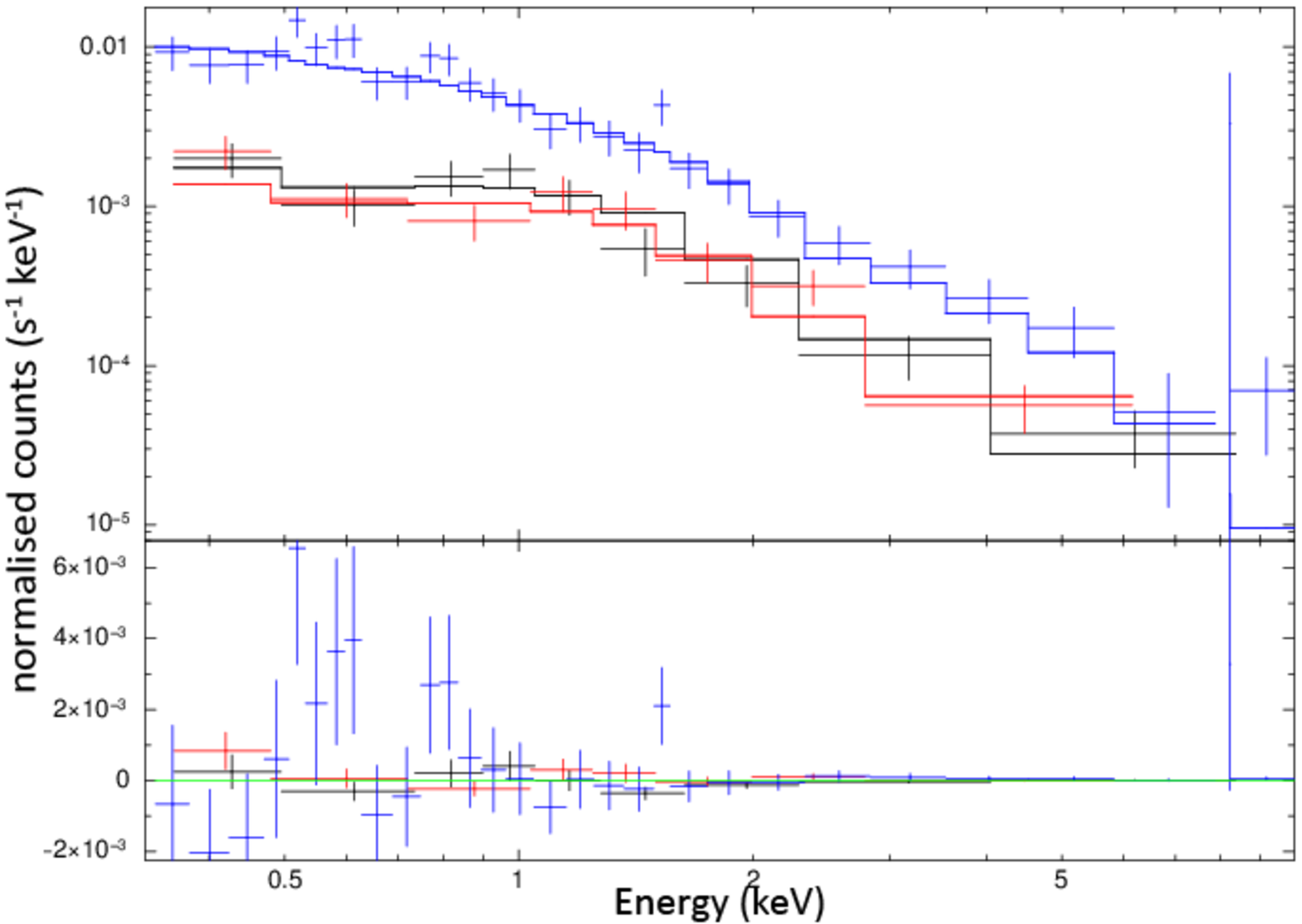}  
\end{subfigure}
\begin{subfigure}[b]{0.5\textwidth}
  \vskip 0pt
  \centering
  \subcaption{J043022-613201}
  \includegraphics[width=1.0\linewidth]{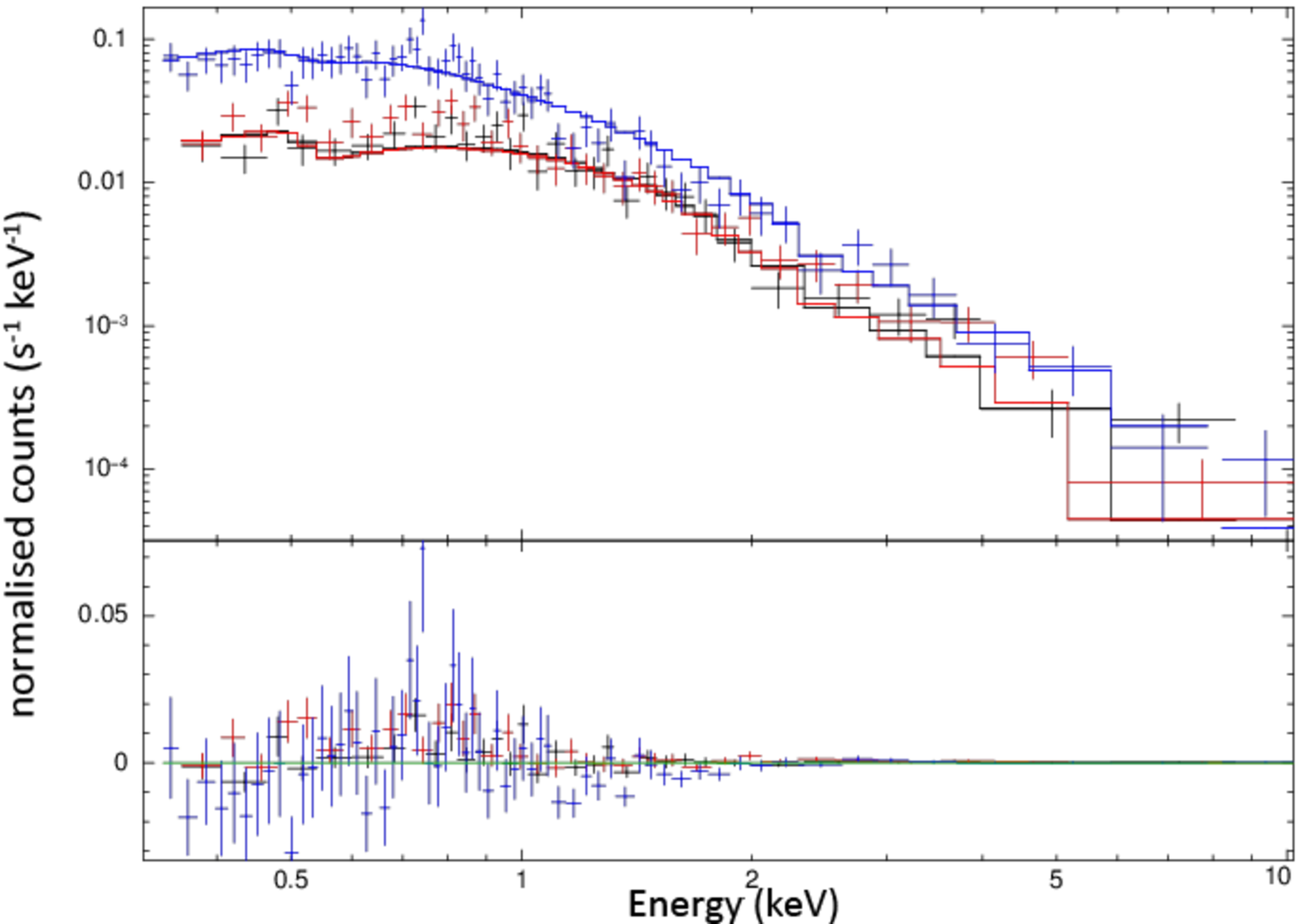}  
\end{subfigure}%
\begin{subfigure}[b]{0.5\textwidth}
  \vskip 0pt
  \centering
  \subcaption{J062706-352916}
  \includegraphics[width=1.0\linewidth]{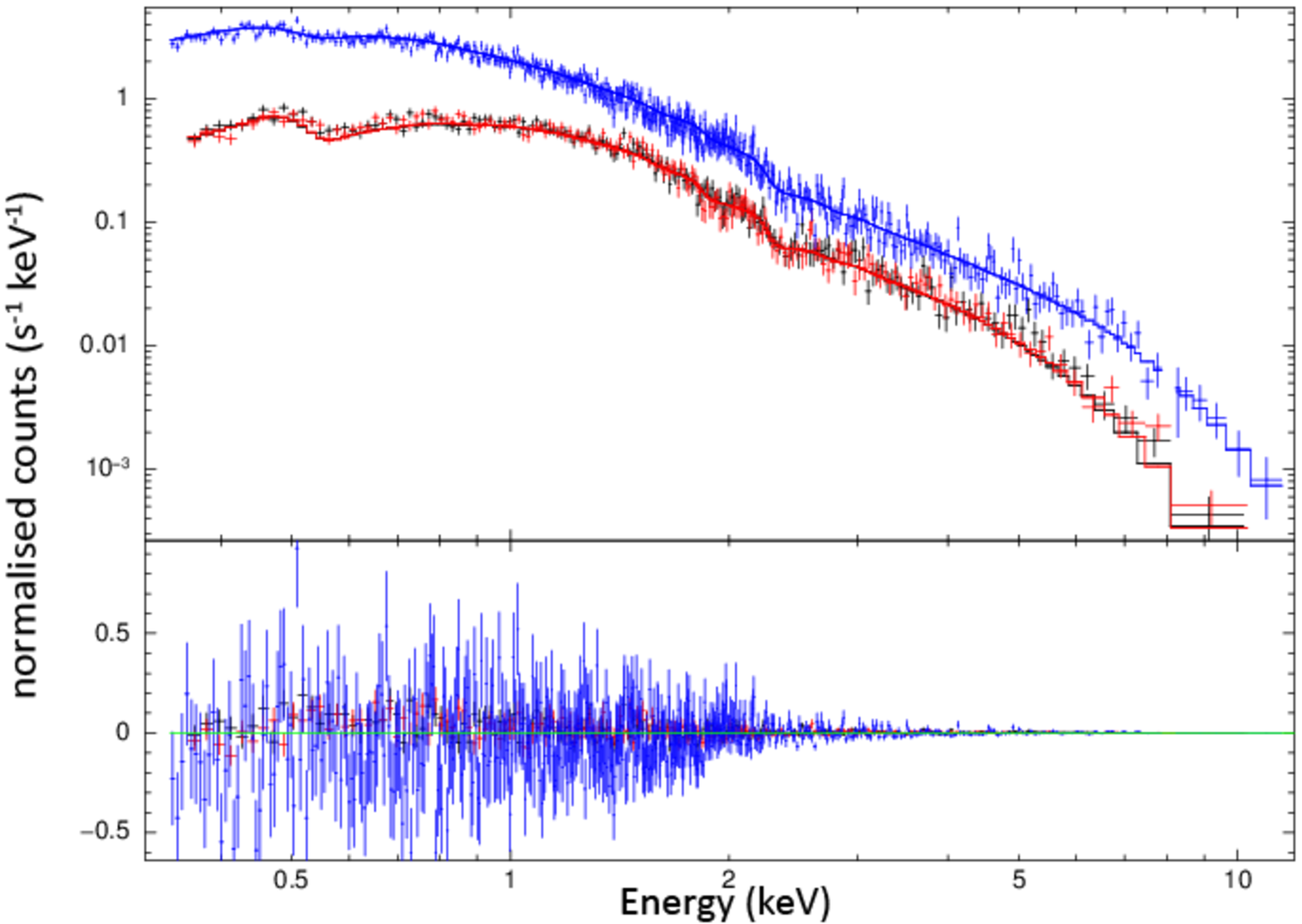}  
\end{subfigure}
\begin{subfigure}[b]{0.5\textwidth}
  \vskip 0pt
  \centering
  \subcaption{J092338-213544}
  \includegraphics[width=1.0\linewidth]{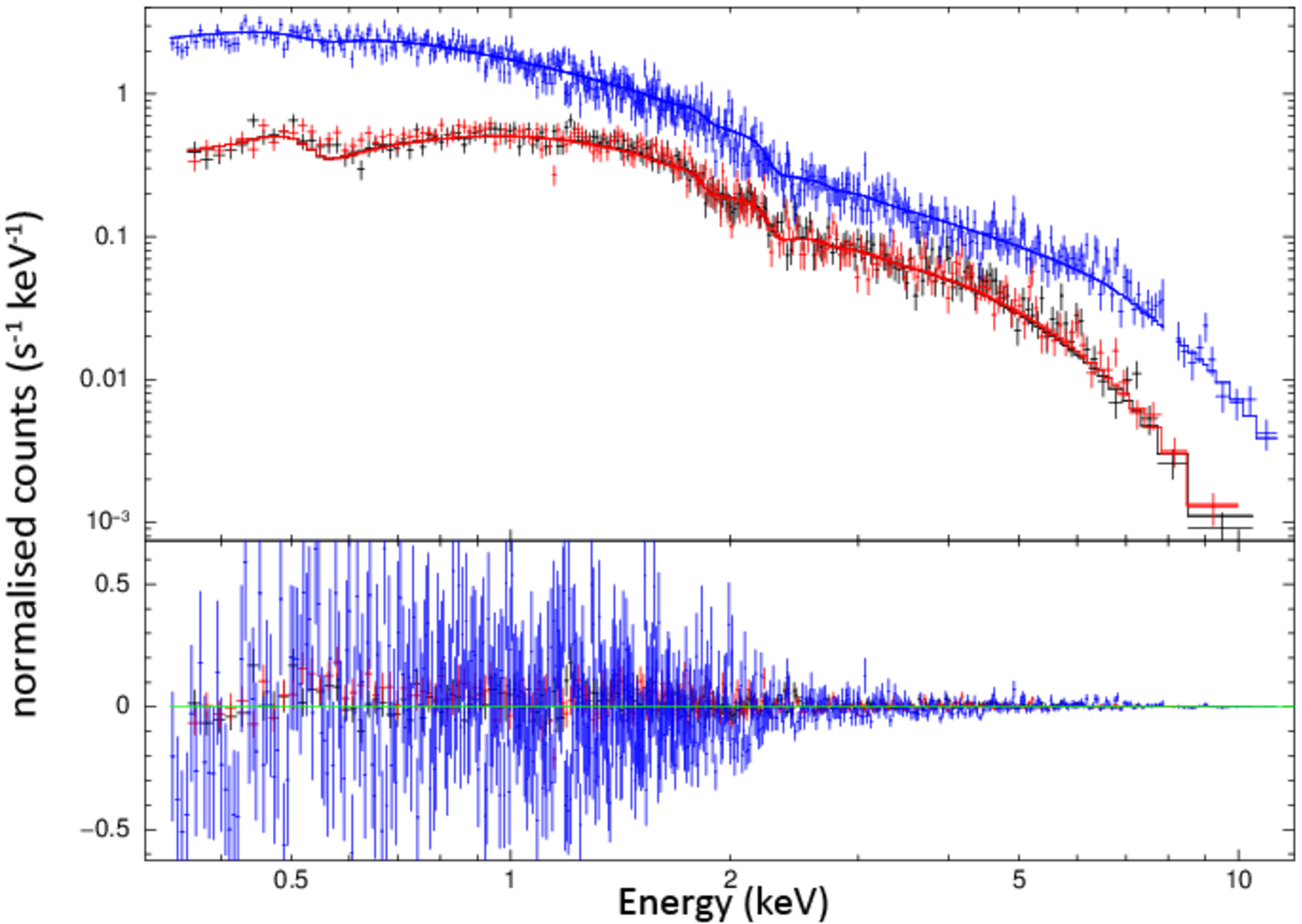}  
\end{subfigure}%
\begin{subfigure}[b]{0.5\textwidth}
  \vskip 0pt
  \centering
  \subcaption{J164416-771548}
  \includegraphics[width=1.0\linewidth]{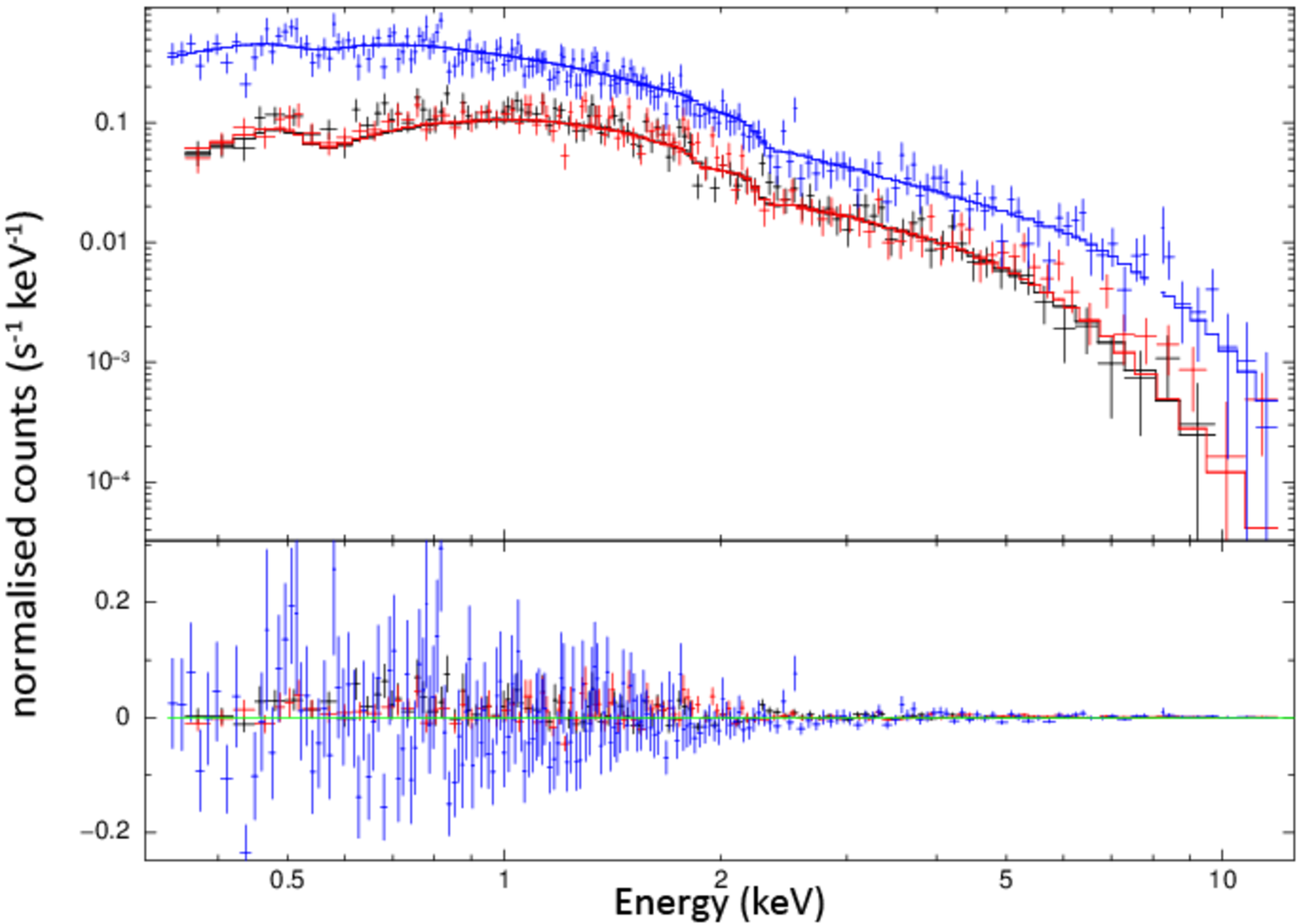}
\end{subfigure}
\caption{Continued. Note that the spectra on this page, and that of J231905-420648 on the following page, do not exhibit a strong downturn at the soft X-ray (low-energy) end and have lower measured values of $\mathrm{N}_\mathrm{H,X}$ (Table 5).}
\end{minipage}
\end{figure*}
\begin{figure}
\captionsetup[subfigure]{aboveskip=-1pt,belowskip=5pt}
\ContinuedFloat
\begin{minipage}{0.5\textwidth}
\centering
\small
\begin{subfigure}[b]{1\textwidth}
  \vskip 0pt
  \centering
  \subcaption{J231905-420648}
  \includegraphics[width=1.0\linewidth]{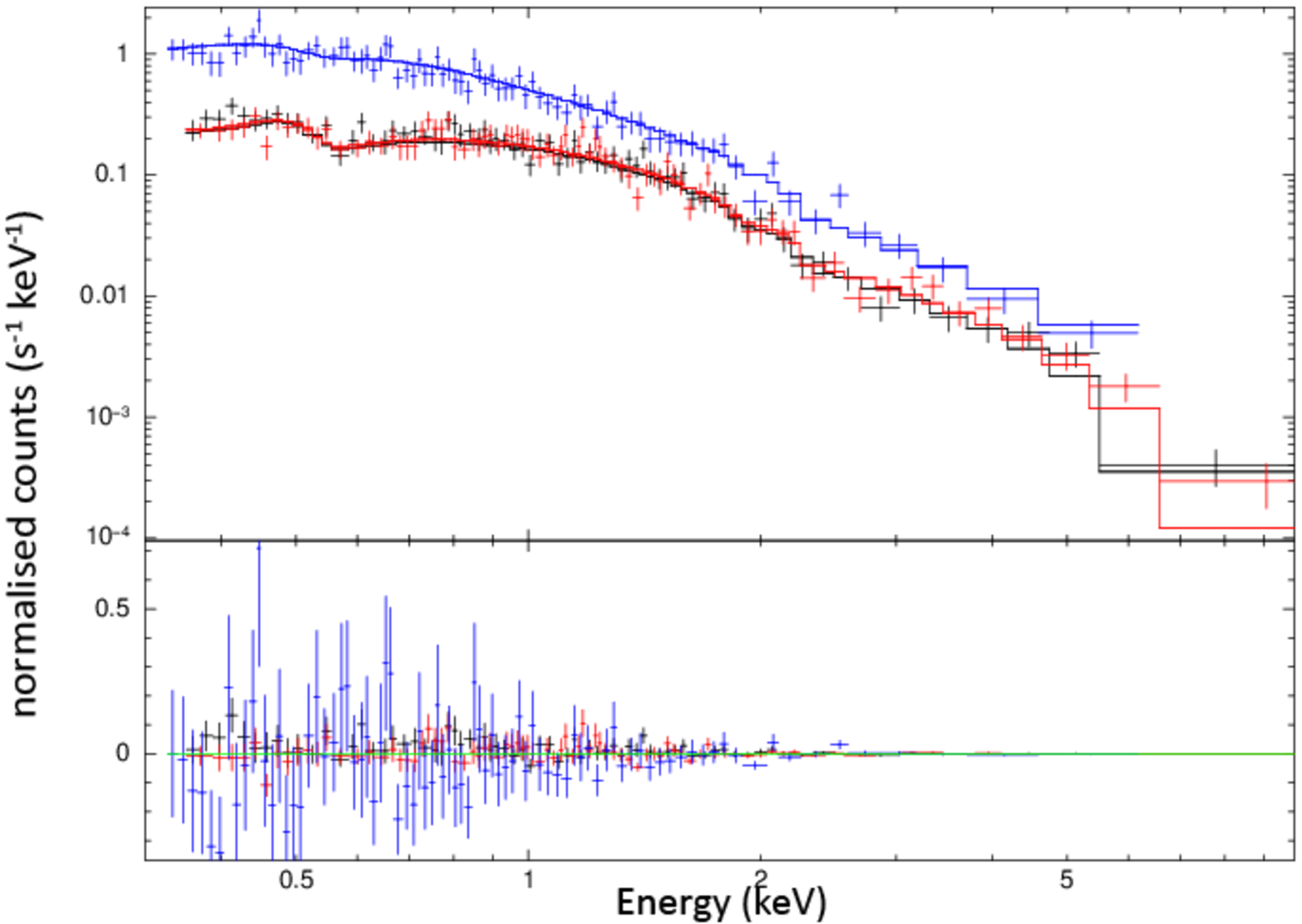}  
\end{subfigure}
\end{minipage}
\caption{Continued.}
\label{figure:xray_spectra}
\end{figure}

As in the study by \cite{Allison2012}, we searched for a statistical detection of H{\sc i} absorption by stacking the 57 individual spectra which were not found to individually exhibit absorption. We also separately stacked spectra of two subsets - the steep spectrum (SS) and flat spectrum (FS) sources, as defined by their spectral index (Table 2, columns 4-8). As before, a velocity bin size was also used to match the uncertainty in the optical spectroscopic redshift (typically less than z$_{\rm{err}}$ = 0.0002, or 0.5\% of the known redshift) for the sample to account for potential smearing of any spectral line. 

Using the spectral-line finding method as outlined in Section 3, we find that there is no evidence of a statistical detection of 21-cm associated H{\sc i} absorption in the stacked non-detections, nor in the separate subsets. Our optical depth upper limit of $\tau <$ 0.004 is consistent with the stacking result by \cite{Gereb2014}, of $\tau <$ 0.002 for 66 sources (Fig.~5). This supports their argument that some of these sources may be genuinely depleted of cold gas, while orientation effects of a disk-like distribution of H{\sc i} may be also partially responsible for this non-detection.

\section{Relationship between X-ray and H{\sc i}}

From here on, we consider all sources listed in Table 2, including those observed by \cite{Allison2012}.

\subsection{Absorbed X-ray spectra}

\begin{table*}
  \caption{Radio and X-ray luminosities, and column densities for hydrogen and specifically neutral hydrogen for our sample found to have X-ray emission. These include J125711-172434 and J181934-634548, our two H{\sc i} absorption detections, for which we have a X-ray spectrum (indicated with a *). $\mathrm{N}_\mathrm{H,X}$ is the elemental hydrogen column density estimate as given by {\sc xspec} from the XMM spectra; ${\left(\mathrm{N}_\mathrm{HI}f\over T_\mathrm{spin}\right)}$ is our measured H{\sc i} column density with no assumed value for the covering factor f and gas spin temperature of $T_\mathrm{spin}$; we then give an upper limit for $T_\mathrm{spin}$ given $\mathrm{N}_\mathrm{H,X}$ and assuming a covering factor of f = 1 (Section  4.2); $\Gamma$ is the measured photon index of the X-ray spectra; $L_\mathrm{5 GHz}$ the luminosity at 5 GHz \citep[from the AT20G flux;][]{Murphy2010}; $L_\mathrm{X}$ the X-ray luminosity, and log$_{10}$($L_\mathrm{X}$/$L_\mathrm{5 GHz}$) the logarithmic ratio of the X-ray and 5 GHz luminosities. A value of log$_{10}$($L_\mathrm{X}$/$L_\mathrm{5 GHz}$) $<$ 1.8 matches with observed values for FR-I sources \citep{Tengstrand2009}, suggesting these may be FR-I radio galaxies, or GPS sources for which we haven't yet observed a turnover in the spectrum at lower frequencies. Source J033114-524148 had an exceedingly low estimate of its hydrogen column density ($<$~1.00~$\times$~10$^{14}$~cm$^{-2}$), and so was excluded from the $T_\mathrm{spin}$ given the non-physical upper limit this value provides.}
  \begin{tabular}{lcrrcccc}
  \hline
 Name & $\mathrm{N}_\mathrm{H,X}$ & ${\left(\mathrm{N}_\mathrm{HI}f\over T_\mathrm{spin}\right)}$ & $T_\mathrm{spin}$ & $\Gamma$ & $L_\mathrm{5 GHz}$ & $L_\mathrm{X}$ & log$_{10}$($L_\mathrm{X}\over L_\mathrm{5 GHz}$)\\
  & cm$^{-2}$ & cm$^{-2}$ K$^{-1}$ & K & & erg s$^{-1}$ & erg s$^{-1}$ & \\
 \hline
J011132-730209 & 3.2 $\pm$ 0.6 $\times$ 10$^{21}$ & $<$ 7.94 $\times$ 10$^{18}$ & $<$ 410 & 2.13 & 3.55 $\times$ 10$^{40}$ & 6.36 $\times$ 10$^{41}$ & 1.25 \\
J031757-441416 & 1.3 $\pm$ 0.1 $\times$ 10$^{21}$ & $<$ 5.89 $\times$ 10$^{17}$ & $<$ 2254 & 2.21 & 3.98 $\times$ 10$^{41}$ & 8.34 $\times$ 10$^{43}$ & 2.32 \\
J033114-524148 & - & $<$ 1.58 $\times$ 10$^{19}$ & - & 1.83 & 2.76 $\times$ 10$^{40}$ & 6.21 $\times$ 10$^{41}$ & 1.35 \\
J043022-613201 & 8.7 $\pm$ 0.1 $\times$ 10$^{20}$ & $<$ 3.16 $\times$ 10$^{18}$ & $<$ 274 & 2.68 & 5.24 $\times$ 10$^{40}$ & 4.99 $\times$ 10$^{41}$ & 0.98 \\
J062706-352916 & 9.6 $\pm$ 0.3 $\times$ 10$^{20}$ & $<$ 3.70 $\times$ 10$^{17}$ & $<$ 2582 & 2.56 & 4.33 $\times$ 10$^{41}$ & 1.90 $\times$ 10$^{43}$ & 1.64 \\
J092338-213544 & 3.5 $\pm$ 0.3 $\times$ 10$^{20}$ & $<$ 3.39 $\times$ 10$^{18}$ & $<$ 102 & 1.75 & 9.46 $\times$ 10$^{40}$ & 4.16 $\times$ 10$^{43}$ & 2.64 \\
J125711-172439* & 3.1 $\pm$ 0.3 $\times$ 10$^{21}$ & 3.39 $\times$ 10$^{19}$ & $<$ 92 & 3.00 & 4.20 $\times$ 10$^{40}$ & 1.20 $\times$ 10$^{42}$ & 1.46 \\
J164416-771548 & 8.2 $\pm$ 0.8 $\times$ 10$^{20}$ & $<$ 2.59 $\times$ 10$^{18}$ & $<$ 315 & 1.83 & 4.86 $\times$ 10$^{40}$ & 5.91 $\times$ 10$^{42}$ & 2.09 \\
J181934-634548* & 2.1 $\pm$ 0.1 $\times$ 10$^{22}$ & 2.17 $\times$ 10$^{19}$ & $<$ 961 & 1.41 & 2.59 $\times$ 10$^{42}$ & 4.31 $\times$ 10$^{43}$ & 1.22 \\
J220538-053531 & 3.1 $\pm$ 0.1 $\times$ 10$^{21}$ & $<$ 3.16 $\times$ 10$^{18}$ & $<$ 944 & 3.08 & 5.66 $\times$ 10$^{40}$ & 6.31 $\times$ 10$^{42}$ & 2.05 \\
J231905-420648 & 6.8 $\pm$ 0.9 $\times$ 10$^{20}$ & $<$ 2.19 $\times$ 10$^{18}$ & $<$ 306 & 2.59 & 1.71 $\times$ 10$^{41}$ & 6.64 $\times$ 10$^{42}$ & 1.59 \\

\hline
\end{tabular}
\label{table:xray}
\end{table*}

AGN activity can be traced by X-ray emission produced by Compton scattering from a corona of hot electrons close to the super-massive black hole. X-ray spectra exhibit a dip in the lower energy end (soft X-rays) when there is a high column density of opaque material within the host galaxy due to photoelectric absorption and Compton scattering effects. 

There is observational evidence that X-ray absorption may be correlated with H{\sc i} absorption, and previous work has examined whether X-rays and radio trace gas in the spatially co-located regions \cite[e.g.][Moss et al., in prep.]{Siemiginowska2008, Siemiginowska2009, Ostorero2010, Ostorero2016}.

The radio galaxy PKS 1740-517 \citep{Allison2015} at redshift z = 0.44 was found to be a X-ray bright source whose X-ray spectrum was well modelled by a standard absorbed power-law. Investigating the potential of X-ray absorption as a proxy for H{\sc i} absorption is a work in progress using ASKAP commissioning data (Moss et al., in prep).

In order to investigate the link between X-ray and H{\sc i} absorption in our sample, we cross-matched our sample against the 3XMM-DR5 catalogue \cite[Third XMM-Newton Serendipitous Source Catalog, Fifth Data Release;][]{Rosen2015}. In total, 14 galaxies were within the XMM fields, with 11 detected in X-ray emission. Two of these sources also have detections of H{\sc i} absorption through our radio observations. Some sources also had noticeable Mg emission lines in their X-ray spectrum at $\sim$~1.2~keV, and J031757-441416 has a strong Fe emission line at $\sim$~6~keV, a typical AGN line attributed to a torus region around the AGN seen to match the source redshift.

A standard absorbed power law model was fitted and compared to all X-ray spectra in the program {\sc xspec}. These models were compared against non-absorbed models and were consistently preferred by {\sc xspec}. From the absorbed power law models, we obtained estimates for the column density of hydrogen (neutral, elemental and ionised), with the assumption made of there being some hydrogen at the known optical redshifts for the host galaxies.

Four of these sources have a decrease in their normalised counts of at least a factor of two between the model's lowest-energy value (typically $\sim$ 0.5 keV) and the peak value (Fig.~6, first set). The remaining seven sources lacked a matching visible dip in the soft X-ray end (Fig.~6, second and third set); unlike the other galaxies, these sources' absorbed power law models have a very minor dip in normalised count from their peak to the lowest energy value, if any. In line with these observations, the four absorbed X-ray sources have a correspondingly higher estimate for their hydrogen column densities relative to the other X-ray emitting sources with no visible dip (Table 5).

Of these soft X-ray absorbed galaxies, two (J011132-730209 and J220538-053531) have total hydrogen column densities of N$_{\rm{H}} \sim$ 3 $\times$ 10$^{21}$ cm$^{-2}$ (Table 5), but no H{\sc i} detected in their radio spectra. At these densities the hydrogen could be mostly molecular \citep{Schaye2001}, ionised, or located (at least partially) in an intervening system. It could also be due to a geometric effect, where we miss the H{\sc i} content but not the X-ray emission. This is discussed further in Section 5.1.3. Of the remaining two galaxies with H{\sc i} absorption detected, J125711-172434 has a hydrogen column density of N$_{\rm{H}}$ = 3.1 $\times$ 10$^{21}$ cm$^{-2}$. This value is similar to the highest N$_{\rm{H}}$ values found for the H{\sc i} non-detections. J181934-634548 meanwhile has a considerably higher value of N$_{\rm{H}}$~=~2.1~$\times$~10$^{22}$~cm$^{-2}$. This supports the idea that absorbed X-ray sources are potentially good tracers of H{\sc i} content within the host galaxy or near the AGN.

\subsection{Upper limits on H{\sc i} spin temperature from X-ray data}

We can compare the estimated H{\sc i} and X-ray column densities to learn about the composition of the overall hydrogen content within these galaxies with X-ray emission. We obtained upper limits on the H{\sc i} spin temperature, by assuming all of the hydrogen gas is atomic, through the relationship

\begin{equation}\label{equation:spin_temperature}
T_{\rm spin} < 5485 \left[{N_{\rm H,X}\over10^{22}\,\mathrm{cm}^{-2}}\right] \left[{\int{\tau_{21}(v)\mathrm{d}v}\over\mathrm{1\,km\,s}^{-1}}\right]^{-1} \left[{1}\over f \right] \mathrm{K},
\end{equation}

where $N_{\rm H,X}$ is the total column density of hydrogen (in atomic, ionised and molecular form) estimated from the \emph{XMM-Newton} spectra, and \textit{f} is the covering factor, assumed here to be a complete covering factor of 1 (motivated by the compact radio source sizes of our sample resulting in higher covering fractions). A lower covering factor would further decrease this upper limit. It is also assumed that the X-ray emission is from the host AGN and not an intervening system. 

Most of our upper limits give $T_{\rm spin} < 1000 $ K, and hence the gas seen in H{\sc i} absorption is unlikely to lie in a hot halo (Table 5). In considering the two sources with H{\sc i} detections, J181934-634548 has the largest column density of hydrogen in our sample. We obtain a high upper limit on its spin temperature of $T_{\rm spin} <$ 961 K for its neutral hydrogen content. J125711-172439 has a comparable column density of hydrogen and H{\sc i}, with an upper spin temperature of $T_{\rm spin} <$ 92 K, indicating this source has cold neutral gas close to its AGN. 

\subsection{Radio and X-ray luminosities}

In order to learn more about the sources with X-ray emission, particularly those also with H{\sc i} detections, we compared the X-ray and radio luminosities (Table 5). GPS galaxies - which are typically compact - have a higher radio luminosity than their X-ray luminosity. The reverse is true for extended FR-II radio galaxies (those with brighter radio lobes than their core component) and blazars \citep{Fossati1998}. Such luminosity ratios seen for GPS sources also match those observed in FR-I galaxies \citep[for which the radio brightness decreases with distance from the core;][]{Tengstrand2009}. A comparison between their X-ray luminosities, calculated through values obtained from {\sc xspec} and their 5 GHz radio luminosity suggests that seven of our sources, including both sources with H{\sc i} absorption detections, have ratios consistent with the FR-I population \cite[where log($L_\mathrm{X}$/$L_\mathrm{5 GHz}$)~$<$~1.8;][]{Tengstrand2009}. Not all these sources appear to be GPS from their SEDs in the literature, such as with J125711-172434, which we classify as a flat-spectrum source; see Table 2 and Section 5.2.3. It is possible that a turn-over exists at frequencies lower than the minimum observed. 

\section{Discussion}

\subsection{Host galaxy properties}

\begin{figure*}
\begin{subfigure}{0.49\textwidth}
\subcaption{(a) Our AT20G sample}
\includegraphics[width=1.0\linewidth]{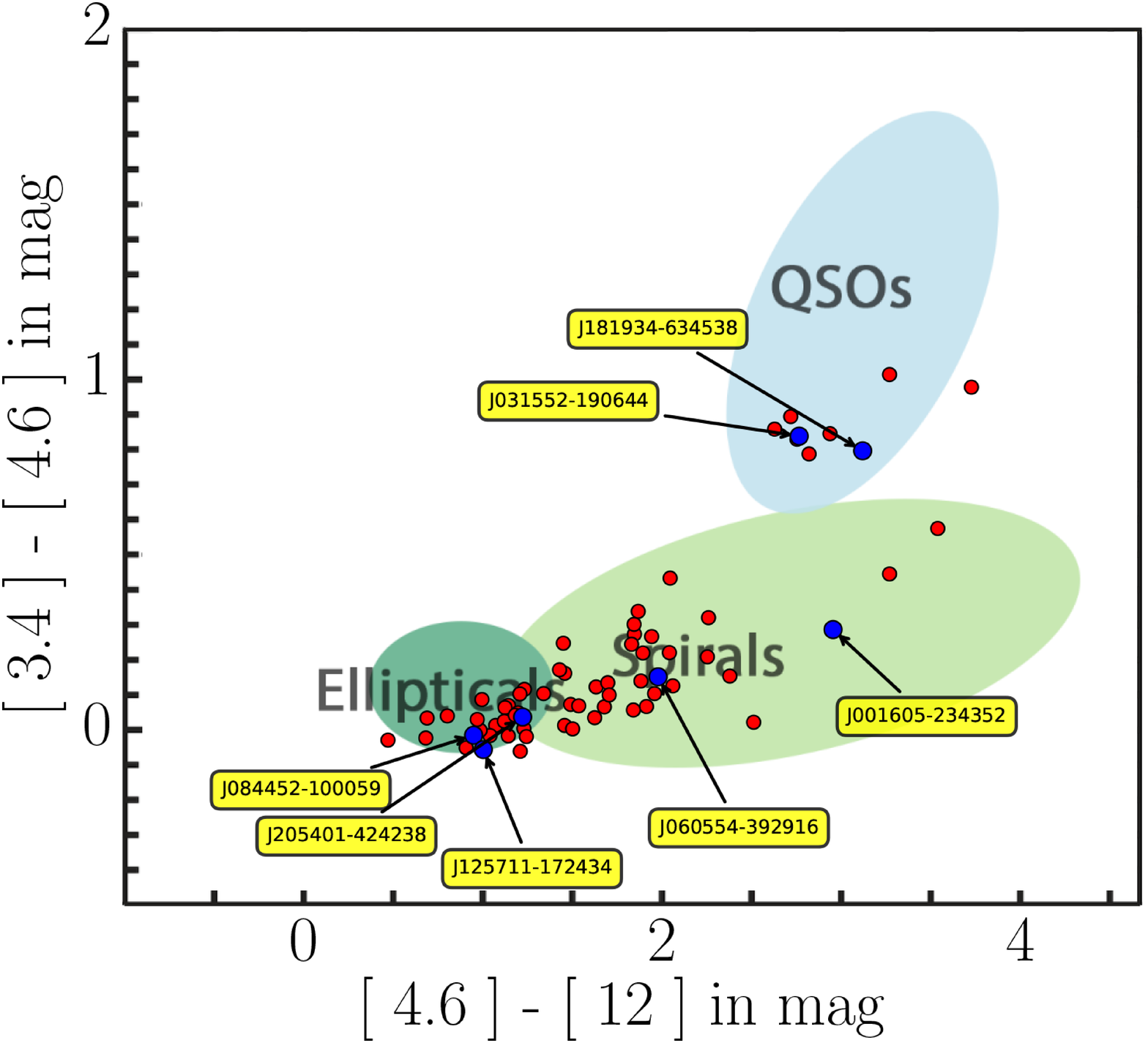}
\end{subfigure}
\begin{subfigure}{0.49\textwidth}
\subcaption{(b) Ger\'{e}b et al. (2015)}
\includegraphics[width=1.0\linewidth]{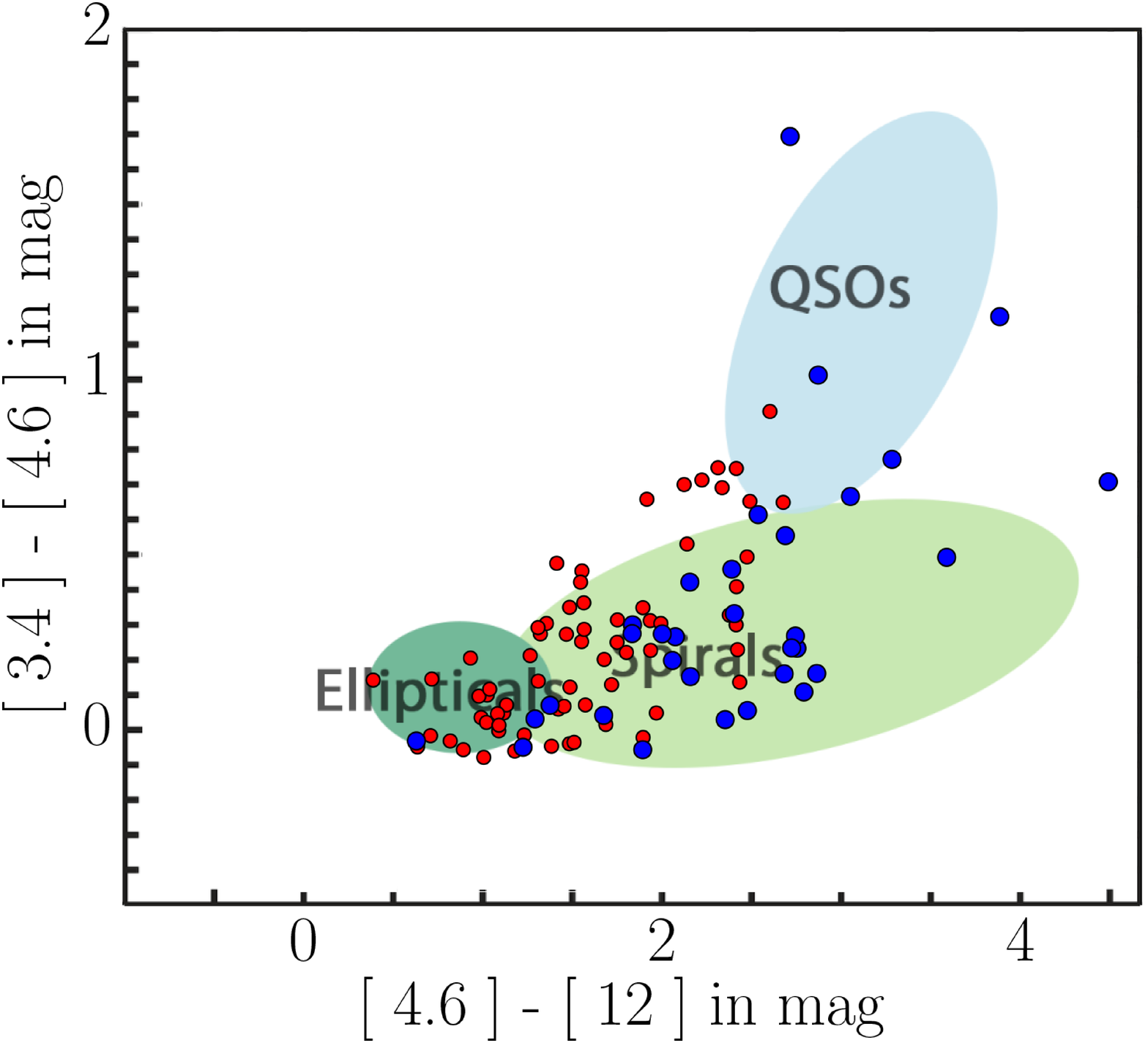}
\end{subfigure}
\caption{WISE colour-colour diagram for (a) our sample and (b) the sample studied by \citet{Gereb2015}, with WISE bands W1-W2 against bands W2-W3. The 4.6 - 12 micron colour increases with star formation rate, and the 3.4 - 4.6 micron colour changes with AGN gas accretion rate. The overlay is a schematic indicating the general location of classes of objects by infrared colour \citep[based on fig. 12,][]{Wright2010}. Typical early IR type galaxies hence tend to have colours near zero, late IR type galaxies redder in the W2-W3 colour and hence a higher star formation rate, and QSO IR types red in both colours and correspondingly higher AGN feeding rates. Blue points are sources with H{\sc i} absorption features detected their spectra, and red points are non-detections.}
\label{figure:WISE}
\end{figure*}

Our sample was selected based on the radio properties of each galaxy. While we considered optical information, it was primarily to select targets within an appropriate redshift range for our choice of telescope. We now look at other host galaxy properties that can not be considered from radio data alone, in particular the information provided by mid-infrared data.

\subsubsection{Mid-infrared colour information}

The mid-infrared wavelength colour information from the WISE survey \cite[Wide-field Infrared Survey Explorer;][]{Wright2010} can provide an indication of both the AGN gas accretion rate and the star formation rate. 

Fig.~7 shows a two-colour plot from the first three WISE bands, overlaid with a schematic which gives the general mid-infrared colour characteristics of classes of extragalactic objects \cite[derived from fig. 12,][]{Wright2010}. Higher values of W1-W2 (WISE band 1 [3.4 $\mathrm{\mu}$m] - WISE band 2 [4.6 $\mathrm{\mu}$m]) typically corresponds to a larger AGN gas accretion rate, while higher magnitudes on the W2-W3 axis correspond to a higher star formation rate \cite[e.g.][]{Donoso2012}. For example, spiral (late IR type) galaxies, typically with higher star formation rates than the quiescent elliptical (early IR type) galaxies, are redder in W2-W3, and quasi-stellar objects (QSO IR type) are red in both axes. 

It should be stressed that the schematic only applies to typical galaxies; the optical host galaxy morphology and that indicated by infrared colours does not necessarily match in individual cases. For example, J031552-190644 is known to have a spiral host galaxy \citep{Ledlow1998} and found to have H{\sc i} gas by \cite{Ledlow2001} within a galactic disc. However, the WISE data indicate that this galaxy lies in the QSO segment, with a higher AGN gas accretion rate than more `typical' late IR type galaxies.

The left panel of Fig.~7 shows our sample, and Table~6 lists the H{\sc i} absorption detection rate within each sector of the figure. The detection rate is highest for QSO IR type sources, which makes sense, as an increase in both rates (particularly the star formation rate) means a greater amount of gas either within the galactic disc or fuelling the AGN, leading to a larger column density of H{\sc i}.

Surprisingly we find that the detection rate is higher in early IR type (3/23) versus late IR type (2/34) galaxies, albeit with low number statistics. We discuss some possible reasons in Section 5.1.2. 

\begin{table}
\centering
  \caption{H{\sc i} detection rates for sources within each sector of the WISE two-colour plot (Fig.~7). Both rates for our AT20G compact core sample and the sample of \citet{Gereb2015} are given. As measured by the mid-infrared, objects in the elliptical (early-type) sector have lower values of star formation rates than typical spiral (late-type) galaxies. QSOs have both higher star formation and AGN gas accretion rates, and were found to have the highest H{\sc i} detection rate. The standard error for each IR type is given.}
  \begin{tabular}{llll}
  \hline
 Study & Early & Late & QSO\\
 & IR type & IR type & IR type\\
 \hline
This work & 13\% $\pm$ 7\% & 6\% $\pm$ 4\% & 22\% $\pm$ 14\%\\\\
\citealt{Gereb2015} & 13\% $\pm$ 7\% & 33\% $\pm$ 6\% & 53\% $\pm$ 12\%\\
\hline
\end{tabular}
\label{table:rates}
\end{table}

J084452-100059 and J205401-424238 (both found with only a weak, broad absorption line), as well as J125711-172434 (which has both a narrow and broad absorption), all fall within the early IR type area of the WISE colour diagram. These had optical depths in the range of only 3\% to 11\%. The host galaxies with only narrow absorption and a range of higher optical depths (17\% to 87\%) do not fall into this region. Host galaxies with small star formation rates may have a corresponding lacking fuel supply (H{\sc i} gas) that is too low to detect. The H{\sc i} gas we do trace in these early IR type galaxies may be close to the AGN rather than within the host galaxy's disc, and hence this gas may be influenced by the feedback or fuelling of such newly born or triggered AGN (supported by their asymmetric features and the broad widths, some $>$ 400 km s$^{-1}$, of these lines). 

\subsubsection{Mid-infrared comparison with the Ger\'{e}b et al. study}

The right panel of Fig.~7 shows the WISE two-colour information for the survey by \cite{Gereb2015}. This study includes both compact and extended sources and so does not directly compare to our work featuring mainly compact core radio galaxies; their source selection was only redshift and radio flux density limited. Nonetheless, it is also useful to investigate whether the mid-infrared information could have predicted the findings of this study. In studying the WISE colours of the \cite{Gereb2015} sample, \cite{Chandola2016} found that the detection rate of H{\sc i} absorption was highest in compact galaxies with WISE infrared colours W2-W3 $>$ 2, i.e. those with high star formation rates. Meanwhile, extended radio sources in the sample and those with W2-W3 $<$ 2 were found to have very low H{\sc i} detection rates.

Unlike our sample, there is a clear increase in H{\sc i} absorption detection rate from the late IR type to the early IR type galaxies in the study by \cite{Gereb2015}. This may be attributed to the fact that their study was more sensitive to lower H{\sc i} column densities than our work (Section 5.2; Fig.~9). Comparing the non-detections, \cite{Gereb2015} reached average upper limits of N$_{\rm{HI}}$~=~4.70~$\times$~10$^{18}$~$\rm{\frac{T_{S}}{f}}$~cm$^{-2}$,  compared to our average upper limit of N$_{\rm{HI}}$~=~6.88~$\times$~10$^{18}$~$\rm{\frac{T_{S}}{f}}$~cm$^{-2}$. Scaling our value for the different velocity resolution and FWHM assumed between the two surveys \citep{Curran2012b}, our average upper limit becomes N$_{\rm{HI}}$~=~8.31~$\times$~10$^{18}$~$\rm{\frac{T_{S}}{f}}$~cm$^{-2}$. These limits and our lowest detection made at N$_{\rm{HI}}$~$=$~2.17~$\times$~10$^{19}$~$\rm{\frac{T_{S}}{f}}$~cm$^{-2}$ suggests we would not have detected gas in the discs of other late IR type galaxies. The low H{\sc i} detection rate of 13\% for early IR type galaxies is seen in both surveys.

It is noted that the finding by \cite{Chandola2016} applies to our sample. 4 out of 44 (9\%) are detected in H{\sc i} for sources with W2-W3~$<$~2, while above this colour limit 3 out of 22 (14\%) sources are detected. As two-thirds of our sample are found to have W2-W3~$<$~2, this may contribute to the lower overall detection rate of our survey. Furthermore, we find a weak positive correlation between the column density values and upper limits, and the W2-W3 magnitude (Spearman rank correlation coefficient of 0.225 $\pm$ 0.078).

\subsubsection{HI detection rate versus AGN gas accretion rate}

All \cite{Gereb2015} sources with W2-W3 $\geq$ 2.7 are detected in H{\sc i} absorption. However, at any fixed W2-W3 value below 2.7, H{\sc i} absorption was detected more frequently in galaxies with a lower W1-W2 value (Fig. 7). The linear fit equations for the H{\sc i} detections at W2-W3~$<$~2.7 is W1-W2~=~0.23~($\pm$ 0.07)~$\times$~W2-W3~-~0.25~($\pm$~0.14), and non-detections W1-W2~=~0.31~($\pm$~0.04)~$\times$~W2-W3~-~0.26~($\pm$~0.06). This suggests that it may be harder to detect H{\sc i} gas in galaxies with higher W1-W2 colour, i.e. higher AGN activity.

There is no one clear definitive explanation for this result. A higher AGN gas accretion rate in the host galaxy would correspond to a higher AGN feedback rate, which could ionise nearby neutral gas at sufficient luminosities \citep{Curran2012}. However, this is at odds with the high H{\sc i} detection rate made for the QSO IR type galaxies which have high W1-W2 colour. We consider the other potential reasons below.

From fig. 2 of \cite{Chandola2016}, the extended sources in the \cite{Gereb2015} sample have higher magnitudes of W1-W2 for any particular value of W2-W3, compared to the compact sources. This is consistent with the scenario of a larger covering factor for compact radio sources, claimed to lead to higher detection rates \cite[e.g. Section 1, and][]{Curran2013}.

Additionally, fig. 9 of \cite{Pace2016} shows a correlation between the W1-W2 colour and the log of the specific radio luminosity (SRL; radio luminosity divided by the galaxy's stellar mass) for low-excitation radio galaxies (LERGs). Furthermore, fig. 5 and 6 of \cite{Chandola2016} shows that the H{\sc i} non-detections have higher specific star formation rates (sSFR) than the H{\sc i} detections, albeit only at 1$\sigma$ significance. 

Hence, the non-detections in the study by \cite{Gereb2015} with higher W1-W2 magnitudes may be LERGs depleted of gas, due to higher radio AGN activity, higher SFR, or a combination (also see Section 5.5). Given the various possible explanations, a study of a larger homogeneous sample is required to further explore both this result and these relations.

\begin{figure}
\includegraphics[width=1.0\linewidth]{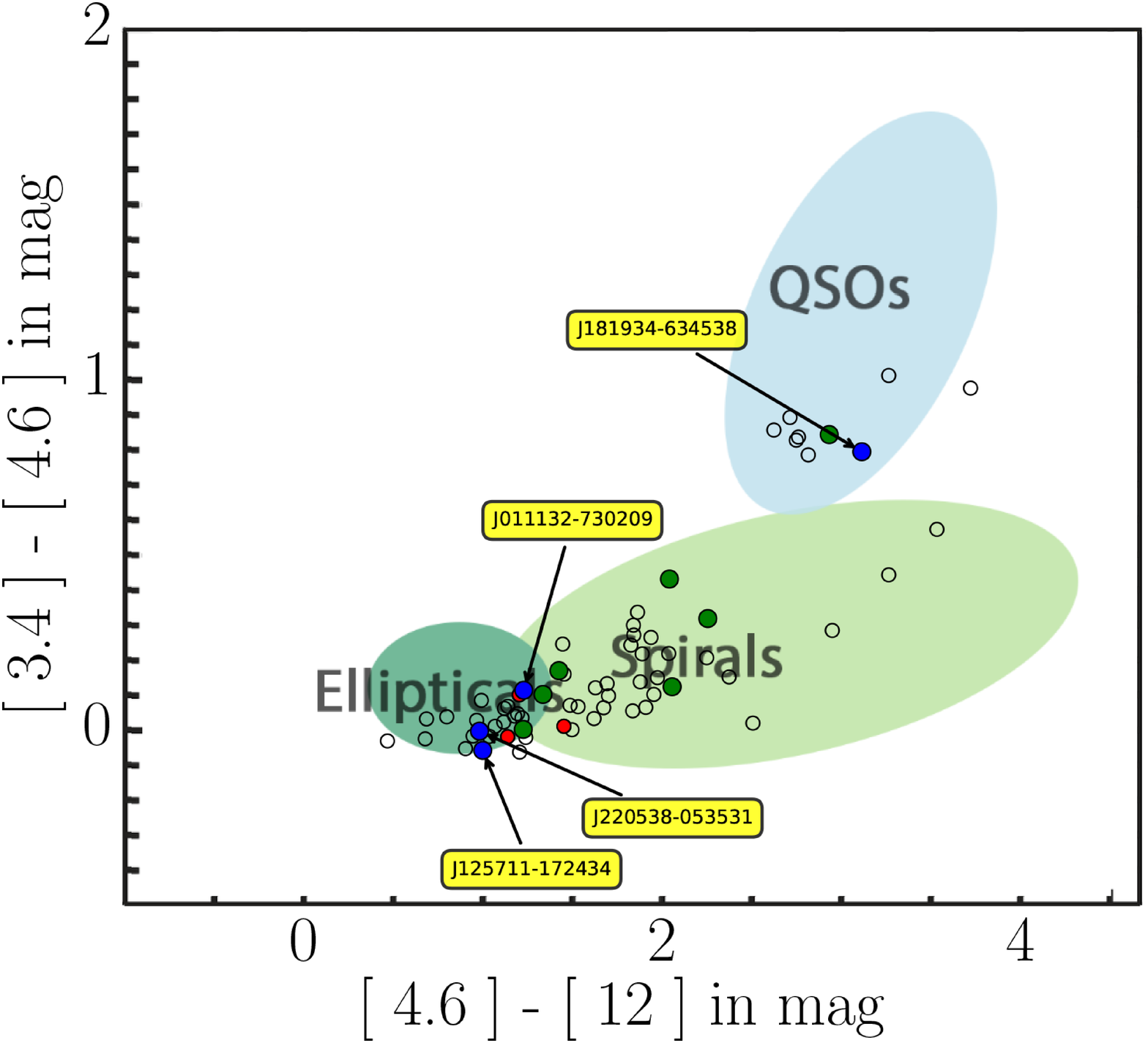}
\caption{WISE two-colour diagram for our sample, indicating where those found to have X-ray emission lie. Unfilled points are those not covered by the XMM survey, red points those searched but with no X-ray emission detections, green points are those with X-ray emission, and blue points are sources with an indication of a high column density of hydrogen from their XMM spectra (Fig.~6).}
\label{figure:WISE}
\end{figure} 

\subsubsection{Mid-infrared colours with X-ray emitting sources}

Fig.~8 shows the mid-infrared colours for our sample, where those found to be X-ray emitting sources are indicated in green (with low hydrogen column densities) and blue points (with hydrogen column densities of N$_{\rm{H}} >$ 2 $\times$ 10$^{21}$ cm$^{-2}$; Section 4). One of the two X-ray emitting sources in the QSO sector has a high column density of hydrogen (and a H{\sc i} detection). The other three sources with high hydrogen column densities are found at the lower ends of both AGN gas accretion and star formation rates. These are low number statistics; however, like with our radio observations, it appears that a range of extragalactic sources can be probed using X-ray information in tracing the gas content within the host galaxy, despite different rates of AGN feeding or star formation. 

\subsection{Detection rate of H{\sc i} absorption}

\begin{figure*}
\begin{minipage}{0.75\textwidth}
\centering
\captionsetup[subfigure]{aboveskip=-1pt}
\begin{subfigure}{1\textwidth}
\subcaption{(a) Column density vs. flux density}
\includegraphics[width=1.0\linewidth]{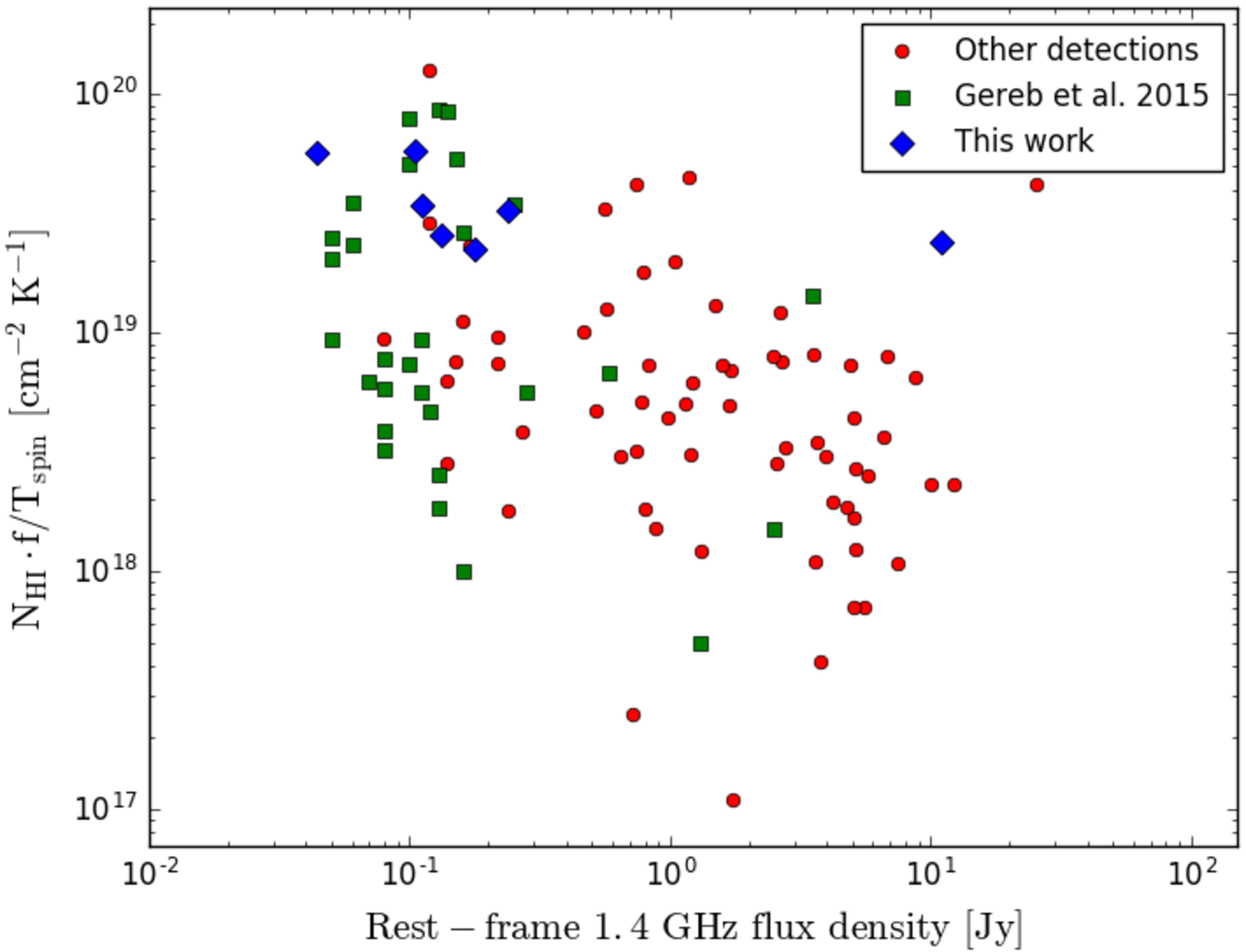}
\end{subfigure}
\begin{subfigure}{1\textwidth}
\subcaption{(b) Flux density vs. redshift}
\includegraphics[width=1.0\linewidth]{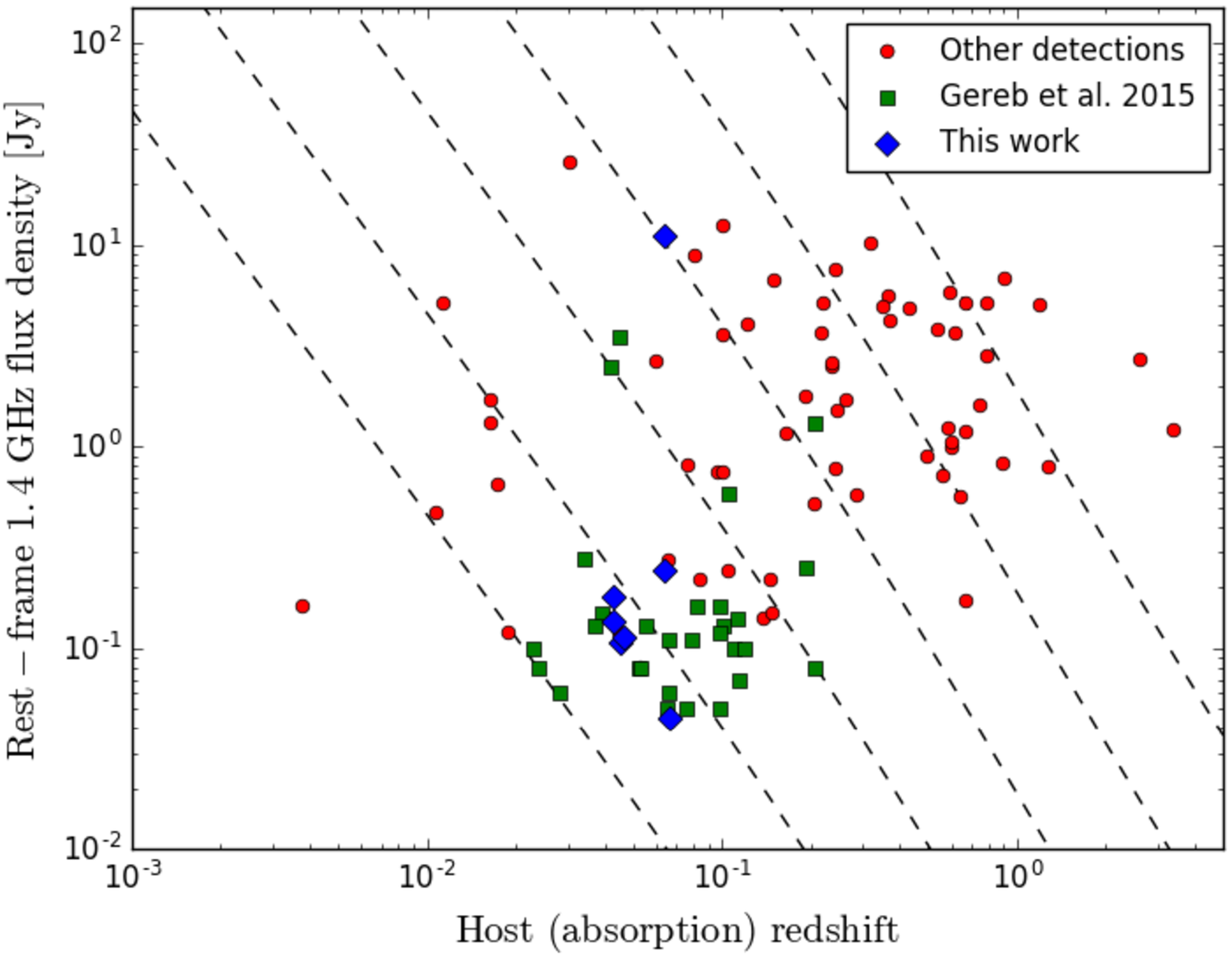}\end{subfigure}
\caption{A comparison of our H{\sc i} absorption detections with those of other surveys\protect\footnotemark. The top panel gives the normalised \mbox{H\,{\sc i}} column density ($1.823\times10^{18}\int{\tau_\mathrm{obs} \mathrm{d}v}$) versus the rest-frame 1.4\,GHz flux density. The bottom gives the flux density against the redshift of the absorption feature, with lines depicting constant radio luminosity and their change with redshift (starting from 10$^{23}$ W Hz$^{-1}$, with each line to the right an increase in magnitude). The red circles represent detections from previous surveys, green squares those from the \citet{Gereb2015} study, and the blue diamonds our seven sources with detections.}
\label{figure:N_vs_S}
\end{minipage}
\end{figure*}

The findings (and detection rates) of other H{\sc i} absorption surveys vary, due to differing source selection methods of radio galaxies and hence different studied samples. For instance, there is mounting evidence in the literature that the detection rate of H{\sc i} absorption in compact `FR-{\sc 0}' radio galaxies is higher than that of their older, extended FR-{\sc I}/{\sc II} counterparts \cite[e.g.][]{Pihlstrom2003, Curran2013, Gereb2015, Aditya2016}. Curran et al. proposed that this is due to a higher covering factor for compact sources, although different levels of sensitivity reached between surveys is another consideration. We compare our study on compact core radio galaxies with these other H{\sc i} absorption surveys to see how our sample selection compares with other studies. 

\subsection{Flux density of sources with associated H{\sc i} absorption}

Fig.~9 plots integrated optical depth against rest-frame flux density, and the flux density against the absorption feature redshift for existing H{\sc i} absorption surveys. In the top panel, our detections have comparatively higher H{\sc i} column densities than other surveys. In the bottom panel, lines of constant radio luminosity strengths, depicting the associated flux density change with redshift, are overlaid, from luminosities of 10$^{23}$ to 10$^{28}$~W~Hz$^{-1}$. Our study probes a lower range in radio luminosity ($<$~10$^{25}$~W~Hz$^{-1}$ for most detected sources) than other samples. This results in a potentially lower sensitivity to column densities of H{\sc i} for our work, and explains the comparatively higher column densities of our detections. \footnotetext{References: \cite{mir89,VanGorkom1989,ubc91,cps92,cmr+98,Morganti2001,Vermeulen2003,Orienti2006, Gupta2006,Emonts2010,ssm+10,cwm+10,Chandola2011,cwwa11,cwt+12,Gereb2015,Aditya2016,ysd+16}.}

\subsection{Disturbed gas kinematics in compact galaxies}

\cite{Gereb2015} had a flux-selected sample with a mixture of host galaxy types within their sample of compact and extended radio sources. They found 32 detections in their sample of 101. Their compact sources exhibited the broadest line features and were more often seen to have asymmetric lines. \cite{Gereb2014} suggest that this trend in compact sources is due to the presence of unsettled gas, which is typically traced by asymmetric and redshifted/blueshifted lines. These properties point to interactions between the AGN and the circumnuclear medium in such compact galaxies - the feedback from the newly triggered AGN which is still propagating through this medium would be the cause of the unsettled gas.

We find a similar trend in the line features; 5 of our 7 detections have a broad feature (where ${v}_{\mathrm{FWHM}}$ $>$ 100 km~s$^{-1}$). Furthermore, an exception in J031552-190644, a source with only a narrow feature, is a known FR-I galaxy with extended low-frequency radio emission. 

\subsection{Distinguishing AGN accretion}

There exists two classes of radio galaxies; the low-excitation radio galaxies (LERGs) and high-excitation radio galaxies (HERGs). LERGs lack AGN emission lines in their optical spectra, and are believed to have inefficient accretion processes from the hot-gas halo. HERGs however have efficient AGN accretion processes, and are found to have emission lines in their optical spectra. 

21-cm absorption preferentially traces the colder gas. \cite{Chandola2016} investigated the difference in the cold gas in LERGs and HERGs for the sample studied in H{\sc i} by \cite{Gereb2015} of compact and extended radio sources. The LERGs yielded a lower H{\sc i} detection rate. The H{\sc i} features that were detected had large velocity shifts to the optical spectroscopic redshift (blueshifts of $\sim$ -300 km s$^{-1}$) and broad profiles. This H{\sc i} could hence be within accreting circumnuclear gas instead of being within a gas-rich galaxy disc for most of these LERGs. Meanwhile, \cite{Chandola2016} found that almost all HERGs in the same sample also had high star formation rates. The three compact HERGs in the study (all detected in H{\sc i}) each had absorption profiles with small ($<$~100~km~s$^{-1}$) velocity shifts. In these cases, the H{\sc i} was attributed to a gas and dust-rich ISM in the host galaxy. Another potential source for the H{\sc i} is a central dusty torus in HERGs.

\begin{table}
\centering
  \caption{Optical spectroscopic identifications made by \citet{Mahony2011} for our sample (see Table 2). Aa represents AGN absorption lines only, while Aae have both absorption and emission lines - both fall into the low-excitation radio galaxy (LERG) population. High-excitation radio galaxies (HERGs) are seen as Ae (with only emission lines) and AeB (broad emission lines). 12 sources did not have sufficient information to identify any optical absorption or emission lines. The standard error for each identification is given.}
  \begin{tabular}{lccrc}
  \hline
 Type & Sources & H{\sc i} detections & Detection rate\\
 \hline
Aa & 25 & 3 & 12\% $\pm$ 6\%\\
Aae & 18 & 1 & 6\% $\pm$ 5\%\\
\hline
LERG & 43 & 4 & 9\% $\pm$ 4\%\\
\hline
Ae & 10 & 3 & 30\% $\pm$ 14\%\\
AeB & 1 & 0 & 0\%\\
\hline
HERG & 11 & 3 & 27\% $\pm$ 13\%\\
\hline
N/A & 12 & 0 & 0\%\\
\hline
\end{tabular}
\label{table:spectro}
\end{table}

Following this analysis, a similar examination can be applied to our sample. We consider the optical spectroscopic information available for our sample in Table 7 \citep{Mahony2011}. We find 25 with only absorption lines (Aa) and 18 with both absorption and emission lines (Aae). We classify both sets as LERGs. 10 sources were found to only have emission lines (Ae) and one with broad emission lines (AeB), with these considered to be HERGs. We have a higher H{\sc i} detection rate in HERGs, but the small-number statistics should be noted.

Of the 12 sources in our sample with no optical spectroscopic information, only one had a WISE W1-W2 colour~$>$~0.8, corresponding to a high AGN gas accretion rate (Section 5.1). This source can hence be attributed to a HERG, and the remaining 11 sources to LERGs. As no H{\sc i} was detected in these 12 sources, if these assumptions are made the H{\sc i} detection rates would drop to 7\% for LERGs and 25\% for HERGs.

Four H{\sc i} detections were made in LERGs. Three of these had broad line widths ($\Delta{v}_{\mathrm{FWHM}}$~$>$~100~km~s$^{-1}$) and shallow features significantly offset from the optical spectroscopic redshift were found in LERGs. Our proposed explanations for the gas kinematics for these sources \cite[][and Section 3.2] {Allison2012, Allison2013} include gas inflows/outflows and gas rotating in a disc around the AGN. J060555-392905 is the only Aa source for which the narrow absorption feature seen at the optical spectroscopic redshift is thought to be within a galactic disc rather than in a dusty torus around the AGN. This source also has a high star formation rate (W2-W3~$>$~2.5) unlike the other three sources. Therefore, most of our LERGs detected in H{\sc i} appear to have disturbed circumnuclear gas. 

Furthermore, all of our HERGs (Ae sources) were found to have narrow deep absorption features close to or at the optical spectroscopic redshift. Therefore, our observed absorption features found within HERGs and LERGS appears to match the findings by \cite{Chandola2016}. 

It has been demonstrated \cite[][and references therein]{Gereb2015, Chandola2016} that sources with higher radio luminosities are found to have more asymmetric H{\sc i} features at greater offsets. While most of our sources with H{\sc i} detections have similar radio luminosities (Fig. 9), we note that the two sources with just narrow absorption features near the optical spectroscopic redshift have relatively lower radio luminosities, and hence gives some support for this result. We note that it is easier to detect H{\sc i} absorption features against brighter radio galaxies, including shallow broad components with low optical depth typically offset from the spectroscopic redshift. This is seen in J181934-634548 (Fig. 4), which has the highest flux density of our sample at 1.4 GHz and has both deep and shallow H{\sc i} components. This factor hence may contribute to this finding.

\section{Summary}

We have searched for H{\sc i} absorption associated with 66 compact, core-dominated sources selected from the AT20G catalogue within the redshift range 0.04 $<$ z $<$ 0.096, with the aim to examine cold gas in the youngest and most recently triggered radio AGN within our local Universe. 

We detect four associated absorption systems in addition to that by \cite{Allison2012}, three of which were previously unknown. Two of these new absorption systems only exhibit narrow deep components close to the optical spectroscopic redshift which we attribute to gas within the galactic disc. The third contains broader, offset velocity components at a lower optical depth, which may represent disturbed gas close to the AGN. The fourth detection we make is tentative, and of a rare weak, broad absorption feature with no associated narrow component. This is the second such feature we find in our survey, with the other verified by \cite{Allison2013}. Our survey probed a source population with lower radio luminosities than many other literature surveys, reflected in the flux densities of sources with H{\sc i} detections (Fig.~9).

We find that the absorbed X-ray sources, with high column densities of elemental hydrogen, appear to correlate with galaxies with detectable amounts of H{\sc i} gas in absorption. This connection between X-ray absorption and H{\sc i} absorption is currently being investigated at higher redshifts using ASKAP (Moss et al., in prep). From the mid-infrared colour information, we find a higher detection rate in objects that host a QSO-like source (those with higher AGN gas accretion rates and a covering factor of f $\approx$ 1). This trend is also seen in \cite{Curran2016b}. 

The H{\sc i} detection rate in our sample is higher for early-type host galaxies than the late-type galaxies despite their generally higher star formation rates, suggesting we can probe different gas kinematics in various host galaxies and may not be sensitive enough to detect sufficiently lower column densities of H{\sc i} in star-forming discs. Spectra with only narrow, deep absorption features, which can be attributed to a galactic disc, are not found within the early IR type galaxy sector of the WISE colour plot, while the broader components ($\Delta{v}_{\mathrm{FWHM}}$ $>$ 100 km s$^{-1}$) are within the early IR type host sector and thought to be attributed to e.g. AGN feedback driving an outflow of gas. This shows that the mid-infrared colours can indicate the expected distribution of H{\sc i} for a given galaxy. We also find the sources with broader offset (by $>$ 100 km s$^{-1}$) components to be LERGs, and those with narrow features near the optical spectroscopic redshift in HERGs, in line with the result by \cite{Chandola2016}. 

We see mostly disturbed gas kinematics among our detections of H{\sc i} absorption in compact core radio galaxies, indicating that young or recently triggered AGN create outflows (jets) that disrupt the interstellar medium within these galaxies. We see shallow, broad components occur more often in early-type galaxies, as modelled in \cite{Curran2016}. Two of our detections against such early-type galaxies lack any narrow and deep component [here and \cite{Allison2013}], an attribute not seen in e.g. \cite{Gereb2015}. Hence, given the H{\sc i} gas kinematics, and combining with multi-wavelength information, we can better understand the dynamics of a host system to a young AGN. 

The upcoming ASKAP-FLASH survey will provide a significantly larger sample size to any previous H{\sc i} survey (an expected number of $\sim$1,000 new H{\sc i} detections in the southern sky in the redshift space 0.4 $<$ z $<$ 1.0, against $\sim$150,000 sightlines), and hence will offer a greater wealth of information about the classes of radio objects with H{\sc i} absorption systems. The expected minimum flux density in which FLASH is expected to detect H{\sc i} absorption is $\sim$ 50 mJy \citep{Morganti2015}; however, at greater redshifts than our sample (z $>$ 0.1), such sources will have a higher radio luminosity (above radio luminosity strengths of 10$^{25}$~W~Hz$^{-1}$; Fig.~9). This means that FLASH will probe more high excitation (high luminosity) objects than what we explored in our sample. Direct comparisons between results of this work and that of FLASH will need to take this into consideration.

\section*{Acknowledgements}

We thank the anonymous referee for useful comments that helped to improve the paper. The Australia Telescope Compact Array is part of the Australia Telescope National Facility which is funded by the Australian Government for operation as a National Facility managed by CSIRO. This research was conducted by the Australian Research Council Centre of Excellence for All-sky Astrophysics (CAASTRO), through project number CE110001020. This research has made use of the NASA/IPAC Extragalactic Database (NED) which is operated by the Jet Propulsion Laboratory, California Institute of Technology, under contract with the National Aeronautics and Space Administration. This research has also made use of NASA’s Astrophysics Data System Bibliographic Services. This research made use of Astropy, a community-developed core Python package for Astronomy (Astropy Collaboration, 2013).

\footnotesize{
  \bibliographystyle{mn2e}
  \bibliography{bibliography}

\begin{thebibliography}{96}
\expandafter\ifx\csname natexlab\endcsname\relax\def\natexlab#1{#1}\fi

\bibitem[{{Aditya} {et~al}\mbox{.}(2016){Aditya}, {Kanekar}, \&
  {Kurapati}}]{Aditya2016}
{Aditya} J.~N.~H.~S., {Kanekar} N., {Kurapati} S., 2016, MNRAS, 455, 4000

\bibitem[{Allison {et~al}\mbox{.}(2012{\natexlab{a}})Allison, Curran, Emonts,
  Ger\'{e}b, Mahony, Reeves, Sadler, Tanna, Whiting, \& Zwaan}]{Allison2012}
Allison J.~R. {et~al.}, 2012{\natexlab{a}}, MNRAS, 423, 2601

\bibitem[{Allison {et~al}\mbox{.}(2013)Allison, Curran, Sadler, \&
  Reeves}]{Allison2013}
Allison J.~R., Curran S.~J., Sadler E.~M., Reeves S.~N., 2013, MNRAS, 430, 157

\bibitem[{Allison {et~al}\mbox{.}(2015)Allison, Sadler, Moss, Whiting,
  Hunstead, Pracy, Curran, Croom, Glowacki, Morganti, Shabala, Zwaan, Allen,
  Amy, Axtens, Ball, Bannister, Barker, Bell, Bock, Bolton, Bowen, Boyle,
  Braun, Broadhurst, Brodrick, Brothers, Brown, Bunton, Cantrall, Chapman,
  Cheng, Chippendale, Chung, Cooray, Cornwell, DeBoer, Diamond, Edwards, Ekers,
  Feain, Ferris, Forsyth, Gough, Grancea, Gupta, Guzman, Hampson, Harvey-Smith,
  Haskins, Hay, Hayman, Heywood, Hotan, Hoyle, Humphreys, Indermuehle, Jacka,
  Jackson, Jackson, Jeganathan, Johnston, Joseph, Kendall, Kesteven, Kiraly,
  Koribalski, Leach, Lenc, Lensson, Mackay, Macleod, Marquarding, Marvil,
  McClure-Griffiths, McConnell, Mirtschin, Norris, Neuhold, Ng, O'Sullivan,
  Pathikulangara, Pearce, Phillips, Popping, Qiao, Reynolds, Roberts, Sault,
  Schinckel, Serra, Shaw, Shields, Shimwell, Storey, Sweetnam, Troup, Turner,
  Tuthill, Tzioumis, Voronkov, Westmeier, \& Wilson}]{Allison2015}
Allison J.~R. {et~al.}, 2015, MNRAS, 453, 1249

\bibitem[{Allison {et~al}\mbox{.}(2012{\natexlab{b}})Allison, Sadler, \&
  Whiting}]{Allison2012b}
Allison J. R.~A., Sadler E. M.~A., Whiting M. T.~C., 2012{\natexlab{b}}, PASA,
  29, 221

\bibitem[{Baldi \& Capetti(2009)}]{Baldi2009}
Baldi R.~D., Capetti A., 2009, A\&A, 508, 603

\bibitem[{{Carilli} {et~al}\mbox{.}(1998){Carilli}, {Menten}, {Reid}, {Rupen},
  \& {Yun}}]{cmr+98}
{Carilli} C.~L., {Menten} K.~M., {Reid} M.~J., {Rupen} M.~P., {Yun} M.~S.,
  1998, ApJ, 494, 175

\bibitem[{{Carilli} {et~al}\mbox{.}(1992){Carilli}, {Perlman}, \&
  {Stocke}}]{cps92}
{Carilli} C.~L., {Perlman} E.~S., {Stocke} J.~T., 1992, ApJ, 400, L13

\bibitem[{Catinella {et~al}\mbox{.}(2008)Catinella, Haynes, Giovanelli,
  Gardner, \& Connolly}]{Catinella2008}
Catinella B., Haynes M.~P., Giovanelli R., Gardner J.~P., Connolly A.~J., 2008,
  ApJ, 685, L13

\bibitem[{Cava {et~al}\mbox{.}(2009)Cava, Bettoni, Poggianti, Couch, Moles,
  Varela, Biviano, D'Onofrio, Dressler, Fasano, Fritz, Kj\ae~rgaard, Ramella,
  \& Valentinuzzi}]{Cava2009}
Cava A. {et~al.}, 2009, A\&A, 495, 707

\bibitem[{{Chandola} \& {Saikia}(2017)}]{Chandola2016}
{Chandola} Y., {Saikia} D.~J., 2017, MNRAS, 465, 997

\bibitem[{Chandola {et~al}\mbox{.}(2011)Chandola, Sirothia, \&
  Saikia}]{Chandola2011}
Chandola Y., Sirothia S.~K., Saikia D.~J., 2011, MNRAS, 418, 1787

\bibitem[{Chhetri {et~al}\mbox{.}(2013)Chhetri, Ekers, Jones, \&
  Ricci}]{Chhetri2013}
Chhetri R., Ekers R.~D., Jones P.~A., Ricci R., 2013, MNRAS, 434, 956

\bibitem[{Colless {et~al}\mbox{.}(2003)Colless, Peterson, Jackson, Peacock,
  Cole, Norberg, Baldry, Baugh, Bland-Hawthorn, Bridges, Cannon, Collins,
  Couch, Cross, Dalton, De~Propris, Driver, Efstathiou, Ellis, Frenk,
  Glazebrook, Lahav, Lewis, Lumsden, Maddox, Madgwick, Sutherland, \&
  Taylor}]{Colless2003}
Colless M. {et~al.}, 2003, preprint (arXiv:astro-ph/0306581)

\bibitem[{Condon {et~al}\mbox{.}(1998)Condon, Cotton, Greisen, Yin, Perley,
  Taylor, \& Broderick}]{Condon1998}
Condon J.~J., Cotton W.~D., Greisen E.~W., Yin Q.~F., Perley R.~A., Taylor
  G.~B., Broderick J.~J., 1998, AJ, 115, 1693

\bibitem[{{Curran}(2012)}]{Curran2012b}
{Curran} S.~J., 2012, ApJL, 748, L18

\bibitem[{Curran {et~al}\mbox{.}(2013{\natexlab{a}})Curran, Allison, Glowacki,
  Whiting, \& Sadler}]{Curran2013}
Curran S.~J., Allison J.~R., Glowacki M., Whiting M.~T., Sadler E.~M.,
  2013{\natexlab{a}}, MNRAS, 431, 3408

\bibitem[{{Curran} {et~al}\mbox{.}(2016{\natexlab{a}}){Curran}, {Allison},
  {Whiting}, {Sadler}, {Combes}, {Pracy}, {Bignell}, \& {Athreya}}]{caw+16}
{Curran} S.~J., {Allison} J.~R., {Whiting} M.~T., {Sadler} E.~M., {Combes} F.,
  {Pracy} M.~B., {Bignell} C., {Athreya} R., 2016{\natexlab{a}}, 457, 3666

\bibitem[{{Curran} {et~al}\mbox{.}(2016{\natexlab{b}}){Curran}, {Duchesne},
  {Divoli}, \& {Allison}}]{Curran2016}
{Curran} S.~J., {Duchesne} S.~W., {Divoli} A., {Allison} J.~R.,
  2016{\natexlab{b}}, \mnras

\bibitem[{{Curran} {et~al}\mbox{.}(2016{\natexlab{c}}){Curran}, {Reeves},
  {Allison}, \& {Sadler}}]{Curran2016b}
{Curran} S.~J., {Reeves} S.~N., {Allison} J.~R., {Sadler} E.~M.,
  2016{\natexlab{c}}, \mnras, 459, 4136

\bibitem[{Curran \& Whiting(2010)}]{cw10}
Curran S.~J., Whiting M.~T., 2010, 712, 303

\bibitem[{{Curran} \& {Whiting}(2012)}]{Curran2012}
{Curran} S.~J., {Whiting} M.~T., 2012, ApJ, 759, 117

\bibitem[{Curran {et~al}\mbox{.}(2011{\natexlab{a}})Curran, Whiting, Murphy,
  Webb, Bignell, Polatidis, Wiklind, Francis, \& Langston}]{cwm+10}
Curran S.~J. {et~al.}, 2011{\natexlab{a}}, MNRAS, 413, 1165

\bibitem[{Curran {et~al}\mbox{.}(2013{\natexlab{b}})Curran, Whiting, Sadler, \&
  Bignell}]{cwsb12}
Curran S.~J., Whiting M.~T., Sadler E.~M., Bignell C., 2013{\natexlab{b}}, 428,
  2053

\bibitem[{Curran {et~al}\mbox{.}(2013{\natexlab{c}})Curran, Whiting, Tanna,
  Sadler, Pracy, \& Athreya}]{cwt+12}
Curran S.~J., Whiting M.~T., Tanna A., Sadler E.~M., Pracy M.~B., Athreya R.,
  2013{\natexlab{c}}, MNRAS, 429, 3402

\bibitem[{Curran {et~al}\mbox{.}(2011{\natexlab{b}})Curran, Whiting, Webb, \&
  Athreya}]{cwwa11}
Curran S.~J., Whiting M.~T., Webb J.~K., Athreya A., 2011{\natexlab{b}}, MNRAS,
  414, L26

\bibitem[{Curran {et~al}\mbox{.}(2008)Curran, Whiting, Wiklind, Webb, Murphy,
  \& Purcell}]{cww+08}
Curran S.~J., Whiting M.~T., Wiklind T., Webb J.~K., Murphy M.~T., Purcell
  C.~R., 2008, 391, 765

\bibitem[{{de Vaucouleurs} {et~al}\mbox{.}(1991){de Vaucouleurs}, {de
  Vaucouleurs}, {Corwin}, {Buta}, {Paturel}, \&
  {Fouqu{\'e}}}]{deVaucouleurs1991}
{de Vaucouleurs} G., {de Vaucouleurs} A., {Corwin}, Jr. H.~G., {Buta} R.~J.,
  {Paturel} G., {Fouqu{\'e}} P., 1991, {Third Reference Cat. of Bright
  Galaxies. Volume I: Explanations and references. Volume II: Data for galaxies
  between 0$^{h}$ and 12$^{h}$. Volume III: Data for galaxies between 12$^{h}$
  and 24$^{h}$.}

\bibitem[{Deboer {et~al}\mbox{.}(2009)Deboer, Gough, Bunton, Cornwell,
  Beresford, Johnston, Feain, Schinckel, Jackson, Kesteven, Chippendale,
  Hampson, Sullivan, Hay, Jacka, Sweetnam, Storey, Ball, \& Boyle}]{Deboer2009}
Deboer B. D.~R. {et~al.}, 2009, Proc. IEEE, 97, 1507

\bibitem[{{Donoso} {et~al}\mbox{.}(2012){Donoso}, {Yan}, {Tsai}, {Eisenhardt},
  {Stern}, {Assef}, {Leisawitz}, {Jarrett}, \& {Stanford}}]{Donoso2012}
{Donoso} E. {et~al.}, 2012, \apj, 748, 80

\bibitem[{Drinkwater {et~al}\mbox{.}(2001)Drinkwater, Gregg, Holman, \&
  Brown}]{Drinkwater2001}
Drinkwater M.~J., Gregg M.~D., Holman B.~a., Brown M. J.~I., 2001, MNRAS, 326,
  1076

\bibitem[{Emonts {et~al}\mbox{.}(2010)Emonts, Morganti, Struve, Oosterloo, van
  Moorsel, Tadhunter, van~der Hulst, Brogt, Holt, \& Mirabal}]{Emonts2010}
Emonts B.~H.~C. {et~al.}, 2010, MNRAS, 406, 987

\bibitem[{Fanaroff \& Riley(1974)}]{Fanaroff1974}
Fanaroff B.~L., Riley J.~M., 1974, MNRAS, 167, 31P

\bibitem[{Fanti {et~al}\mbox{.}(1995)Fanti, Fanti, Dallacasa, Schilizzi,
  Spencer, \& Stanghellini}]{Fanti1995}
Fanti C., Fanti R., Dallacasa D., Schilizzi R.~T., Spencer R.~E., Stanghellini
  C., 1995, A\&A, 302, 317

\bibitem[{Ferris \& Wilson(2002)}]{Ferris2002}
Ferris R., Wilson W., 2002, in URSI XXVIIth General Assembly, poster 1629.

\bibitem[{{Fossati} {et~al}\mbox{.}(1998){Fossati}, {Maraschi}, {Celotti},
  {Comastri}, \& {Ghisellini}}]{Fossati1998}
{Fossati} G., {Maraschi} L., {Celotti} A., {Comastri} A., {Ghisellini} G.,
  1998, MNRAS, 299, 433

\bibitem[{Freudling {et~al}\mbox{.}(2011)Freudling, Staveley-Smith, Catinella,
  Minchin, Calabretta, Momjian, Zwaan, Meyer, \& O'Neil}]{Freudling2011}
Freudling W. {et~al.}, 2011, ApJ, 727, 40

\bibitem[{Ger\'{e}b {et~al}\mbox{.}(2015)Ger\'{e}b, Maccagni, Morganti, \&
  Oosterloo}]{Gereb2015}
Ger\'{e}b K., Maccagni F.~M., Morganti R., Oosterloo T.~A., 2015, A\&A, 575,
  A44

\bibitem[{{Ger{\'e}b} {et~al}\mbox{.}(2015){Ger{\'e}b}, {Maccagni}, {Morganti},
  \& {Oosterloo}}]{gmmo14}
{Ger{\'e}b} K., {Maccagni} F.~M., {Morganti} R., {Oosterloo} T.~A., 2015, 575,
  44

\bibitem[{Ger\'{e}b {et~al}\mbox{.}(2014)Ger\'{e}b, Morganti, \&
  Oosterloo}]{Gereb2014}
Ger\'{e}b K., Morganti R., Oosterloo T.~A., 2014, A\&A, 569, 1

\bibitem[{Ghisellini(2011)}]{Ghisellini2011}
Ghisellini G., 2011, in AIP Conf. Ser. 25th Texas Symp. on Relativ. Astrophys.,
  Aharonian F.~A., Hofmann W., Rieger F.~M., eds., Vol. 1381, Am. Inst. Phys.,
  New York, p. 180

\bibitem[{{Giovanelli} \& {Haynes}(2016)}]{Giovanelli2016}
{Giovanelli} R., {Haynes} M.~P., 2016, A\&AR, 24, 1

\bibitem[{{Grasha} \& {Darling}(2011)}]{gd11}
{Grasha} K., {Darling} J., 2011, in American Astronomical Society Meeting
  Abstracts, Vol.~43, p. 345.02

\bibitem[{Gupta {et~al}\mbox{.}(2006)Gupta, Salter, Saikia, Ghosh, \&
  Jeyakumar}]{Gupta2006}
Gupta N., Salter C.~J., Saikia D.~J., Ghosh T., Jeyakumar S., 2006, MNRAS, 373,
  972

\bibitem[{Hambly {et~al}\mbox{.}(2001)Hambly, MacGillivray, Read, Tritton,
  Thomson, Kelly, Morgan, Smith, Driver, Williamson, Parker, Hawkins, Williams,
  \& Lawrence}]{Hambly2001}
Hambly N. {et~al.}, 2001, MNRAS, 326, 1279

\bibitem[{Hancock {et~al}\mbox{.}(2009)Hancock, Tingay, Sadler, Phillips, \&
  Deller}]{Hancock2009}
Hancock P., Tingay S., Sadler E., Phillips C., Deller A., 2009, MNRAS, 397,
  2030

\bibitem[{{Heckman} \& {Best}(2014)}]{Heckman2014}
{Heckman} T.~M., {Best} P.~N., 2014, ARAA, 52, 589

\bibitem[{Hoessel {et~al}\mbox{.}(1980)Hoessel, Gunn, \& Thuan}]{Hoessel1980}
Hoessel J.~G., Gunn J.~E., Thuan T.~X., 1980, ApJ, 241, 486

\bibitem[{Jauncey {et~al}\mbox{.}(1978)Jauncey, Wright, Peterson, \&
  Condon}]{Jauncey1978}
Jauncey D.~L., Wright A.~E., Peterson B.~A., Condon J.~J., 1978, ApJ, 219, L1

\bibitem[{Johnston {et~al}\mbox{.}(2009)Johnston, Feain, \&
  Gupta}]{Johnston2009}
Johnston S., Feain I.~J., Gupta N., 2009, The Low-Frequency Radio Universe ASP
  Conference Series, 407, 446

\bibitem[{Jones {et~al}\mbox{.}(2009)Jones, Read, Saunders, Colless, Jarrett,
  Parker, Fairall, Mauch, Sadler, Watson, Burton, Campbell, Cass, Croom, Dawe,
  Fiegert, Frankcombe, Hartley, Huchra, James, Kirby, Lahav, Lucey, Mamon,
  Moore, Peterson, Prior, Proust, Russell, Safouris, Wakamatsu, Westra, \&
  Williams}]{Jones2009}
Jones D.~H. {et~al.}, 2009, MNRAS, 399, 683

\bibitem[{Katgert {et~al}\mbox{.}(1998)Katgert, Mazure, den Hartog, Adami,
  Biviano, \& Perea}]{Katgert1998}
Katgert P., Mazure A., den Hartog R., Adami C., Biviano A., Perea J., 1998,
  A\&ASS, 129, 399

\bibitem[{Ledlow {et~al}\mbox{.}(2001)Ledlow, Owen, Yun, \& Hill}]{Ledlow2001}
Ledlow M., Owen F., Yun M., Hill J., 2001, ApJ, 552, 120

\bibitem[{Ledlow {et~al}\mbox{.}(1998)Ledlow, Owen, \& Keel}]{Ledlow1998}
Ledlow M.~J., Owen F.~N., Keel W.~C., 1998, ApJ, 495, 227

\bibitem[{{Maccagni} {et~al}\mbox{.}(2014){Maccagni}, {Morganti}, {Oosterloo},
  \& {Mahony}}]{Maccagni2014}
{Maccagni} F.~M., {Morganti} R., {Oosterloo} T.~A., {Mahony} E.~K., 2014, A\&A,
  571, A67

\bibitem[{{Maccagni} {et~al}\mbox{.}(2016){Maccagni}, {Santoro}, {Morganti},
  {Oosterloo}, {Oonk}, \& {Emonts}}]{Maccagni2016}
{Maccagni} F.~M., {Santoro} F., {Morganti} R., {Oosterloo} T.~A., {Oonk}
  J.~B.~R., {Emonts} B.~H.~C., 2016, Astronomische Nachrichten, 337, 154

\bibitem[{Mahony {et~al}\mbox{.}(2011)Mahony, Sadler, Croom, Ekers, Bannister,
  Chhetri, Hancock, Johnston, Massardi, \& Murphy}]{Mahony2011}
Mahony E.~K. {et~al.}, 2011, MNRAS, 417, 2651

\bibitem[{Mauch {et~al}\mbox{.}(2003)Mauch, Murphy, Buttery, Curran, Hunstead,
  Piestrzynski, Robertson, \& Sadler}]{Mauch2003}
Mauch T., Murphy T., Buttery H.~J., Curran J., Hunstead R.~W., Piestrzynski B.,
  Robertson J.~G., Sadler E.~M., 2003, MNRAS, 342, 1117

\bibitem[{{Mirabel}(1989)}]{mir89}
{Mirabel} I.~F., 1989, ApJ, 340, L13

\bibitem[{Morganti {et~al}\mbox{.}(2011)Morganti, Holt, Tadhunter,
  Ramos~Almeida, Dicken, Inskip, Oosterloo, \& Tzioumis}]{Morganti2011}
Morganti R., Holt J., Tadhunter C., Ramos~Almeida C., Dicken D., Inskip K.,
  Oosterloo T., Tzioumis T., 2011, A\&A, 535, A97

\bibitem[{Morganti {et~al}\mbox{.}(2001)Morganti, Oosterloo, Tadhunter, van
  Moorsel, Killeen, \& Wills}]{Morganti2001}
Morganti R., Oosterloo T.~A., Tadhunter C.~N., van Moorsel G., Killeen N.,
  Wills K.~A., 2001, MNRAS, 323, 331

\bibitem[{{Morganti} {et~al}\mbox{.}(2015){Morganti}, {Sadler}, \&
  {Curran}}]{Morganti2015}
{Morganti} R., {Sadler} E.~M., {Curran} S., 2015, Advancing Astrophysics with
  the Square Kilometre Array (AASKA14), 134

\bibitem[{Morganti {et~al}\mbox{.}(2005)Morganti, Tadhunter, \&
  Oosterloo}]{Morganti2005}
Morganti R., Tadhunter C.~N., Oosterloo T.~A., 2005, A\&A, 444, L9

\bibitem[{Murphy {et~al}\mbox{.}(2010)Murphy, Sadler, Ekers, Massardi, Hancock,
  Mahony, Ricci, Burke-Spolaor, Calabretta, \& Chhetri}]{Murphy2010}
Murphy T. {et~al.}, 2010, MNRAS, 402, 2403

\bibitem[{Orienti {et~al}\mbox{.}(2006)Orienti, Morganti, \&
  Dallacasa}]{Orienti2006}
Orienti M., Morganti R., Dallacasa D., 2006, A\&A, 457, 531

\bibitem[{Ostorero {et~al}\mbox{.}(2010)Ostorero, Moderski, Diaferio, Kowalska,
  Cheung, Kataoka, Begelman, \& Wagner}]{Ostorero2010}
Ostorero L., Moderski R., Diaferio a., Kowalska I., Cheung C.~C., Kataoka J.,
  Begelman M.~C., Wagner S.~J., 2010, ApJ, 715, 1071

\bibitem[{{Ostorero} {et~al}\mbox{.}(2016){Ostorero}, {Morganti}, {Diaferio},
  {Siemiginowska}, {Stawarz}, {Moderski}, \& {Labiano}}]{Ostorero2016}
{Ostorero} L., {Morganti} R., {Diaferio} A., {Siemiginowska} A., {Stawarz}
  {\L}., {Moderski} R., {Labiano} A., 2016, Astronomische Nachrichten, 337, 148

\bibitem[{Owsianik \& Conway(1998)}]{Owsianik1998}
Owsianik I., Conway J.~E., 1998, A\&A, 337, 69

\bibitem[{{Pace} \& {Salim}(2016)}]{Pace2016}
{Pace} C., {Salim} S., 2016, ApJ, 818, 65

\bibitem[{Peterson {et~al}\mbox{.}(1979)Peterson, Wright, Jauncey, \&
  Condon}]{Peterson1979}
Peterson B.~A., Wright A.~E., Jauncey D.~L., Condon J.~J., 1979, ApJ, 232, 400

\bibitem[{Pihlstr\"{o}m {et~al}\mbox{.}(2003)Pihlstr\"{o}m, Conway, \&
  Vermeulen}]{Pihlstrom2003}
Pihlstr\"{o}m Y.~M., Conway J.~E., Vermeulen R.~C., 2003, A\&A, 404, 871

\bibitem[{Postman \& Lauer(1995)}]{Postman1995}
Postman M., Lauer T.~R., 1995, ApJ, 440, 28

\bibitem[{Readhead {et~al}\mbox{.}(1996)Readhead, Taylor, Xu, Pearson,
  Wilkinson, \& Polatidis}]{Readhead1996}
Readhead A.~C.~S., Taylor G.~B., Xu W., Pearson T.~J., Wilkinson P.~N.,
  Polatidis A.~G., 1996, ApJ, 460, 612

\bibitem[{Reynolds(1994)}]{Reynolds1994}
Reynolds J., 1994, A revised flux scale for the at compact array. Tech. Rep. AT
  Technical Document AT/39.3/040.)

\bibitem[{Rosen {et~al}\mbox{.}(2015)Rosen, Webb, Watson, Ballet, Barret,
  Braito, Carrera, Ceballos, Coriat, {Della Ceca}, Denkinson, Esquej, Farrell,
  Freyberg, Gris\'{e}, Guillout, Heil, Law-Green, Lamer, Lin, Martino, Michel,
  Motch, Gomez-Moran, Page, Page, Page, Pakull, Pye, Read, Rodriguez, Sakano,
  Saxton, Schwope, Scott, Sturm, Traulsen, Yershov, \& Zolotukhin}]{Rosen2015}
Rosen S.~R. {et~al.}, 2015, preprint (arxiv:1504.07051)

\bibitem[{Sadler {et~al}\mbox{.}(2014)Sadler, Ekers, Mahony, Mauch, \&
  Murphy}]{Sadler2014}
Sadler E.~M., Ekers R.~D., Mahony E.~K., Mauch T., Murphy T., 2014, MNRAS, 438,
  796

\bibitem[{{Salter} {et~al}\mbox{.}(2010){Salter}, {Saikia}, {Minchin}, {Ghosh},
  \& {Chandola}}]{ssm+10}
{Salter} C.~J., {Saikia} D.~J., {Minchin} R., {Ghosh} T., {Chandola} Y., 2010,
  ApJ, 715, L117

\bibitem[{Sault {et~al}\mbox{.}(1995)Sault, Teuben, \& Wright}]{Sault1995}
Sault R., Teuben P., Wright M., 1995, RA Shaw, HE Payne, \& JJE Hayes (San
  Francisco, CA: ASP), 433

\bibitem[{{Schaye}(2001)}]{Schaye2001}
{Schaye} J., 2001, \apjl, 562, L95

\bibitem[{Schinckel {et~al}\mbox{.}(2012)Schinckel, Bunton, Cornwell, Feain, \&
  Hay}]{Schinckel2012}
Schinckel A.~E., Bunton J.~D., Cornwell T.~J., Feain I., Hay S.~G., 2012, Proc.
  SPIE 8444, Ground-based and Airborne Telescopes IV, 8444, 84442A

\bibitem[{Siemiginowska(2009)}]{Siemiginowska2009}
Siemiginowska A., 2009, Astron. Nachr., 330, 264

\bibitem[{Siemiginowska {et~al}\mbox{.}(2008)Siemiginowska, LaMassa, Aldcroft,
  Bechtold, \& Elvis}]{Siemiginowska2008}
Siemiginowska A., LaMassa S., Aldcroft T.~L., Bechtold J., Elvis M., 2008, ApJ,
  684, 811

\bibitem[{{Simpson} {et~al}\mbox{.}(1993){Simpson}, {Clements}, {Rawlings}, \&
  {Ward}}]{Simpson1993}
{Simpson} C., {Clements} D.~L., {Rawlings} S., {Ward} M., 1993, MNRAS, 262, 889

\bibitem[{Skrutskie {et~al}\mbox{.}(2006)Skrutskie, Cutri, Stiening, Weinberg,
  Schneider, Carpenter, Beichman, Capps, Chester, Elias, Huchra, Liebert,
  Lonsdale, Monet, Price, Seitzer, Jarrett, Kirkpatrick, Gizis, Howard, Evans,
  Fowler, Fullmer, Hurt, Light, Kopan, Marsh, McCallon, Tam, Van~Dyk, \&
  Wheelock}]{Skrutskie2006}
Skrutskie M.~F. {et~al.}, 2006, AJ, 131, 1163

\bibitem[{Smith {et~al}\mbox{.}(2004)Smith, Hudson, Nelan, Moore, Quinney,
  Wegner, Lucey, Davies, Malecki, Schade, \& Suntzeff}]{Smith2004}
Smith R.~J. {et~al.}, 2004, AJ, 128, 1558

\bibitem[{Tengstrand {et~al}\mbox{.}(2009)Tengstrand, Guainazzi, Siemiginowska,
  {Fonseca Bonilla}, Labiano, Worrall, Grandi, \& Piconcelli}]{Tengstrand2009}
Tengstrand O., Guainazzi M., Siemiginowska A., {Fonseca Bonilla} N., Labiano
  A., Worrall D.~M., Grandi P., Piconcelli E., 2009, A\&A, 501, 89

\bibitem[{Uson {et~al}\mbox{.}(1991)Uson, Bagri, \& Cornwell}]{ubc91}
Uson J.~M., Bagri D.~S., Cornwell T.~J., 1991, PRL, 67, 3328

\bibitem[{van Gorkom {et~al}\mbox{.}(1989)van Gorkom, Knapp, Ekers, Ekers,
  Laing, \& Polk}]{VanGorkom1989}
van Gorkom J.~H., Knapp G.~R., Ekers R.~D., Ekers D.~D., Laing R.~A., Polk
  K.~S., 1989, AJ, 97, 708

\bibitem[{Verheijen {et~al}\mbox{.}(2007)Verheijen, van Gorkom, Szomoru,
  Dwarakanath, Poggianti, \& Schiminovich}]{Verheijen2007}
Verheijen M., van Gorkom J.~H., Szomoru a., Dwarakanath K.~S., Poggianti B.~M.,
  Schiminovich D., 2007, ApJ, 668, L9

\bibitem[{Vermeulen {et~al}\mbox{.}(2003)Vermeulen, Pihlstr\"{o}m, Tschager,
  de~Vries, Conway, Barthel, Baum, Braun, Bremer, \& Miley}]{Vermeulen2003}
Vermeulen R.~C. {et~al.}, 2003, A\&A, 404, 861

\bibitem[{Vettolani {et~al}\mbox{.}(1989)Vettolani, Cappi, Chincarini, Focardi,
  Garilli, Gregorini, \& Maccagni}]{Vettolani1989}
Vettolani G., Cappi A., Chincarini G., Focardi P., Garilli B., Gregorini L.,
  Maccagni D., 1989, A\&AS, 79, 147

\bibitem[{Wagner {et~al}\mbox{.}(2012)Wagner, Bicknell, \&
  Umemura}]{Wagner2012}
Wagner A.~Y., Bicknell G.~V., Umemura M., 2012, ApJ, 757, 136

\bibitem[{Wills \& Wills(1976)}]{Wills1976}
Wills D., Wills B.~J., 1976, ApJS, 31, 143

\bibitem[{Wolfe \& Burbidge(1975)}]{Wolfe1975}
Wolfe A., Burbidge G., 1975, ApJ, 200, 548

\bibitem[{Wright {et~al}\mbox{.}(2010)Wright, Eisenhardt, Mainzer, Ressler,
  Cutri, Jarrett, Kirkpatrick, Padgett, McMillan, Skrutskie, Stanford, Cohen,
  Walker, Mather, Leisawitz, III, McLean, Benford, Lonsdale, Blain, Mendez,
  Irace, Duval, Liu, Royer, Heinrichsen, Howard, Shannon, Kendall, Walsh,
  Larsen, Cardon, Schick, Schwalm, Abid, Fabinsky, Naes, \& Tsai}]{Wright2010}
Wright E.~L. {et~al.}, 2010, AJ, 140, 1868

\bibitem[{Yan {et~al}\mbox{.}(2016)Yan, Stocke, Darling, Momjian, Sharma, \&
  Kanekar}]{ysd+16}
Yan T., Stocke J.~T., Darling J.~K., Momjian E., Sharma S., Kanekar N., 2016,
  preprint (arXiv:1512.07707)

\end{thebibliography}
}
\label{lastpage}

\end{document}